\documentclass[manuscript]{aastex}

\newcommand{\myemail}{demello@on.br}
\newcommand{\CC}{C$_2$}

\slugcomment{To be published on Astrophysical Journal Supplements.}

\shorttitle{NSCC - A New Scheme of Classification of C-rich Stars}
\shortauthors{De Mello et al}

\begin{document}


\title{NSCC - A New Scheme of Classification of C-rich Stars\\
		Devised from Optical and Infrared Observations}

\author{A. B. De Mello}
\affil{Observat\'orio Nacional - MCT, Brazil}
\email{\myemail}

\author{S. Lorenz-Martins}
\affil{Observat\'orio do Valongo, Universidade Federal do Rio de Janeiro, Brazil}

\author{F. X. de Ara\'ujo}
\affil{Observat\'orio Nacional - MCT, Brazil}

\author{C. Bastos Pereira}
\affil{Observat\'orio Nacional - MCT, Brazil}

\and

\author{S. J. Codina Landaberry}
\affil{Observat\'orio Nacional - MCT, Brazil}




\begin{abstract}
A new classification system for carbon-rich stars is presented based on an analysis of 51 AGB carbon stars through the most relevant classifying indices available \citep{Keenan93,Morgan04}. The extension incorporated, that also represents the major advantage of this new system, is the combination of the usual optical indices that describe the photospheres of the objects, with new infrared ones, which allow an interpretation of the circumstellar environment of the carbon-rich stars. This new system is presented with the usual spectral subclasses and \CC -, j-, MS- and temperature indices,  and also with the new SiC- (SiC/C.A. abundance estimation) and $\tau$- (opacity) indices. The values for the infrared indices were carried out through a Monte Carlo simulation of the radiative transfer in the circumstellar envelopes of the stars. The full set of indices, when applied to our sample, resulted in a more efficient system of classification, since an examination in a wide spectral range allows us to obtain a complete scenario for carbon stars. \end{abstract}
 

\section{Introduction}
The asymptotic giant branch stars with a ratio C$/$O $> 1$ have their optical spectra ruled by bands of carbon compounds, which obscure many atomic features. The green-red optical spectrum is dominated by Swan bands $^{12}$C$^{12}$C, Red System bands $^{12}$C$^{14}$N and sometimes present isotopic bands, e.g. $^{13}$C$^{12}$C and $^{13}$C$^{14}$N. As a result, a classification carried out through the optical atomic data on carbon stars is troublesome. During their ascent on the AGB phase, the mass loss from the star creates a circumstellar envelope of gas and dust. The compounds of this shell have their maximum emission on the infrared spectral region. The main infrared feature for carbon stars is the 11.3$\mu$m emission due to the presence of SiC grains, which also represent an evidence of a C-rich dust envelope. Spitzer IRS spectra also confirmed two absorptions at 7 and 13.7 $\mu$m, whose origin is still not well understood \citep{speck06}. 

The first classification system for carbon-rich stars was the Henry Draper Catalogue \citep{HD}, in which the stars were presented in two spectral classes: type R and N, that were also divided into temperature subclasses. These subclasses were based on lines in the blue spectral region, that, although being a good source of information to determine the temperature classes of G and K stars, were not an appropriate choice for the cold carbon stars. Therefore, the distinction between an R8, the redder R-type stars, from an Na, the most blue N-type, was not obvious. \citet{MK41} decided to rearrange all the carbon-rich stars in another classification scheme that presented a single temperature sequence. The numerical temperature type was, at this time, determined based on the atomic and molecular structures, considered more susceptible to temperature variations. They also added a \CC -index, based on the strength of the \CC~bands. Many attempts were made to improve this MK system, while keeping its basic structure. \citet{Yama72,Yama75} with the C-Classification System listed in his tables a very large number of stars with the two main parameters from the MK System, and additional intensity indices of several other atomic and molecular characteristics from carbon stars. The result of this attempt, was a detailed notation for carbon-rich objects, but it was not very practical or compact.

Later, \citet{Keenan93} pointed out several reasons for replacing this old C-Classification System, since much importance was given to the Na D-line in determining the numerical indices of temperature. The assumption that the Na D-line could be a good tracer of temperature in this case, however, was revealed to be quite unfortunate, specially in the case of N-type stars, as they have an enormous molecular opacity in the same spectral region due to the (7,2) CN band. Furthermore, the N-type and the R-type stars, in fact, describe two different populations that should not be classified under the same spectral class. The new classification system proposed by Keenan93, the MK~Revised System, re-established the spectral subclasses for the carbon rich objects and the temperature indices based on infrared intensities. Additionally, this Revised MK System listed four abundance indices from \citet{Yama72}: the intensity of the \CC~band, the isotopic carbon ratio, the Si\CC\ band and the CH band strength. These indices and system are the most widely accepted and used nowadays in the study of carbon stars \citep{Morgan03,Morgan04,deroo07}. 

Concerning the circumstellar analysis, \citet{sloan98} presented the first classification based on an infrared study of the spectra of carbon and oxygen-rich AGB stars. Qualitative classes for the circumstellar material were settled through two indices: one chemical and other a strength indicator of the characteristic structure of the emission spectrum.

In the present paper we present a new scheme of classification for C-rich AGB stars (hereafter NSCC) that includes a notation conceived through a wide analysis of the main spectroscopic features of these carbon-rich objects with circumstellar material in its surroundings. The main difference between this scheme and the others available is that, instead of using a single region of the spectrum, we suggest a classification system based on a large wavelength range, from the blue-optical to the mid-infrared. Thus, this notation draws a more complete scenario of each star. The parameters used in this extended classification system were devised from the analysis of a sample of 51 observed AGB carbon stars.

Section \ref{obs} describes the optical observations and other infrared data used and section \ref{optical} and \ref{ir} details how indices and parameters of classification were obtained. While, section \ref{optical} explains the optical indices, section \ref{ir} contains the infrared ones. Section \ref{disc} shows an analysis of the classification indices, a discussion of possible evolutionary sequences for the AGB carbon stars and also an atlas of optical spectra and infrared radiative transfer models of the stars studied in this work. The overall results are discussed in section \ref{conc}.

\section{Observations and Other Data}
\label{obs}
Most of the stars observed were selected because either they have a dubious or incomplete classification or doesn't even have one defined. Our intent was to study these targets in a large spectral range and, therefore, stars with data on the Atlas of Low Resolution Infrared Astronomical Satellite Spectra \citep[hereafter LRS-IRAS]{iras} \defcitealias{iras}{LRS-IRAS} or on the Short-Wave Spectrometer of the Infrared Space Observatory \citep[hereafter SWS-ISO]{iso}, \defcitealias{iso}{SWS-ISO} were preferred. In few cases, the selected target lacked any available infrared data. We aim to observe those in this spectral region to complete their classification, in the next years.

Another 6 targets, W CMa, RY Mon, V Hya, TW Oph, RY Hya and BE CMa, were used as primary standards in optical range and for the temperature index. These enable a calibration of this new scheme here presented with the old ones selected from \citet{Yama72,Yama75}, \citet{barn96} and \citet{Keenan93}. Although some other targets from our sample have been studied by these authors, the 6 standards selected either figure simultaneously in more than one of the three catalogues or were well studied by others.

The sample comprises 51 stars which can be seen from the South Hemisphere, that are bright enough and not too cold to be observed with an appropriate signal-to-noise ratio. Most objects of the sample, 42 targets, were optically observed at the 1.52m telescope at the European Southern Observatory, La Silla in several observational runs between 1996\ and 1998. It were used the Boller and Chivens spectrograph facility with a Cassegrain f/14.9 focus. Some of the gratings used at that time are no longer operational. The \#11 grating was centered on 6000\AA\, with a dispersion of 66\AA/mm. The \#23 had the central wavelength set to 5600\AA\ with 126\AA/mm dispersion. Grating \#26 was centered on 5800\AA\ and 5900\AA\ with 66\AA/mm dispersion. Finally, the \#32 grating had the central wavelength at 5000\AA\ and 23\AA/mm dispersion. The higher resolution was achieved with the \#32 with 0.48\AA/pix, followed by the \#11 and \#26 with a 0.99\AA/pix and the \#23, with 1.89\AA/pix. The spectral ranges and dispersions were selected in order to obtain as much details as possible in low resolution spectra. Several objects were observed with at least two different gratings and, of those, some were observed in four.

Eight more stars were observed with the 1.60m telescope and the Coud\'{e} spectrograph at the Observat\'{o}rio do Pico dos Dias, Itajub\'{a} - Brazil, during 2006 and 2007. We used the CCD \#098 which provides a low resolution spectra of 0.25\AA/mm. To cover the same spectral region of the ESO spectra, two regions were observed: $\lambda$\ 4300\AA\ - $\lambda$\ 5300\AA\ and $\lambda$~5700\AA\ - $\lambda$\ 6300\AA. 

All optical spectroscopic data were post-processed with the Image Reduction and Analysis Facility (IRAF), including wavelength and absolute flux calibration. For the wavelength calibration, He-Ar and Th-Ar lamps images were used as reference, which were also observed at La Silla and Pico dos Dias, respectively.  Several spectrophotometric standards, such as HR1541, HR3454, HR4468 and HR5501, were observed to allow flux calibration. Both calibration tasks are critical for all the measurements in this work, and these data reduction enabled a detailed analysis on the optical spectral features of each star of the sample.

The infrared spectra used were taken from \citetalias{iso} and \citetalias{iras} databases. These two catalogues together give a spectral range from 2.4 to 45 $\mu$m. The former provides a better resolution (circa 1500 - 2000) than the latter (circa 20 - 60), as well as a wider spectral range. Unfortunately, only two objects from our sample have an \citetalias{iso} spectra. For the other stars we had to rely only on the \citetalias{iras} spectra.

\section{Optical Classification}
\label{optical}
The notation adopted in this work for the optical properties is the one described on the MK Revised System \citep{Keenan93} with some slight improvements. The enhancement at the notation is simple and self-explanatory, and its four indices allow a better sketching of the scenario of an AGB carbon star photosphere. The luminosity classes were not applied, as they are not well defined for most of the objects. As usual, an uncertainty character (:) was added, and it suits indices that were calculated with a poor quality spectrum. 

Most optical parameters are measured relative to the continua, but unlike G- and K-type stars, to establish a continuum for an AGB carbon-rich star is not a trivial task, due to the high opacities on its optical spectrum. Therefore, local continua or pseudo-continua were used instead, defined by regions where \CC\ e CN absorption bands are weak. As a result, some uncertainties to the parameters are added, that are not directly measurable. Nevertheless, all the parameters used to calculate our indices were obtained by averaging measurements from multiple observations of each star, when it was possible. At least two -- and in several cases four -- different spectra were used in this average, decreasing the effects of uncertainties in the measurements.

\begin{figure}
\begin{center}
\plottwo{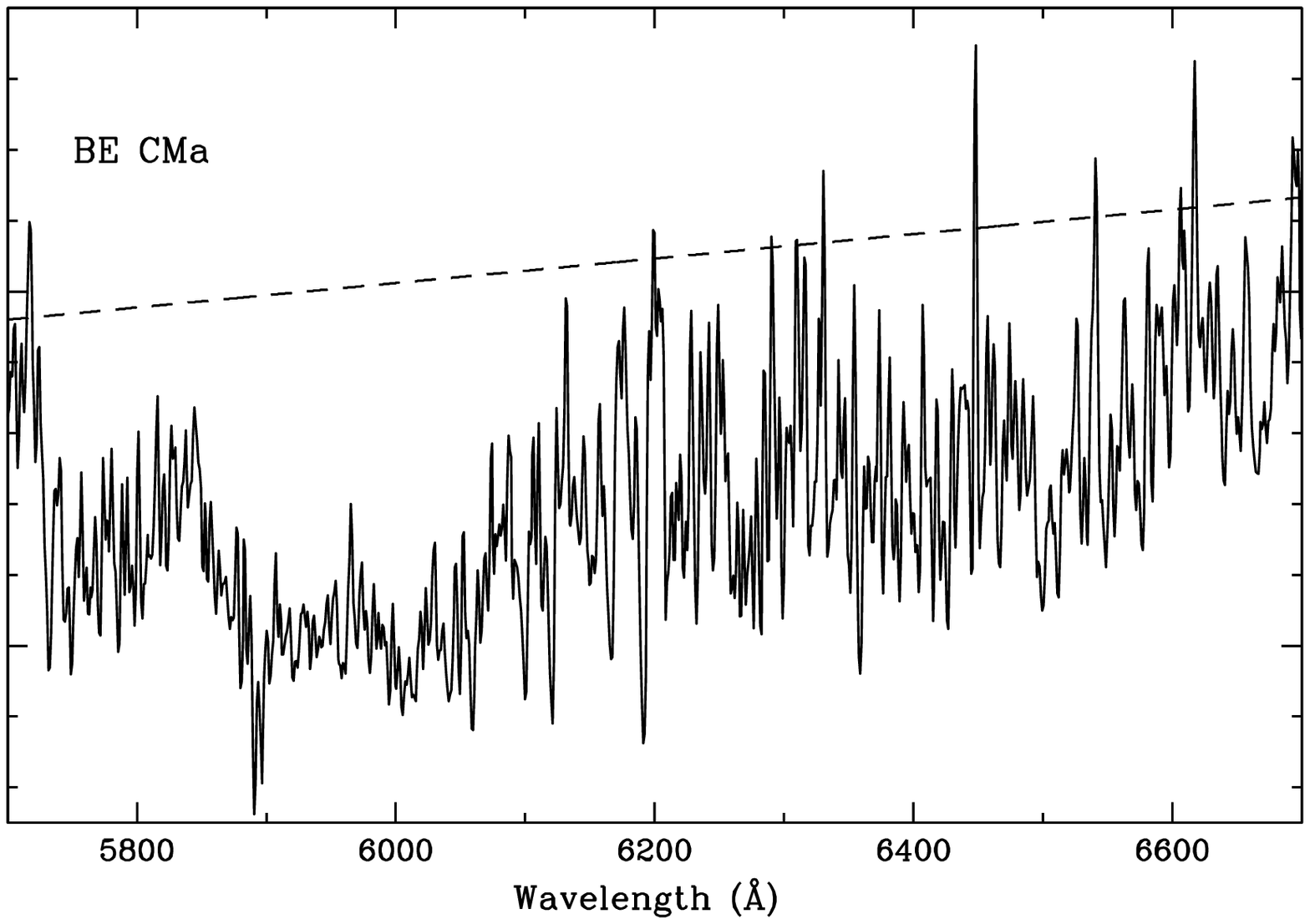}{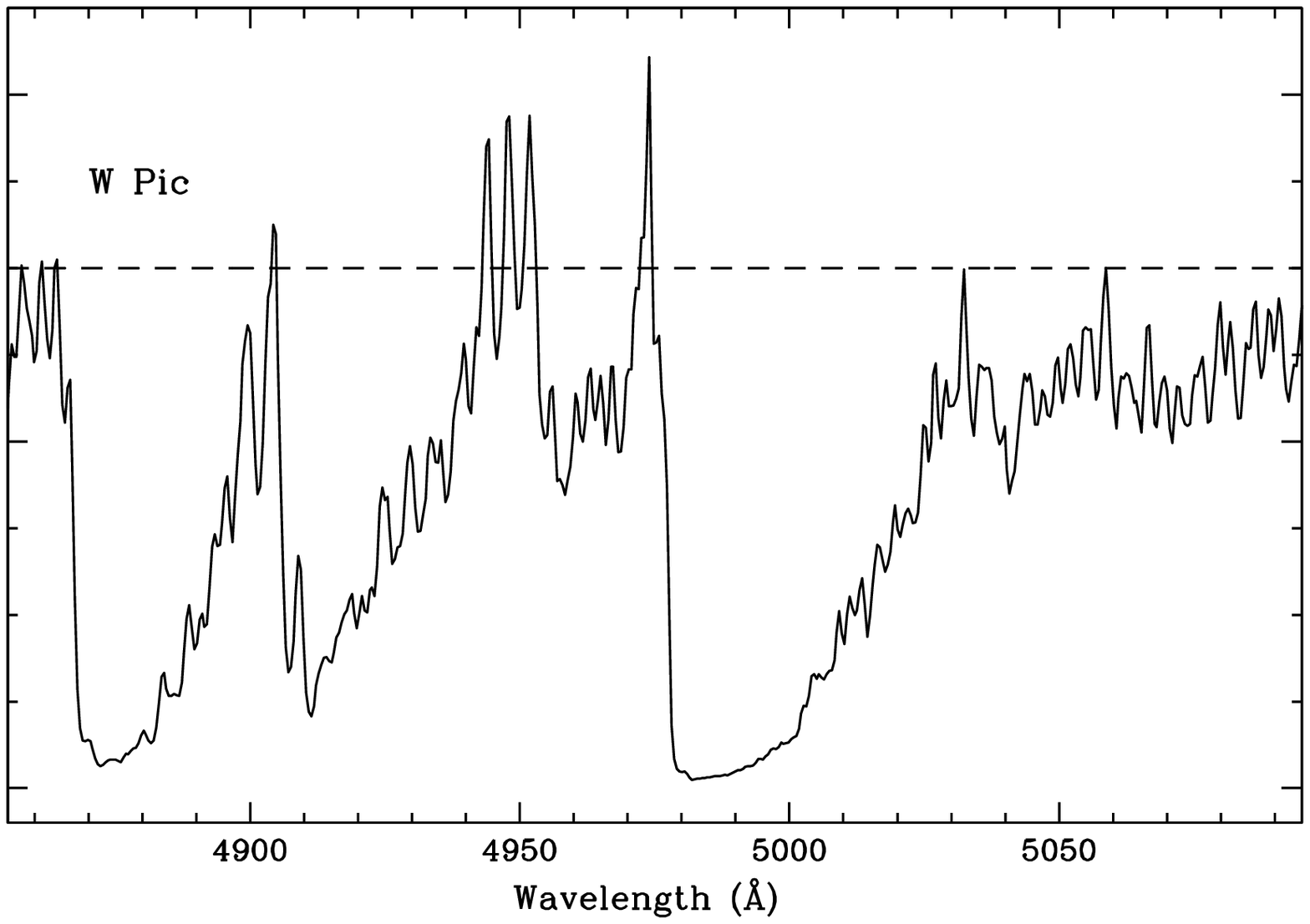}
\caption{On the left is the pseudo-continuum used as reference for \CC- and j-index parameters, set between maxima $\lambda$ 5722\AA, $\lambda$\ 6202\AA\ and $\lambda$\ 6620\AA. On the right is the one defined between maxima $\lambda$\ 4962\AA\ and $\lambda$\ 5030\AA\ used in the measure of the MS-index parameter. \label{continuum}}
\end{center}
\end{figure}

Only the three first indices, \CC -, j- and MS-index, had their parameters measured relative to the continua, and for those, two pseudo-continua were drawn: one for the first two indices and another for the later. The first pseudo-continuum was defined as a legendre function fit between maxima at $\lambda$~5722\AA, $\lambda$\ 6202\AA\ and $\lambda$\ 6620\AA, as can be seen in Figure \ref{continuum} (left). These are points where the absorption band of \CC\ and CN is weakest and have been suggested for AGB carbon-rich stars by \citet{westerlund91} and \citet{Morgan03}. For some stars, the maximum at $\lambda$\ 6180\AA\ was greater than the one at $\lambda$~6202\AA, so the pseudo-continuum was defined without that point. The second, Figure \ref{continuum} (right), was a linear local continuum defined between maxima at $\lambda$\ 4962\AA\ and $\lambda$\ 5030\AA. These maxima were suggested by \citet{sarre00} because no overlap from metallic lines and other bands compromise the fluxes at the points. All this was done after applying a slight smoothing to the spectra using a 3-pixel box and then, the spectrum was normalized related to the pseudo-continuum settled. The strengths were calculated with respect to the pseudo-continuum and also to a local continuum defined by a nearby maximum.

\subsection{The \CC-index} 
\label{CC}
The more essential abundance index is the \CC-index, which is an index of the carbon excess over the oxygen in the carbon-rich photospheres. This index was originally defined in the Revised MK System \citep{Keenan93} as dependent in strength of the Swan Bands $\lambda$\ 5165\AA\ and $\lambda$\ 5635\AA. However, these bands are just suitable for C-R and C-H, since in these types of stars the \CC\ bands are strong; but for the N-type and most of the J-type stars, that have a compromised flux at this spectral range, it is difficult to obtain the \CC-index through analysis of these two specific bands. Fortunately, many Swan bands are widely present at the AGB carbon star spectra, and several may be used for the same purpose.

\begin{figure}
\begin{center}
\plottwo{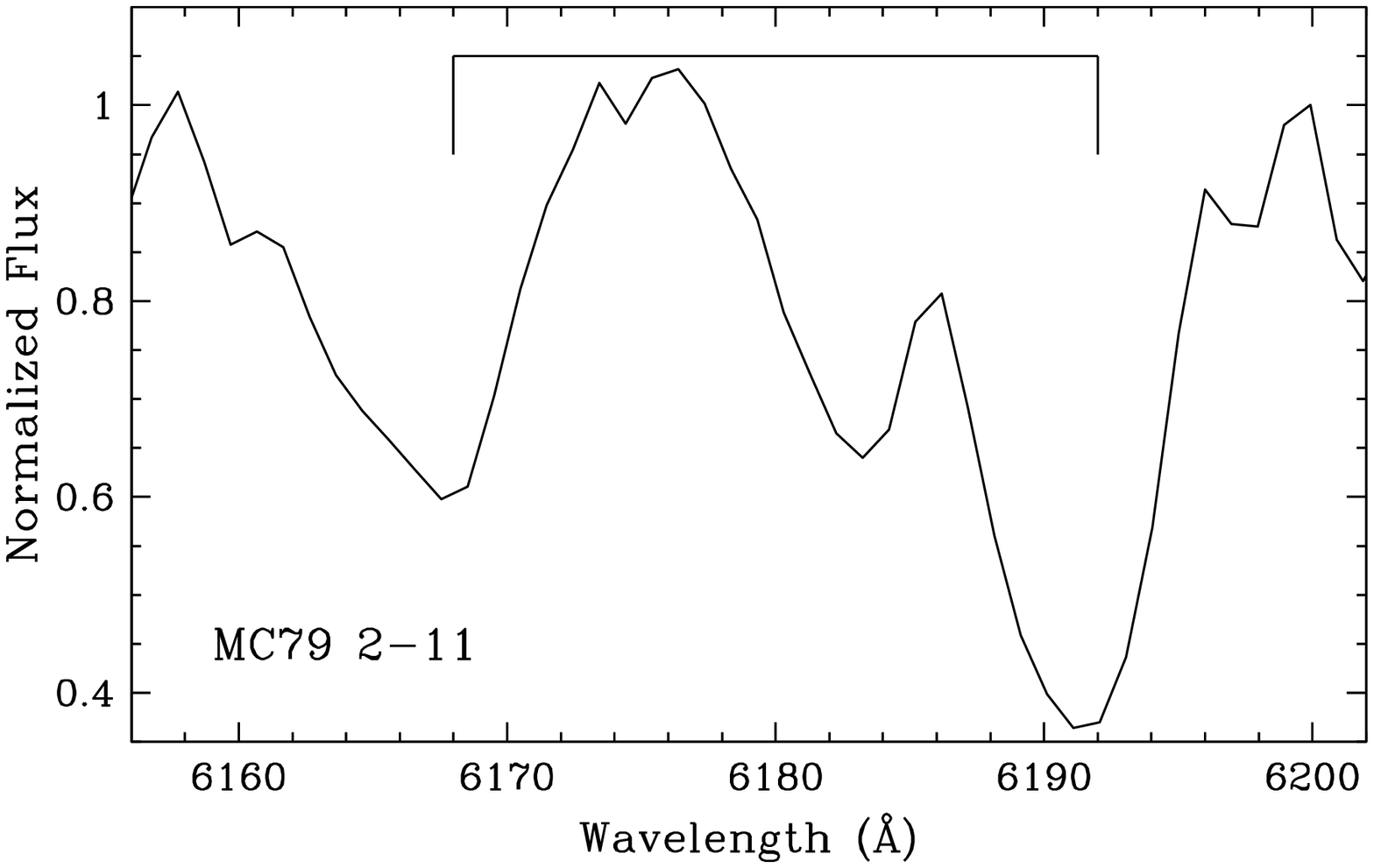}{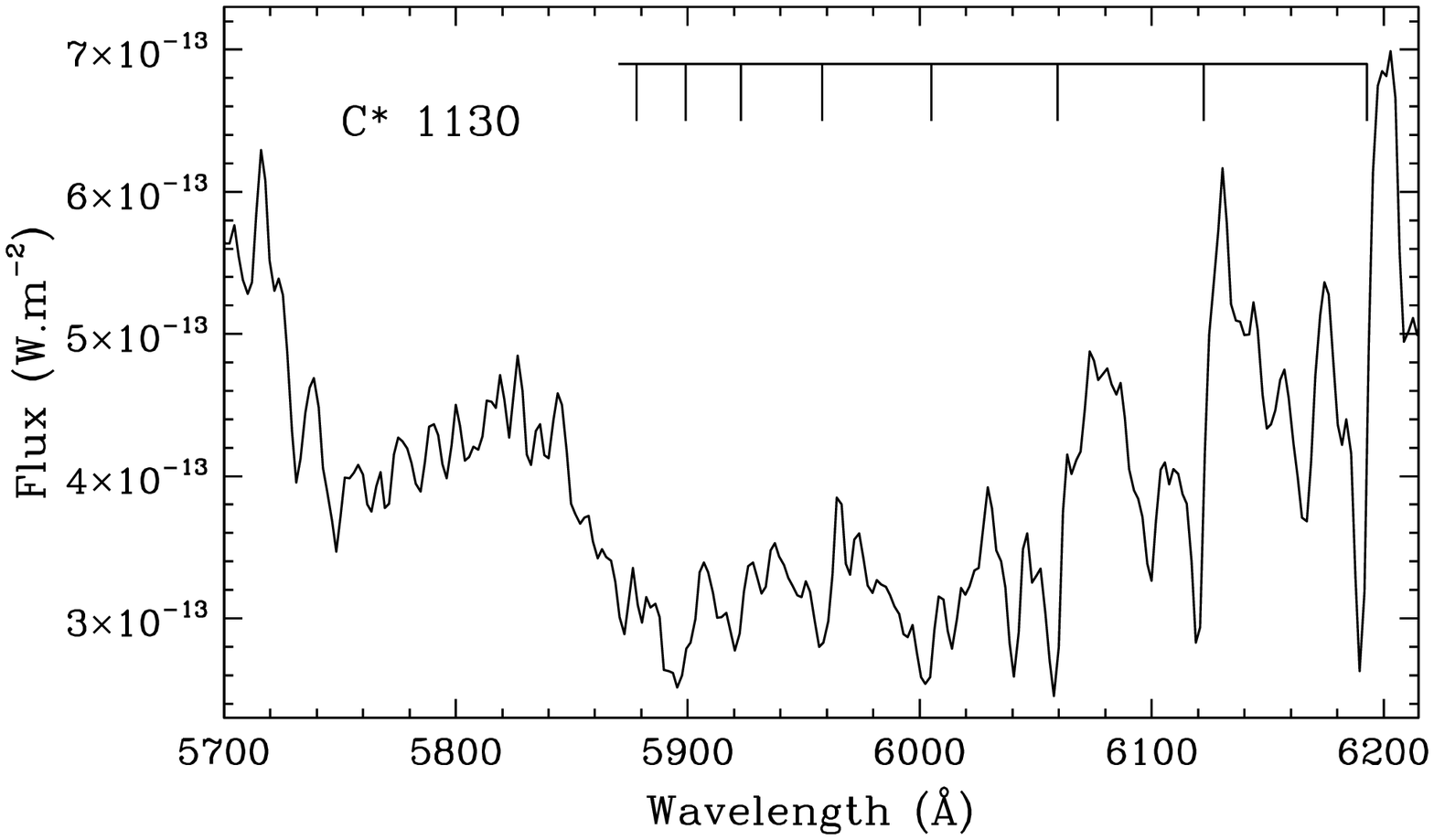}
\caption{On the left is normal and isotopic \CC\ band, $\lambda$\ 6192\AA\ and $\lambda$\ 6168\AA\ respectively from MC79~2-11, and on the right is a set of \CC\ absorptions at $\lambda$\ 5722\AA\ - $\lambda$\ 6202\AA\ from C*~1130. \label{fig1}}
\end{center}
\end{figure}

Thus, we decided to adopt the well correlated parameters suggested by \citet{Morgan03} in their study of a LMC carbon sample: the sum of the strengths of the normal and isotopic \CC\ band, associated with $\lambda$\ 6192\AA\ and $\lambda$\ 6168\AA\ respectively, shown in Figure \ref{fig1} (left); and the equivalent width of a complete absorption in the range of $\lambda$\ 5722\AA\ - $\lambda$\ 6202\AA, shown in Figure \ref{fig1} (right). The strength of the bands, D$_{\lambda\ 6168}$ and D$_{\lambda\ 6192}$, were measured from the minima, at the wavelength described, with respect to the pseudo-continuum for normalized spectra. The complete absorption by C-rich molecules associated with the equivalent width, W$_{\lambda\ 5722\ -\ \lambda\ 6202}$, were measured in the edges at the maxima described.  Both parameters, strength and equivalent width of the bands, are measurements of the same physical property, i.e. the carbon abundance of the photosphere. That means, these are expected to have a positive correlation, since an enhance on the carbon abundance will reflect in a more strong strength of the bands as well as a broader equivalent width.

It is also important to notice that the isotopic band is significant only in stars with $^{13}$C excess, and for those it has to be considered. For the ordinary carbon stars, the spectrum between $\lambda$\ 6159\AA\ and $\lambda$\ 6175\AA\ shows a broad \CC\ absorption feature with a minimum at $\lambda$\ 6162 - $\lambda$\  6164\AA\ and a secondary minimum at $\lambda$\ 6168\AA. On the other hand, the stars with $^{13}$C excess show a single minimum at $\lambda$\ 6168\AA.
An extra difficulty in identifying $^{13}$C$^{12}$C $\lambda$ 6168\AA, when it  is weak, is the proximity of CaI $\lambda$ 6162\AA\ and several CN and $^{12}$C$^{12}$C bands.

The former parameter, D$_{\lambda\ 6192}$ + D$_{\lambda\ 6168}$\ , is obtained in a shorter spectral range, and therefore it is a more reliable measurement, once the local continuum is better defined, and for this reason it was designated as the primary parameter, $p_1$. Thus, the equivalent width of the large range, $\lambda$\ 5722\AA\ to $\lambda$\ 6202\AA, was designated as a secondary parameter, $p_2$. If, for a given star, only $p_1$ can be measured, it is usually reliable enough to be used alone, in place of the average indicator described bellow. 

This hierarchy of trust were used, then, to re-scale $p_2$, by putting it at the same scale of the primary parameter. Some stars studied have multiple spectra and, for those an average of each parameters was calculated and it was also possible to assign an error for $p_1$ and $p_2$, caused by the use of different telescope and grating combinations for the observations. Then, after measuring the both parameters, a linear least-squares fit, $p_2 = a \cdot p_1 + b$, was calculated, wielding $a = 242$ and $b = 38$. The fine correlation found by \citet{Morgan03} can also be seen for our Galactic C stars, as were suggested previously by \citet{Keenan93}, and so the linear fit were adopted. The coefficients $a$ and $b$ were used to apply a linear transformation to the secondary parameter, in order to get a new one, $p_2^{\prime}$ , in the same scale of $p_1$:
\begin{equation}
\label{transf}
p_2^{\prime} = \frac{p_2-b}{a} .
\end{equation}

The next step was the evaluation of the average indicator: 
\begin{equation}
\label{pmed}
\bar{p} = \frac{p_2^{\prime} + p_1}{2}.
\end{equation}
This indicator is used to estimate the ranges of the levels of the carbon abundance index. This simple average, between the parameters previously set to the same scale, allows a further reduction in the uncertainties of the measurements.

In \citet{Keenan93}, the \CC-index has a increasing scale of eight levels, varying from 1 to 8, which we adopted. And the range values given by the average indicator for each index level were set by comparing $\bar{p}$ of the 6 standards objects to the \CC-index set in \citet{Yama72,Yama75} and \citet{barn96}. None of these schemes have given the boundary values of their domains, nevertheless, we were able to make our scale agrees closely with the index of excess of carbon over oxygen given by Yamashita's tables. The ranges of each level of the \CC-index , which have an equally interspace, is given by the average indicators described on Table \ref{tbl-C2}. 

The \CC-index can be calculated for any carbon star, by obtaining $p_1$ and $p_2$ from its spectra, then using eqs. [\ref{transf}] and [\ref{pmed}] to obtain $\bar{p}$ and consulting the Table \ref{tbl-C2}. The results of the \CC-index obtained for our entire sample are discussed in Section \ref{disc}. Therefore, the \CC-index, in this way described, corresponds to a direct measure of the carbon intensity observed on the carbon-rich star, and not one relative to a standard of same temperature, as used in the past.

As multiple spectra for some stars were taken, including different combinations of telescope and gratings, it is possible to expose how consistent is the \CC-index of this new scheme for a same star observed with different resolutions. We manage to have different combinations for 20 targets. The parameters $p_1$ and $p_2$ were obtained for each star spectra, and then, a deviation from the average value for each parameter of each target were calculated. In order to get a global behavior, the parameter's deviations were averaged through the whole sample and it was found an error of $\delta _{p_1} = 0.06$ and $\delta _{p_2} = 13.80$ for the primary and secondary parameters, respectively, which means an error of $\delta _{\bar{p}} = 0.04$ for the average indicator. These errors, compared to the magnitude of ${p_1}$ and ${p_2}$ obtained and $\bar{p}$ calculated, indicate a good reliability in the method and in the bands adopted. In fact, the strongest \CC\ bands of the Swan System can be estimated even on prismatic spectrograms of very low resolution \citep{blanco78,westerlund78,sand77}. On the other hand, a discussion about how consistent our \CC-index is with previous schemes is placed in section \ref{conc}.

\subsection{The j-index} 
\label{j}

Let us discuss now the isotopic carbon abundance. It deserves a specific index, because of its historical importance and it is extremely relevant in stars with low $^{12}$C:$^{13}$C ratio. The expression J-Star, used for the very first time by \citet{bouigue54}, refers to stars that have an unusual strong isotopic carbon bands. The exact evolutionary stage and nature of this $^{13}$C-rich stars, is still unclear \citep{abia03}, since \citet{ut85} found no excess of s-process's elements in J-type stars, as it would be expected for any ordinary AGB carbon star. An inspection in R-type stars reveals that they also do not present an excess of these elements together with an overabundance of $^{13}$C \citep{dom84}. Despite both C-R and C-J may seem to have their carbon-rich photosphere originated with the same mechanism, with an anomalous helium-flash \citep{rstar}, no connection between both types of stars was found yet. So it is not certain whether C-J evolved and cooled from the C-R. The absence of s-process's element enrichment does not fit these stars in the AGB scenario that supposedly already crossed the thermal pulses, discarding the possibility of a $^{13}$C enrichment due to the third dredged-up \citep{pila87}. The Tc detection together with the s-process's element absence in two C-J Peculiar Stars made them even more riddling, as the Tc presence points dredge-up occurrence. Some attempts have been made to explain this apparent incoherence through a cool bottom processing theory \citep{wbs95}. However no conclusion about the nature of the J-Stars is presented, these objects are represented in the NSCC by a C-J notation for the spectral sub-class index. We intend to stress their unusual nature and to distinguish them from stars with normal $^{13}$C abundance.

\begin{figure}
\epsscale{0.5}
\begin{center}
\plotone{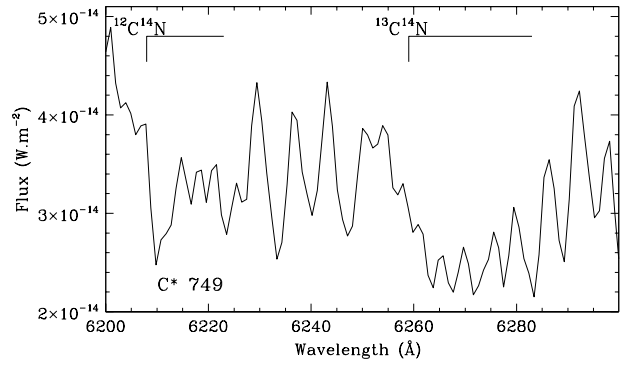}
\caption{The isotopic and normal equivalent width of the CN bands for C*~749, which head-bands are located at $^{13}$C$^{14}$N $\lambda$ 6260\AA\ and $^{12}$C$^{14}$N $\lambda$ 6206\AA. \label{fig3}}
\end{center}
\end{figure}

\citet{Morgan03} found a strong correlation between the ratio of the strength of the isotopic and normal \CC\ band, associated with $\lambda$ 6168\AA\ and $\lambda$ 6192\AA\ respectively, shown in Figure \ref{fig1} (left); and the ratio of the isotopic and normal equivalent width of the CN bands whose band-heads are located at $^{13}$C$^{14}$N $\lambda$ 6260\AA\ and $^{12}$C$^{14}$N $\lambda$ 6206\AA, shown in Figure \ref{fig3}, for LMC stars. These strength ratios and equivalent width of the isotopic and normal bands are the ones suggested in the study of Galactic C-Stars by \citet{Keenan93}, in a previous work. In our scheme of classification these parameters were also used to evaluate the isotopic carbon abundance, namely the j-index.  As done for the \CC-index, the strength of the bands, D$_{\lambda\ 6168}$ and D$_{\lambda\ 6192}$, were measured from the minima of the band with respect to the pseudo-continuum at the normalized spectra, and the equivalent width, W$_{\lambda\ 6206}$ and W$_{\lambda\ 6260}$, were measured at the maxima of the edges of each waveband. Regarding the isotopic and normal CN bands for C-J stars, it can be seen a clear splitting of the waveband in two halves $\lambda$ 6205 - $\lambda$ 6250\AA\ and $\lambda$ 6255 - $\lambda$ 6290\AA; while for normal N stars this splitting are not so obvious.

All measurements for the j-index are taken in short wavelength ranges, being equally reliable. The ratio of the strengths of the \CC\ bands, D$_{\lambda\ 6168}$ / D$_{\lambda\ 6192}$, was designated as parameter $q_1$, and the ratio of the CN bands, W$_{\lambda\ 6260}$~/~W$_{\lambda\ 6206}$, as parameter $q_2$. For the stars of the sample that have multiple spectra were possible to assign an average for each parameter and an error associated with the effects in $q_1$ and $q_2$ caused by the use of different telescope and grating combinations on the observations.

Both parameters are measurements of the $^{13}$C excess, thus it is expected that they have a positive correlation, since an enhance on $^{13}$C will reflect in both measurements. Nevertheless, a fine correlation can only be seen for C-J stars. For this reason, only likely J-stars could be used to calibrate the average indicator formula. We selected those candidates by applying the definition given by \citet{Gor68} for J-stars, as the ones that have the isotopic band $^{13}$C$^{12}$C $\lambda$ 6168\AA\ with at least half the strength of the normal band $^{12}$C$^{12}$C $\lambda$ 6122\AA.

The $q_1$ and $q_2$ obtained for the targets selected through Gordon's definition seem to have a linear correlation as were suggested by \citet{Keenan93}. Then, a linear least-squares fit, $q_2 = c \cdot q_1 + d$, was calculated for this small sub-sample, wielding $c = 3$ and $d = -1$. The coefficients $c$ and $d$ were used again to apply a linear transformation on $q_2$ to acquire two parameters at the same scale, $q_2^{\prime}$ and $q_1$. Thus, a simple average between these two gave the average indicator, $\bar{q}$.

From the targets selected through Gordon's definition only 3 stars of our sample (BE~CMa, BM Gem and SU Sco) already had a j-index available, then, they were defined as the standard objects. Following the MK Revised System, the j-index were divided into eight levels, varying from 0 to 7, which also agree with the isotopic carbon strength index from \citet{Yama72,Yama75}. As presented in Section \ref{CC}, the ranges of the average indicator for each level were settled by comparing the $\bar{q}$ of the standards to the j-index data in \citet{Yama72} and \citet{barn96}. Again, the boundary domains of each catalogue were not described but we were able to make our scale agrees closely with the index of $^{13}$C excess in Yamashita's tables. The ranges of each level of the j-index , which also have an equally interspace, is given by the average indicators described on Table \ref{tbl-j}. 

Usually, C-N stars have j-indices between 1.5 and 3 and C-R between 2.5 and 4 \citep{Keenan93}. Thus, in the NSCC, only targets with j $\geqslant$ 4 are surely considered a J-star and have an explicit j-index. In all other cases, the value of the j-index is omitted. The j-index can be assigned for any carbon star, by obtaining $q_1$ and $q_2$ from its spectra and consulting the Table \ref{tbl-j}. The results of the j-index obtained for our entire sample are discussed in Section \ref{disc}.

Through the multiple spectra, with different combinations of telescope and gratings, taken for 20 targets, it were possible to expose how consistent is the j-index of the NSCC for a same star observed with different resolutions. First, a deviation from the parameters $q_1$ and $q_2$ were calculated as done in the previous Section. Then the global behavior, an average of the parameter's deviations through the hole sample, was evaluated and it was found an error of $\delta _{q_1} = 0.03$ and $\delta _{q_2} = 0.09$, which means an error of $\delta _{\bar{q}} = 0.02$ for the average indicator. These errors, compared to the magnitude of ${q_1}$ and ${q_2}$ obtained and $\bar{q}$ calculated, also indicate a good reliability in the method and in the bands adopted.

\subsection{The MS-index} 
\label{MS}

\begin{figure}
\epsscale{0.5}
\begin{center}
\plotone{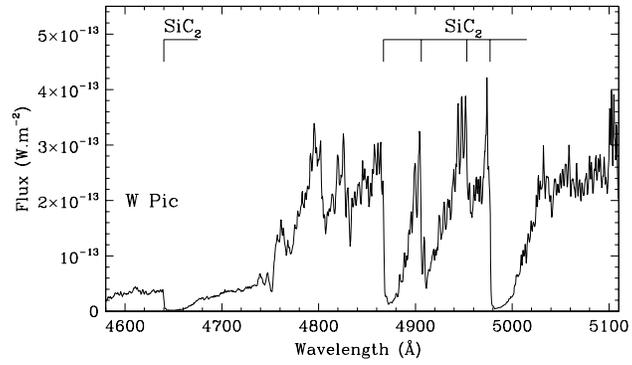}
\caption{The MS-band structures in the $\lambda$ 4600\AA\ - $\lambda$ 5100\AA\ spectral range for W Pic. \label{fig5}}
\end{center}
\end{figure}

\citet{Merrill26} and \citet{Sanford26} observed for the first time a molecular absorption, in the violet-blue region, due to a compound associated with the Si\CC\ molecule. Some Si\CC\ absorptions (the Merrill-Sanford bands, hereafter MS-bands) can be found in the $\lambda$~4300\AA~-~$\lambda$~5000\AA\ spectral range. The MS-bands are not present in all C-rich stars, but when present, they are very strong (see Figure \ref{fig5}) and, therefore, a representative index of the strength of the Si\CC\ molecule in their photospheres can be obtained as suggested by \citet{Keenan93}. The C-J stars offer favorable conditions to their study, since they usually present MS-band structures, and their flux at the violet-blue range are not as compromised as one of a C-N star.

The MS-index is the last optical index added to the NSCC notation for the description of the photosphere of the carbon-rich stars. Unfortunately, not all stars in our sample were observed in the spectral region populated by Si\CC\ structures, thus, it is possible that some targets, which have a strong MS band, do not have a MS-index in the final classification. As it is a complementary index, we considered that there is no need to subdivide it into too many levels as it has been usually done \citep{Keenan93}. Our scheme of carbon star classification presents the strength of the Si\CC\ index divided into only tree levels: weak, medium and strong \citep[as in][]{sarre00}. The parameter employed to obtain the MS-indices were the strength related to a local continuum, $d$, of a Si\CC\ structure with head-band located at $\lambda$ 4977\AA.

The ranges of each level of the MS-index in the NSCC are those described by \citet{Morgan04}, that also used the same band for $d$, and vary from 1 to 3 in a increasing scale. The NSCC has an additional MS0 level, for very weak or absent MS bands, which can be omitted from the final notation.

Using different combinations of telescope and gratings, taken for 16 targets, it were possible to expose how consistent is the MS-index of the NSCC for a same star observed with different resolutions. A deviation from the parameters $d$ could be calculated for these stars, and then, the global behavior error, an average of the parameter's deviations through the hole sample, was evaluated in $\delta _{d} = 0.03$.

\subsection{Spectral Subclasses} 
\label{pop}

Due to the high opacity in the optical spectra of the cool carbon stars, several criteria have been suggested to define each spectral subclass for carbon stars, but each single one did not allow by itself a consistent classification. In this work, we made an inspection through all visible spectrum, by combining the use of some of these criteria. When applied together in a simple yes/no logical analysis for each characteristic feature, they complement each other, giving a more conclusive result.

The first criterion concerns the extension of the spectra to the blue-violet range. Although a high opacity in wavelengths lower than $\lambda$ 4977\AA\ points to a carbon C-N star, it is not a sufficient condition, as cool C-R and C-H stars may also exhibit this effect. 

\begin{figure}
\epsscale{1.0}
\begin{center}
\plottwo{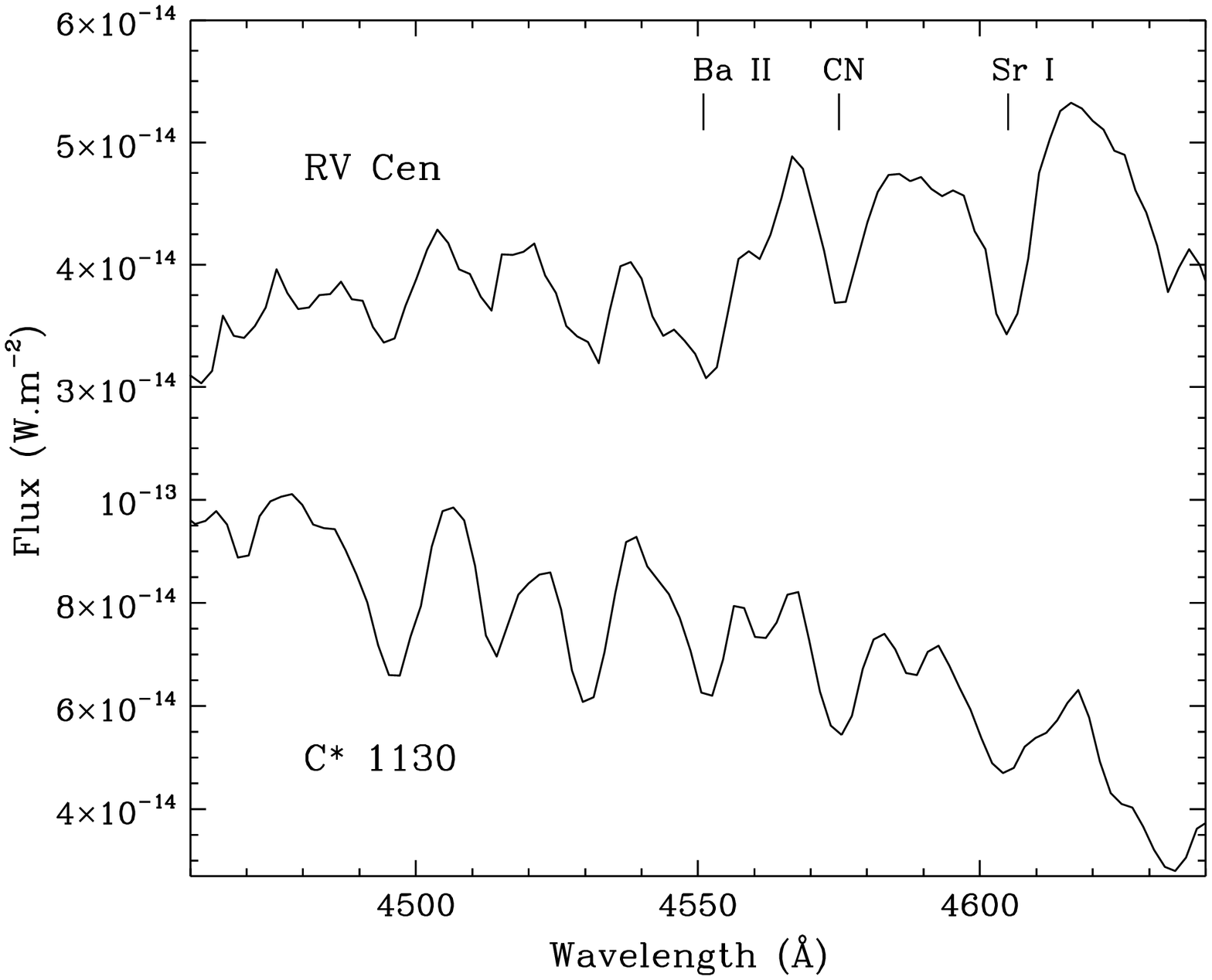}{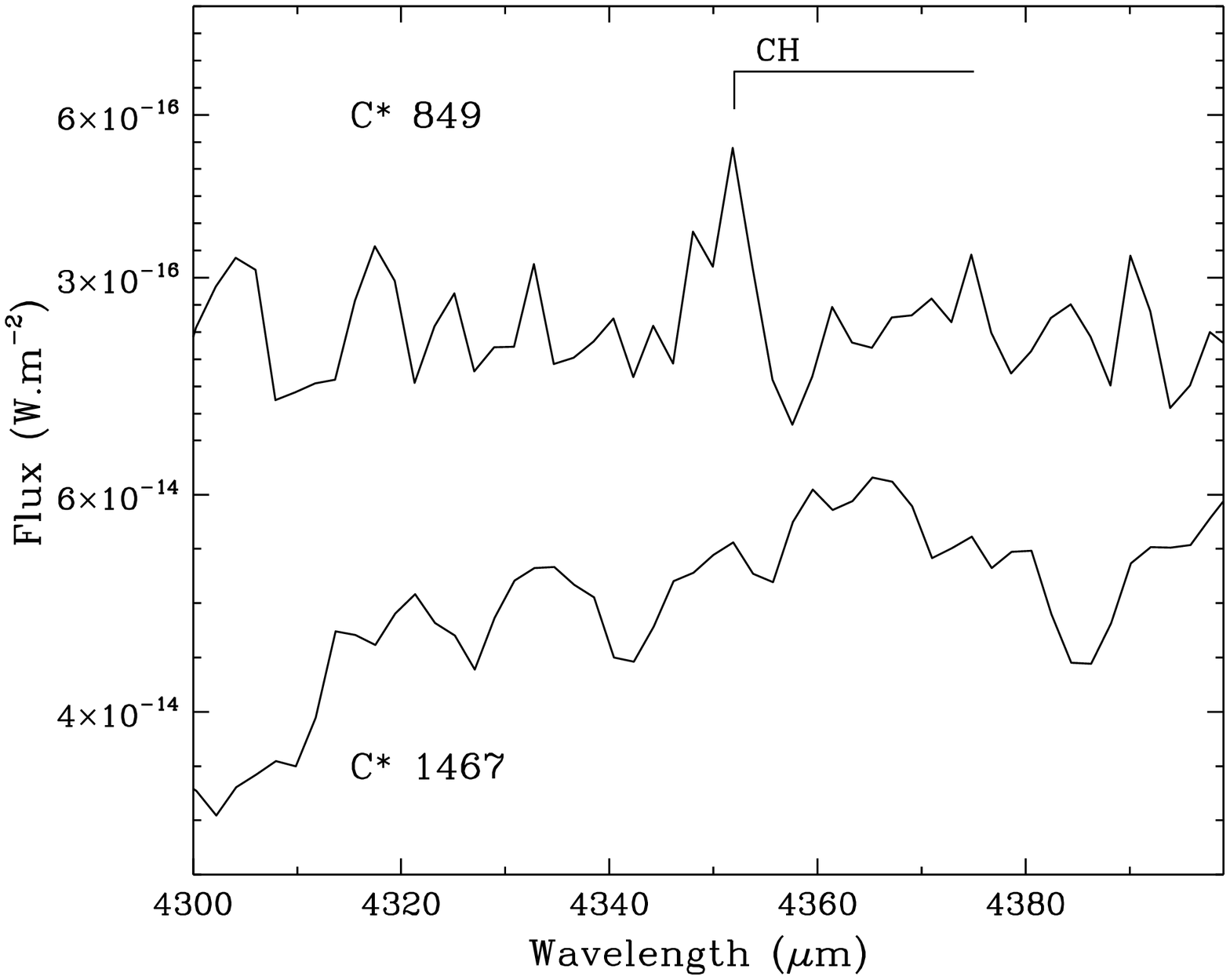}
\caption{On the left, the BaII and SrI lines are stronger as the CN band in RV~Cen spectra, but the same does not happens for C*~1130. On the right, the P-branch at $\lambda$ 4352\AA\ can be seen just in C*~849. \label{fig6}}
\end{center}
\end{figure}

The presence of structures related to s-process elements are easy to identify in C-N and C-H but not in C-R stars, which were not enriched with these elements. A good method to qualitatively detect the presence of these structure lines is: when the intensity of the BaII lines $\lambda$~4554\AA\ and $\lambda$ 4934\AA\ is stronger than the closer CN bands, for instance, the band-head at $\lambda$ 4576\AA; and when the SrI $\lambda$ 4607\AA\ has about the same intensity of the band, this might be a C-N star \citep{barn96}. In Figure \ref{fig6} (left), the RV Cen star has the BaII and SrI lines stronger as the CN band, but the same does not happens for C* 1130. 

Finally, a third inspection concerns the distinction between C-H and C-N stars, as both present strong BaII lines with respect to the CN band. It is possible to distinguish these type of carbon stars by presence of the P-branch, which band-head is located at $\lambda$ 4352\AA. When P-branch is evident, this is typical of C-H's and not expected for C-N's, as it can be seen in Figure \ref{fig6} (right). This branch is also present in most of the C-R stars. A summary of the combined use of these criteria can be seen in Table \ref{tbl-pop}.

All these inspections (the extension to the blue-violet spectral range, the presence of s-process element's lines or the presence of the P-branch, $\lambda$ 4352\AA) are just a qualitative analysis based on a visual inspection. On the other hand, such a study can be applied even to prismatic spectrograms of very low resolution, as no high precision is required to establish the spectral subclass through this method. Anyway, these three inspections together can give an unambiguous way to assign the spectral subclass of each carbon star studied. The spectral subclass goes at the beginning of the NSCC notation.

\section{Circumstellar Envelope Classification}
\label{ir}
During the Asymptotic Giant Branch phase, stars suffer intense mass loss which leads to the formation of circumstellar dust envelopes. In order to describe a complete scenario, three more indices were added to NSCC: two concerning the circumstellar environment itself and one related to the temperature of the star. This infrared classification, which we are introducing in this work, was obtained through a Monte Carlo numerical algorithm which simulates the radiative transfer problem in the circumstellar envelopes.

The numerical treatment of radiative transfer applied to a spherical envelope, used by us here, was described in a previous work of \citet{lml94}. Here, we present it briefly for the sake of completeness. The propagation of stellar and grain radiative energy is simulated photon by photon following a Monte Carlo scheme. For each interaction between a photon and a grain, a fraction of the energy is stored (absorption) and the remaining part is scattered according to the scattering diagram. The stellar radiation leads to a first distribution of dust temperature and the thermal radiation from grains is simulated, giving after several iterations the equilibrium temperature.

When two kinds of grain are present, the path length of photons is defined along any direction by the total opacity
$$\tau_{ext}(\lambda) = \tau_1(\lambda) + \tau_2(\lambda)$$\noindent
where,
$$\tau_i(\lambda) = {\int_0^l N_i(r) \pi a_i^2 Q_{ext_i}(\lambda)dr}$$\noindent
in which $a_i$ is the radius of the grain $i$, $Q_{ext_i}(\lambda)$ its extinction efficiency and $N_i(r)$ its number density at the distance $r$ from the center of the star. Efficiencies and scattering diagrams are computed at each wavelength using the Mie theory. When the position of the interactive grain is defined its nature is determined: the probability of having a grain $i$ is
$$P_i(\lambda) = a_i^2 Q_{ext_i} N_i(r)/ \sum_j a_j^2 Q_{ext_j} N_j(r).$$

The energy absorbed is then calculated with the relevant $Q_{abs}$ and the new direction of scattering is generated according to the corresponding scattering diagram. The energy is stored separately for each kind of grain and then two sequences of temperature are obtained. The origin of the emergent radiation at any wavelength is easily identified (direct stellar radiation, scattered light, emission from each kind of grain).

The mixture is characterized by the ratio $N_1(r)$/ $N_2(r)$. Here, this ratio will be independent of the position point. When one specie is much less abundant than the other, a large number of events have to be simulated to get the good statistics. The total number of grains is defined by the extinction opacity calculated along the radial direction between the inner radius of the shell $R_i$ and its outer radius $R_o$ at $\lambda$ = 1$\mu$m. The number density is assumed to vary as $r^{2}$, corresponding to an expansion at constant velocity. Therefore, the following physical quantities are required:

\begin{itemize}
	\item the effective temperature of the central star $T_{eff}$;
	\item the inner and outer radii of the shell $R_i$ and $R_o$. We shall assume here that both kinds of grain exist within the same limits. This point will be discussed later;
	\item the grain radii and refractive indices at all wavelengths. (A grid of 30
wavelengths is used);
	\item the ratio $N_1$/ $N_2$;
	\item the extinction opacity at $\lambda$ = 1$\mu$m.\noindent
\end{itemize}

The computation gives the spectral distribution of the total flux and of its different components (direct, scattered, emitted), the temperature law for each kind of grain. Best fits ive the effective temperature of the central star, SiC/A.C. ratio, $\tau$ and grain size.

The absorption and scattering efficiencies, as well as the albedo for the grains were calculated by us using the Mie theory and the optical constants tabulated in the literature. For the amorphous carbon (hereafter A.C.) we have used the optical constants published by \citet{rm91} since they were calculated for a wide range in energy ($4.127 \cdot 10^3$ \textit{eV} - $3.500 \cdot 10^3$ \textit{eV}) satisfying the Kramers-Kronig relations. For SiC grains we used the optical constants of silicon carbide (SiC) determined by \citet{p88} and for silicate those obtained by \citet{pd95}. These last were used only to model silicate carbon stars. Finally, the infrared notation adopted is based on parameters and ranges of each index level proposed previously by \citet{lml94}.

Although empirical measures of the size of the emission feature could be applied with some adjusts in the methodology here presented, it is not advised. Anyone interested in the NSCC can use a public radiative transfer code, for instance DUST \citep{dust}, and the quantities needed in the scheme can be obtained.

The circumstellar envelope classification completes the NSCC, and can be filled using any infrared spectra that provides wavelength ranges around the dust features, i.e. 11.3 $\mu$m for normal carbon stars and 9.8 $\mu$m for the peculiar ones. There are about 900 carbon stars observed by IRAS satellite and almost 50 by ISO satellite. Furthermore, it is possible to obtain new fresh IR data with Spitzer satellite and, at ground, using Gemini telescopes with TReCS and Michelle spectrometers which provide wavelength ranges required for the NSCC indices.

\subsection{The $\tau$-index} 
\label{tau}
The $\tau$-index or opacity index is an estimate of the circumstellar envelope's optical depth, and can provide information about the evolutionary stages of the carbon-rich stars, as discussed in section disc. Through the opacity index it is possible to establish how optically thick is the envelope, allowing us to draw a better scenario for the star-envelope system.

The parameter used to define the $\tau$-index is one of the outputs of the radiative transfer model fit, the $\tau$ parameter. We divided this index into four levels, varying from 0 to 3, and the ranges of each level agree with the ones described by \citet{lml94}, as can be seen in Table \ref{tbl-tau}. The $\tau$0 level corresponds to an extremely low opacity value, nevertheless, both $\tau0$ and $\tau1$ correspond to the opacity level I, defined by \citeauthor{lml94}.

The deviation of the optical depth found through the radiative transfer code applied to a spherical envelope was $\delta_{\tau} = 0.03$. This is a zero-point error for all the circumstellar envelope's optical depth obtained with the \citeauthor{lml94} code. When compared $\delta_{\tau}$ to the  ranges of each level of the $\tau$-index, we verify the good reliability in the method adopted.

\subsection{The SiC-index} 
\label{sic}
The SiC-index is a relative abundance of silicon carbide to A.C. (hereafter SiC/A.C.) for the grains in the circumstellar envelope. This ratio abundance is based on the spectral feature at 11.3~$\mu$m, which appears in most of the carbon stars. Modeling their circumstellar envelope by taking into account only the amorphous carbon dust grains is a good first approximation, since it corresponds to the main emission of the envelopes. Nevertheless, it does not reproduces the emission feature at 11.3 $\mu$m, associated with the silicon carbide.

\begin{figure}
\epsscale{1.0}
\begin{center}
\plottwo{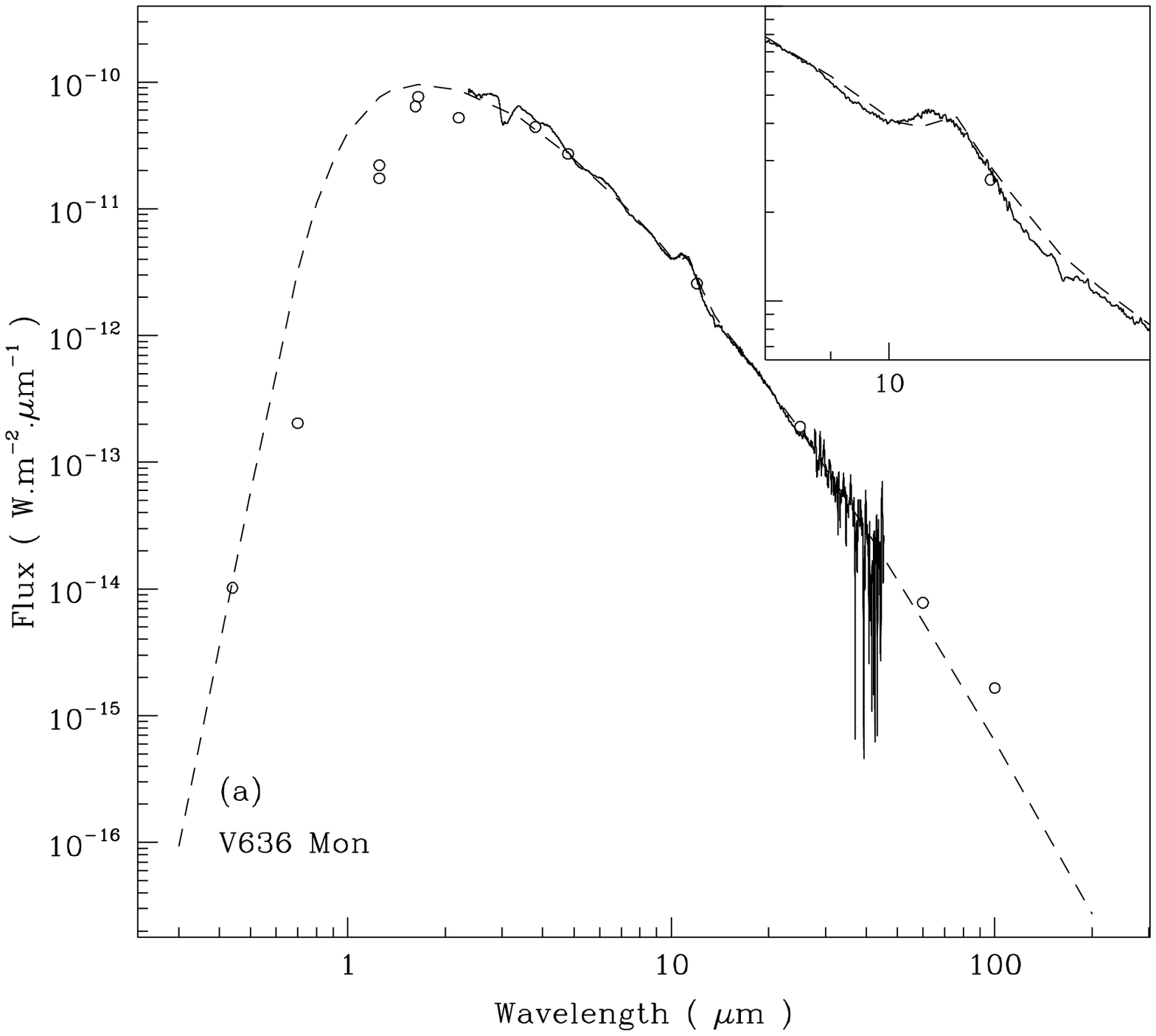}{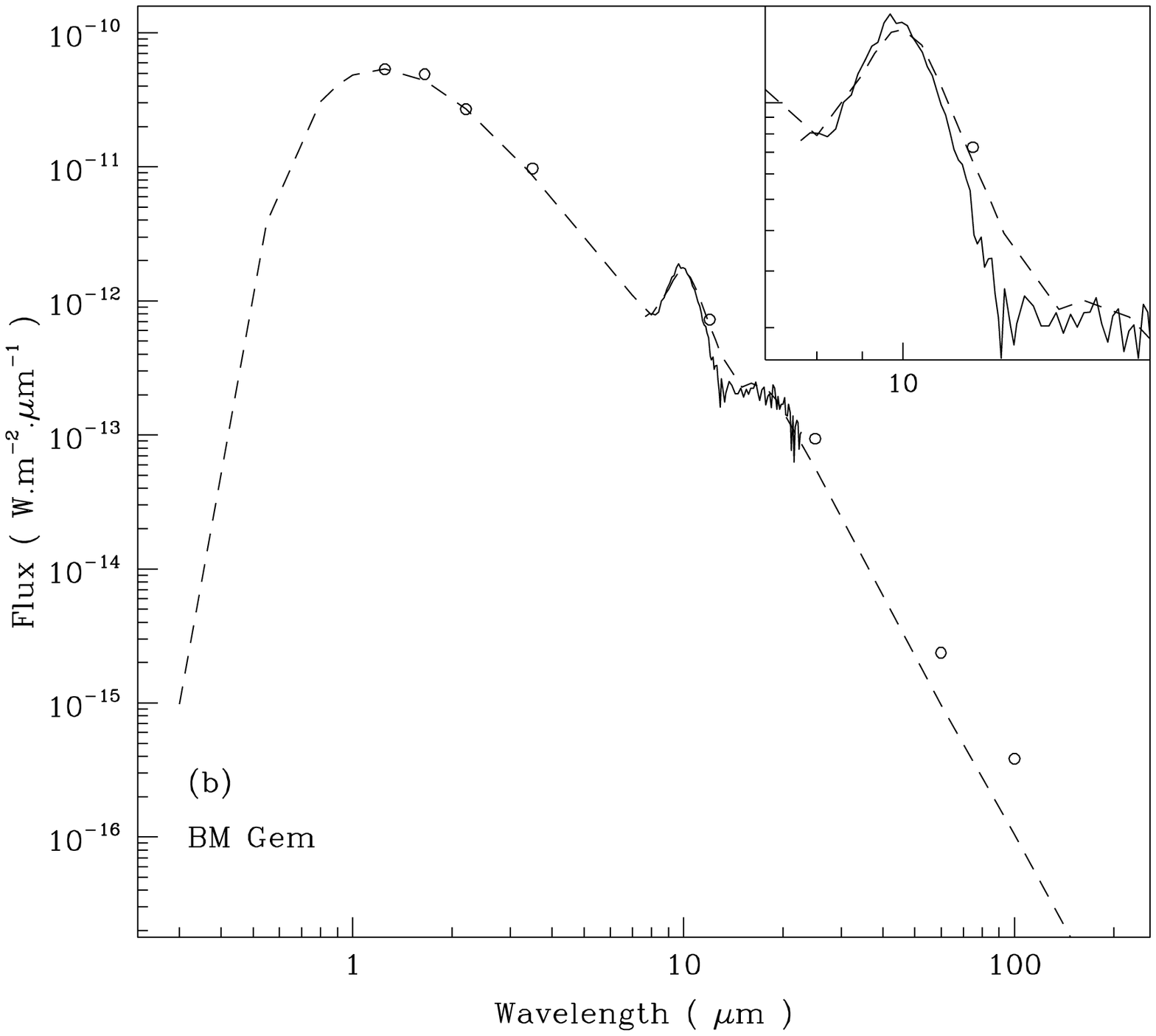}
\caption{This figure presents best fit to the models. The open circles denote photometric data taken from SIMBAD, full line are ISO data (a) and LRS IRAS data (b), dashed lines represent the best models. (a) is the best fit to V636 Mon which presents a 11.3 $\mu$m feature, due to silicon carbide (SiC) grains; (b) is the best fit to BM Gem, which presents silicate emission features at 9.8 $\mu$m and 18 $\mu$m. At the top right of each graphic it is possible to see the emission features in detail.\label{fig7}}
\end{center}
\end{figure}

Therefore, the dust compounds applied to obtain the SiC-index were the SiC/A.C. ratio abundance, which is also an output parameter of the radiative transfer model. A model fit to the spectral feature at 11.3 $\mu$m can be seen in Figure \ref{fig7} (a), in dashed line. The open circles in the figures denote the photometric data taken from SIMBAD database, ranging from optical to infrared wavelengths, while the full line is the \citetalias{iso} data. The deviation of the SiC/A.C. ratio abundance found, for all models calculated, through the radiative transfer code was $\delta_{SiC/A.C.} = 0.02$, which also represents a good trustworthiness in the method.

A simplified SiC index was presented by \citet{sloan98}. They have developed an infrared spectral classification for 96 carbon star based on \citetalias{iras} spectra. Basically they extracted the blackbody contribution using a 2400 K Plank function for their sample. One problem is that the residual spectra include not only the dust features but also the photospheric absorption bands. In addition, they used a flux ratio to obtain their classes. Even so this is an interesting way to classified dust envelopes, we believe that both $\tau$- and SiC- (or Silicate) indices are better obtained using a full modeling including radiative transfer. In this case, we can estimate the contribution of any kind of dust present in the circumstellar envelope.

We set four levels for the SiC-index, varying from 1 to 4, and the domains of each level are listed in the Table \ref{tbl-sic}.  \citet{sloan98} described three indices related to the intensity of the emission feature and others related to the surrounding continuum and to the unusual profiles found by them. When the levels of the SiC-index are compared with the description of the first three indices of \citeauthor{sloan98}, we have, roughly, that: our $SiC1$ and $SiC2$ correspond to their $SiC$ class, our $SiC3$ to their $SiC +$ class and our $SiC4$ to their $SiC ++$ class.

The dust species formed in the circumstellar envelope of carbon stars should, at first, reflect the chemical composition of their photospheres. However, \citet{lm86} and \citet{wj86}, independently, inspecting the \citetalias{iras} catalogue, discovered the existence of a sample of carbon-rich stars which exhibit oxygen-rich dust envelopes. For these peculiar carbon stars, the carbon-rich nature is revealed in the visible spectral region while the oxygen-rich one is exposed in the mid-infrared, where amorphous and crystalline silicates can be detect. \citet{wj86} showed that these stars are, in fact, J-type carbon stars. In our scheme of classification, these silicate carbon stars received a $JPec$ notation instead of a SiC-index. Figure \ref{fig7} (b) presents the best fit model to the spectral feature at 9.8 $\mu$m associated with silicates, typical of a silicate carbon star.

\subsection{Temperature Index} 
\label{temp}
It was only with \citet{Keenan93} that a system of classification of carbon stars incorporated a numerical index of temperature based on infrared fluxes. Those indices were more reliable than the atomic lines commonly used to estimate the temperature for stars of the class M and hotter. 

In the NSCC, the temperature index is also devised from a parameter based on infrared fluxes by way of the radiative transfer simulation. Once the modeled SED which best fits the infrared data is set, the effective temperature of the central star ($T_eff$) is one of the outputs of the simulation. And that is the parameter that should be used to establish the temperature index levels of each target. The temperature index is placed, in the final notation, just after the spectral subclasses, as usual.

By means of this radiative transfer code applied to a spherical envelope we were able to settle an error nearly 100K for the effective temperature of the central star calculated. This is a zero-point error that should be considered for all data modeled with the \citeauthor{lml94} code. Above this error value, the calculated models differs meaningfully of each other.

The range values of the effective temperature for each index level were set by comparing this output parameter of the 6 standards objects to the indices set in \citet{Yama72,Yama75,Keenan93} and \citet{barn96}. These boundaries can be seen in Table \ref{tbl-temp}. Even knowing that some of these previous schemes employed parameters others than based on infrared fluxes to set their temperature scales, we made an effort to agree our indices levels with these old ones, by just reformulating the range values of each level of the temperature index.

\section{Sample Classification}
\label{disc}
As a first application of the NSCC, we calculated all the parameters presented in sections \ref{optical} and \ref{ir} for all stars in our sample. Thus, up to seven indices of classification were established for each target, including for the 6 standards used in the development of the methodology, that can be seen in Table \ref{tbl-final}.

\subsection{Optical Parameters}
Through the spectral subclasses we identified not only C-R and C-N stars in the sample but also four C-H candidates, which should be further investigated, to a final conclusion. Among those, W CMa and V Hya have currently a C-N classification, and as it was not possible to analyze other reliable CH bands, they received a C-N: index in the final notation, denoting an uncertainty. Some stars previously classified as C-N, were identified as actually belonging to the C-R spectral subclass: DH Gem and UW Sgr were reclassified from C-N0 to C-R and NP Pup from C-N4.5 to C-R.

\begin{figure}
\epsscale{1.1}
\begin{center}
\plottwo{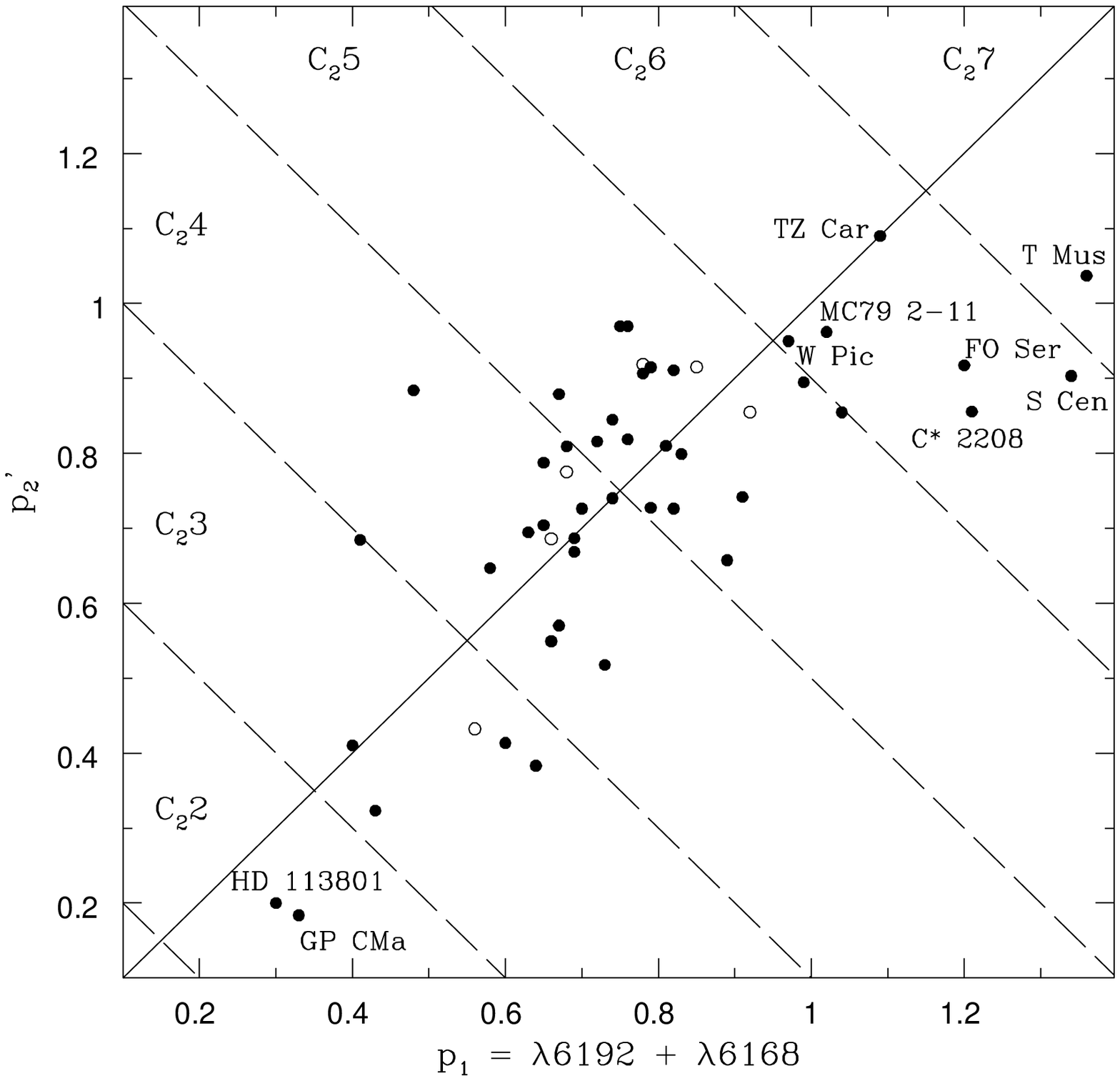}{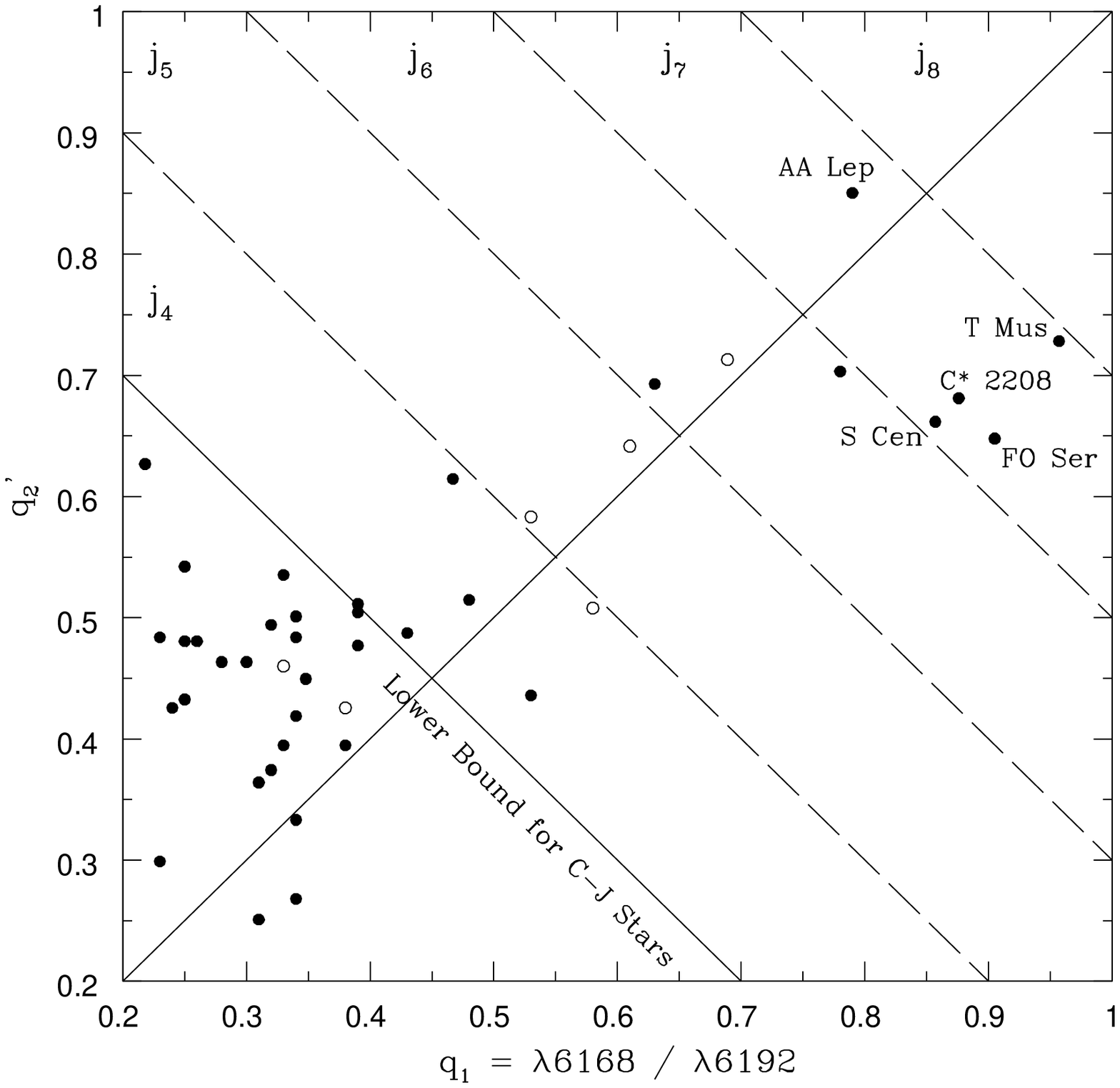}
\caption{On the left, the spread of the sample along the domains of the \CC-index, while on the right it is along the j-index domains. The standards used to determinate the ranges of both index are marked with open circles, and the boundary edges are represented with a dashed lines. The full line represents the situation in which the primary and secondary parameters match. On both figures the targets with higher values of the indices are named. On the right it is possible to see a correlation between the parameters adopted for the J-stars, although the same is not seen for the ones with j $\leqslant$ 4. \label{fig2}}
\end{center}
\end{figure}

The spread of our sample in regards to the parameters $p_1$ and $p_2^{\prime}$ for the \CC-index and the established ranges of the average indicator, $\bar{p}$, can be seen in Figure \ref{fig2} (left). The full line illustrates when primary and secondary parameters match, while dashed lines mark the edges of each level range. It's also possible to see, in this figure marked with open circles, the standard objects, which were used to obtain the ranges of each \CC-index level. The stars with the higher \CC-index, namely \CC6 and \CC7, were named in Figure \ref{fig2} (left) and T Mus has the higher estimate of carbon abundance. On the other hand, HD 113801 and GP CMa have the lower \CC-index calculated, \CC-2.

When the j-index was assigned to all stars, it revealed 14 targets as J-stars, three more than if selected only by \citeauthor{Gor68}'s definition: V971 Cen, \object[IRAS 06529+0626]{CL Mon} and \object[IRAS 08180+0520]{C\* 1130}. Figure \ref{fig2} (right) displays the edges of each level of the index in dashed lines and plots the j-index obtained for the whole sample. The correlation between the parameters $q_1$ and $q_2^{\prime}$ exists only for J-like stars, which means that, this behavior is an evidence of stars with $^{13}$C excess only. It can be seen that, indeed, it do not holds for stars with j $\leqslant$ 4. As done before, the objects used to calibrate the ranges of each level are marked with open circles. The stars with the higher j-index, namely j7, are named in Figure \ref{fig2} (right) and, again, the isotopic carbon abundance estimate for T Mus is the higher one. We confirmed the classification of \citet{chen07} for S Cen, V971 Cen and C* 2208 as C-J stars, in contrary to the C-R classification presented by \citet{barn96} for the first two and by \citet{chan93} for the later, in the past. 

The average parameter $\bar{d}$ were used to obtain the MS-index for the 32 stars of our sample. Comparing the MS-index with \CC-index it is possible to see that the stars with the higher value of the Si\CC\ abundance, index MS3, seems to be also those with the higher values of \CC-index (\CC6 and \CC7): S Cen, TZ Car, W Pic C* 2208, FO Ser and T Mus, for instance.

\begin{figure}
\epsscale{0.88}
\begin{center}
\plotone{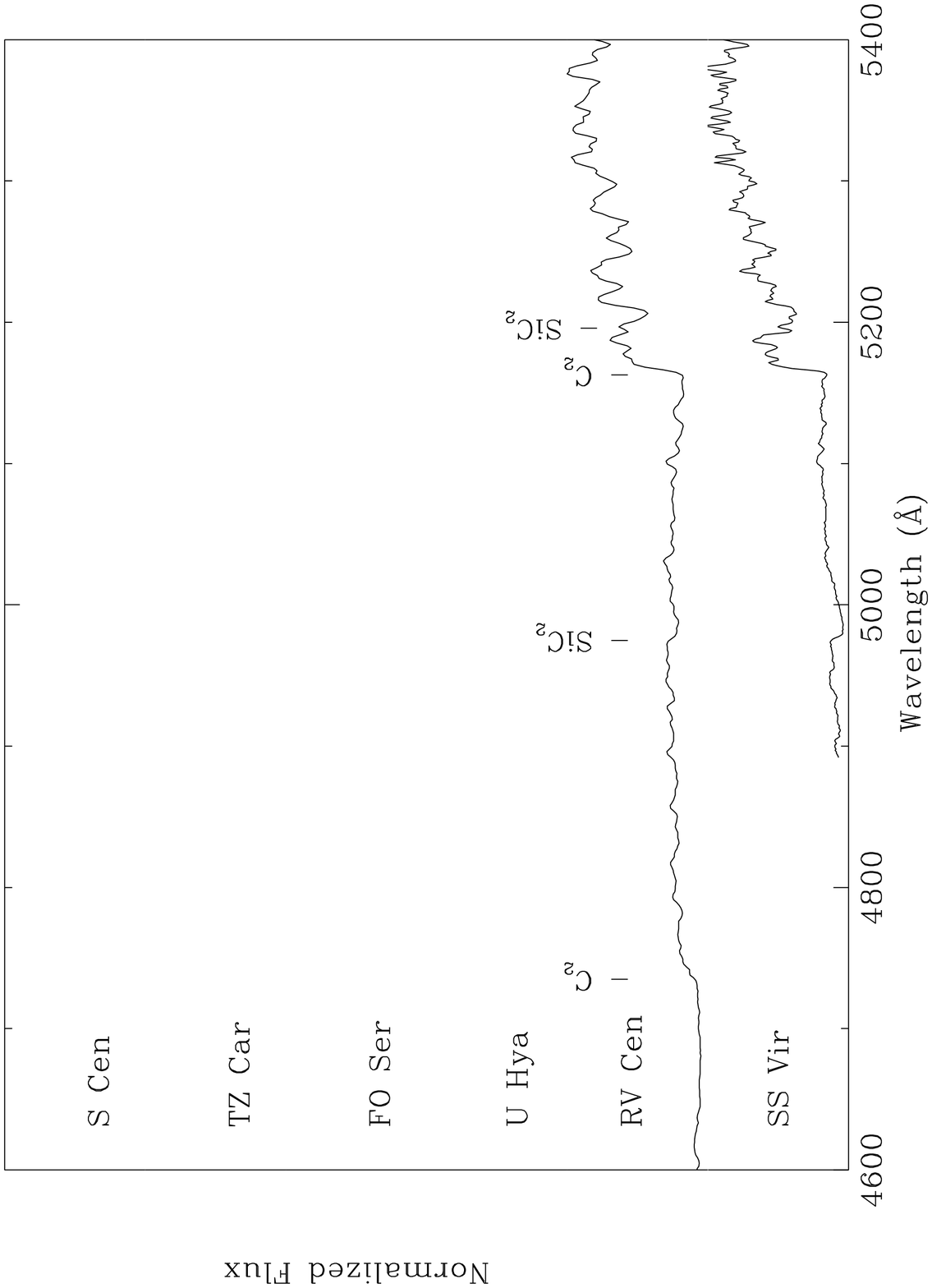}
\caption{C-N stars in a decreasing sequence of their temperature indices.} \label{specn1}
\end{center}
\end{figure}

The optical spectra of the stars classified in this work can be seen in Figures \ref{specn1} to \ref{specn8}. The most significant features for the classification of these objects under the NSCC are marked on the spectra whenever present. The spectra are displayed in a decreasing scale of temperature separated by their spectral subclasses: C-N stars are shown in Figures \ref{specn1} to \ref{specn5}, C-R in Figure \ref{specn6} and C-J in Figures \ref{specn7} and \ref{specn8}. C-H stars are displayed together with the C-N due to their spectral subclass uncertainties.

\subsection{Circumstellar Envelope Parameters}
We also have determined, for the first time, circumstellar envelope opacity index for 37 stars and SiC/A.C. ratio abundance index for 26 stars. C* 1003 and HV CMa obtained the higher values on $\tau$-index, suggesting an optically thick envelope usually associated with tip-AGB stars. It is possible to assume, that these stars already ejected a considerable amount of matter, which condensed and accumulated, creating a denser envelope. Nevertheless, some stars have an extremely low opacity value, $\tau$0, which represents the opposite scenario (e.g. NP Pup and BE CMa). These stars have an optically thin circumstellar envelope which may correspond to early stages of carbon AGB. In regards to the J-type carbon stars, we notice that the star T Mus not only have the higher optical indices, but also the higher opacity index, $\tau$2. It is possible, than, T Mus may be the most evolved J-type star from this sample, while, on the other hand, BE CMa could be the least evolved one. Not all \citetalias{iras} spectra here presented has a good signal to noise ratio, so, in these cases, after the circumstellar envelope classification the object with a poor infrared datum received an uncertainty index (:).

All models calculated for the stars that were classified in the infrared are displayed at the Figures \ref{modnormal} and \ref{modsil}. The stars in Figure \ref{modnormal}, ordinary carbon stars, had their radiative transfer models calculated to fit the 11.3 $\mu$m feature, associated with the ratio SiC/A.C., while Figure \ref{modsil} shows the models calculated to fit the silicate feature at 9.8 $\mu$m for the silicate carbon stars. Both groups are sorted by an increasing scale of their opacity indices.

The fit of the model is not always perfect. It can be due to different effects. For instance, the data of each target were obtained by several observers in different moments, which means that, the photometric data and the \citetalias{iso} or \citetalias{iras} spectra were observed at different time. Knowing that carbon stars are variable stars, it is expected that the flux varies with each luminosity phase. Another effect is that, in some cases, the circumstellar envelope may not be spherically symmetric, and therefore, the fits, which results from the spherically symmetric models, are not always a good first approximation.

SiC grains may nucleate closer to the star, at higher temperatures of about 1300 K to 1500 K, and therefore they may be the first grains to condense in the carbon-rich star envelopes \citep{lml94}, while A.C. grains condense at 1000K \citep{f89}. High values of SiC/A.C. ratio denote, then, early stages of the AGB. When more carbon is ejected to the circumstellar envelope, during the thermal pulse phase, other compounds begin to nucleate and become more abundant than the SiC, for instance amorphous carbon grains. Thus, we can expect that more evolved AGB stars, besides being optically thicker, have lowest SiC/C.A. ratios. 

It is also possible to assume that the circumstellar opacity increases monotonically with time, since as more evolved the star is, the thicker is its envelope. Differently, a given SiC/A.C. ratio decreases gradually during the evolution of an AGB carbon star. Figure \ref{fig9} illustrates both parameter behaviors by the global spread of the circumstellar opacity and SiC/A.C. ratio values calculated for the sample.

We can assume that, by the observed spread, as the envelope opacity grows thicker, the SiC/A.C. ratio decreases with a power-law: $\mathrm{SiC}/\mathrm{A.C.} = \beta \tau^{\alpha}$. Considering the stars that exhibit the greater values of SiC/A.C. at each opacity, i.e. the outer points in Figure \ref{fig9}, we may presume that these stars represent the different evolutionary stages of an SiC/A.C. extreme object. Therefore, through an empirical fit to these stars we established $\alpha$ as approximately $-0.5$. 

Varying the $\beta$ coefficient it is possible to obtain several evolutionary sequences determined by each initial SiC/A.C. ratio. Clearly, we can not consider $\tau = 0$ to compute the initial ratio, since at this stage level, the SiC/A.C. value would be indeterminate by the lack of matter at the circumstellar envelopes. But, assuming an initial opacity of $\tau = 0.25$, which we may consider a stage were it start having a significant density of condensates, it is then possible to calculate the $\beta$ value. Thus, the evolutionary sequence expression that connects circumstellar opacity with SiC/A.C. ratio is:
\begin{equation}
\mathrm{SiC/A.C.} = \frac{\mathrm{SiC/A.C.}_{initial}}{2}\ \tau^{-1/2} .
\end{equation} 

Four examples of evolutionary sequences are plot in Figure \ref{fig9} with the spread of C-N, C-R, and C-J stars of our sample. The evolutionary sequence of a C-J stars is still an open question, but it surely does not follow the same evolution of an ordinary AGB carbon star. These stars are spectroscopically similar to the C-R stars, which present anomalous carbon enrichment during the second dredge-up, to such an extent that it is possible that the later evolved to C-J \citep{evans86}. For fact, as a star go further into the TP-AGB phase, the $^{12}$C:$^{13}$C ratio increase with the thermal pulses, i.e. decreasing the photospheric $^{13}$C abundance. We may, then, consider that as a C-J star evolves into the TP-AGB, it gradually change their photospheric abundance becoming, at the end, a C-N star. It is cautious to point out that not all the C-N stars evolves from a C-J star, most of them should follow the well accepted sequence of M $\longrightarrow$ MS $\longrightarrow$ S $\longrightarrow$ SC $\longrightarrow$ C-N (see e.g. \citet{herwig05}). But, if we consider the previous suggestion for the C-J evolution, the possible sequence that describes it is C-R $\longrightarrow$ C-J $\longrightarrow$ C-N. 

Regarding the circumstellar envelope, we can consider that C-R and C-J star becomes carbon-rich earlier than C-N stars, as they began their C-rich mass loss earlier. Therefore, their envelope should have SiC/A.C. ratios greater than the C-N stars, since C-R and C-J began to nucleate grains before the later ones. Additionally, following our suggested evolutionary sequence, it is reasonable to expect that a C-N star has a thicker envelope than a C-J, and on the other hand, C-J's envelope should be also thicker than the C-R's. Figure \ref{fig9} illustrates this possible evolutionary sequence C-R $\longrightarrow$ C-J $\longrightarrow$ C-N as can be seem in both the sequences with $\beta = 0.02$ and $\beta = 0.24$. Of course, a greater sample is essential to confirm this assumption.

\subsection{Comparison with Other Schemes} 
\label{comp}
A discussion about how consistent is the NSCC compared with previous schemes is important to evaluate its reliability. The seven indices of our new system acquired for the sample were qualitatively compared with the most quoted schemes for carbon stars. Although most of the indices could be examined, the infrared ones are improvements to our scheme and not all of them have been consider by other authors.

The spectral subclasses obtained through the NSCC methodology could be compared with two schemes of classification, \citet{Keenan93, barn96}, that had 4 and 11 stars of our sample already published, respectively. The \CC-index were examined in regards to three schemes: \citet{Yama72,Yama75} that had 16 stars in common with our sample, \citet{Keenan93} with 4 and \citet{barn96} with 11. In total, 18 stars of \citet{Yama72,Yama75}, 1 of \citet{Keenan93} and 3 of \citet{barn96} could be used in comparison with the j-indices obtained through the NSCC methodology. These three schemes were also used to compare the MS-index published with ours, however, only 6 stars of \citet{Yama72,Yama75}, 1 of \citet{Keenan93} and 3 stars of \citet{barn96} could be used. The consistence of the NSCC indices with the ones published by these authors can be seen in Table \ref{consist}. The \citeauthor{barn96}'s classification seems to better agree with the NSCC optical indices.

Concerning the circumstellar envelope indices, only the temperature and the SiC-indices could be discussed, because no other scheme of classification had presented an opacity index. The temperature index obtained through the NSCC methodology could be compared with the ones of three schemes of classification: \citet{Yama72,Yama75} with 18 stars, \citet{Keenan93} with 4 and \citet{barn96} with 11 ones. On the other hand, the SiC-index were just compared with the scheme of \citet{sloan98}. For this last comparison, we considered the equivalence of the indices as described in section \ref{sic} and the authors have 12 stars published in common with the ones in our sample.

\section{Conclusions}
\label{conc}

This New Scheme of Classification of C-Rich AGB Stars can not be applied to any carbon star. It is not a good application for AGB carbon stars that have a very thick envelope, e.g. extreme AGB stars can not be treated this way as their optical spectra are highly obscured. Moreover, the methodology employed is not tied to the sample presented, it has the flexibility to serve to all these kind of carbon stars by using the given coefficients and parameters. The indices correspond to either a direct measurement of the intensities and equivalent widths of features observed in low resolution spectra of carbon-rich stars or obtained through a radiative transfer model fit to infrared data. 

The seven indices presented describe in detail the complex scenario of the carbon rich stars. It is possible just by analyzing the compact final notation to get a full set of basic information about an AGB carbon star. As all calibrations were established based on the well quoted works, the indices and levels employed represent the more successful historical parameters on the study of carbon stars.

Regarding the use of the indices, we demonstrated how the SiC/A.C. ratio and the $\tau$ parameters could together provide information about the evolutionary stages of the carbon-rich stars. The evolutionary sequences here proposed may provide a new insight over the nature of objects such as the odd C-J stars.






\acknowledgments

A.\,B.\,M. acknowledges support from Conselho Nacional de Desenvolvimento Cient\'ifico e Tecnol\'ogico - CNPq - Brazil and S.L.M. acknowledges support from CAPES (AEX 1360/06-0) for this work.

\bibliography{demello}
\clearpage

\clearpage

\clearpage

\clearpage

\clearpage

\clearpage

\clearpage

\clearpage

\begin{figure}
\begin{center}
\plotone{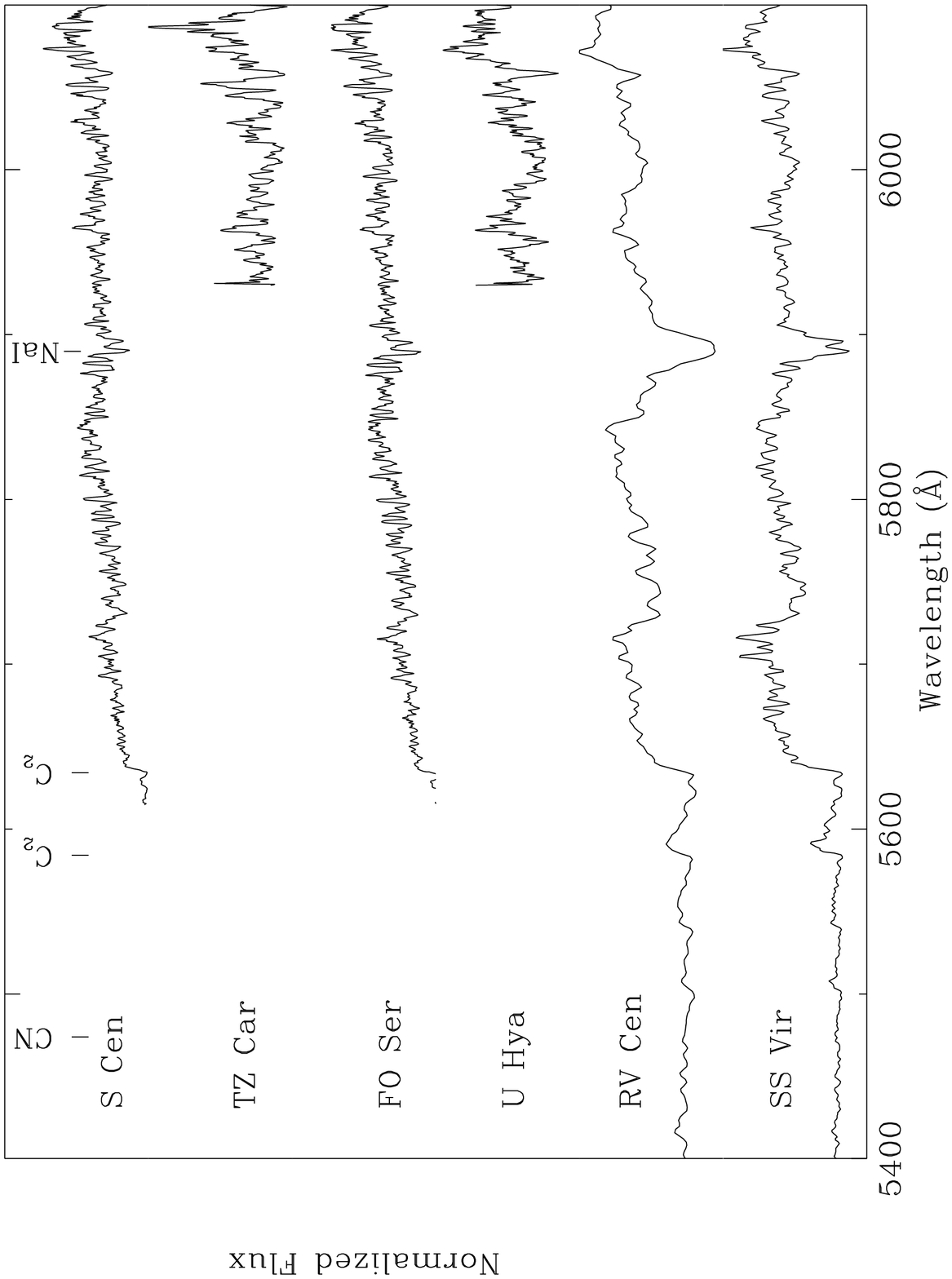}
\end{center}
Fig.~\ref{specn1}.--- continued
\end{figure}
\begin{figure}
\begin{center}
\plotone{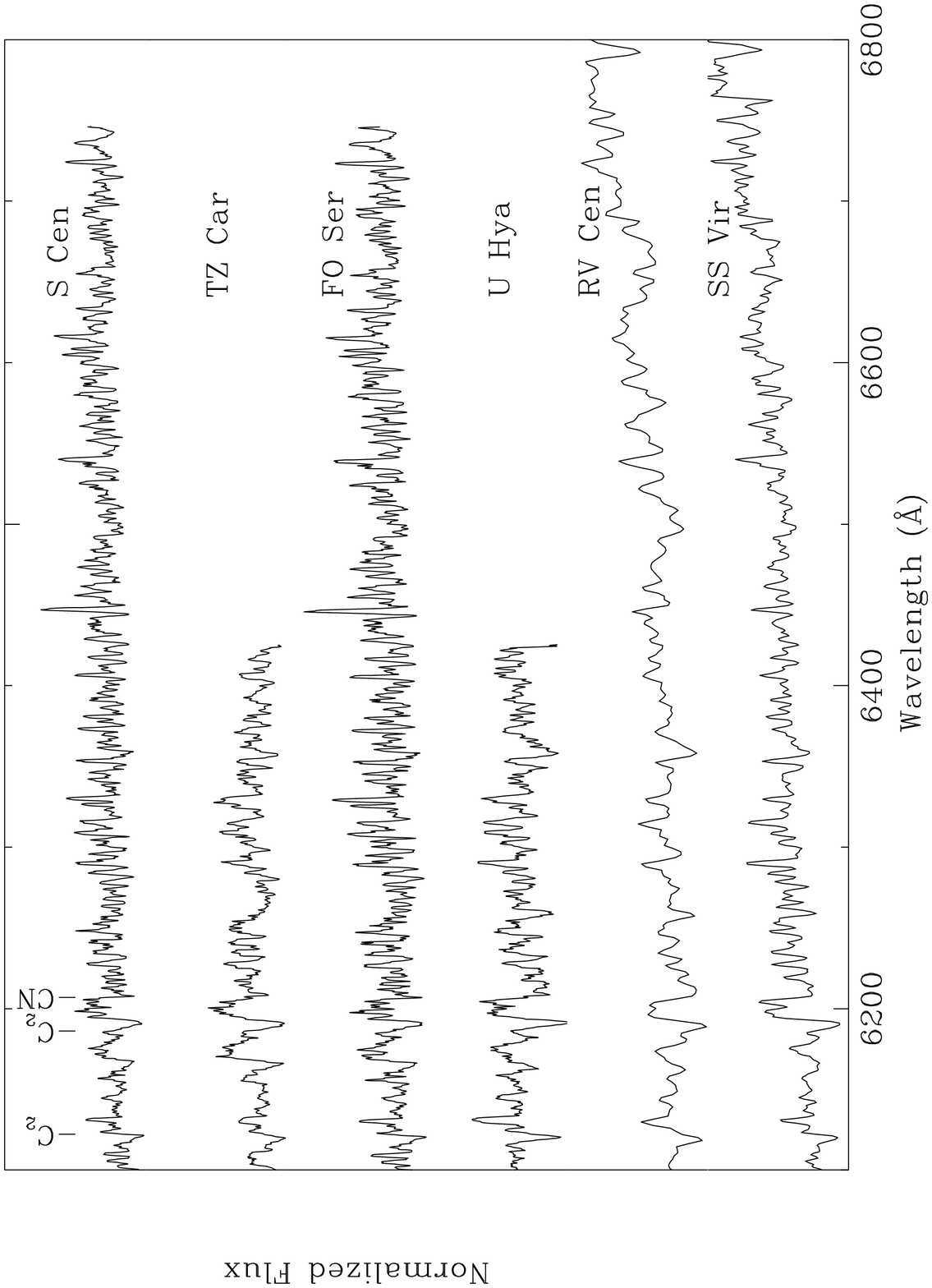}
\end{center}
Fig.~\ref{specn1}.--- continued
\end{figure}

\begin{figure}
\begin{center}
\plotone{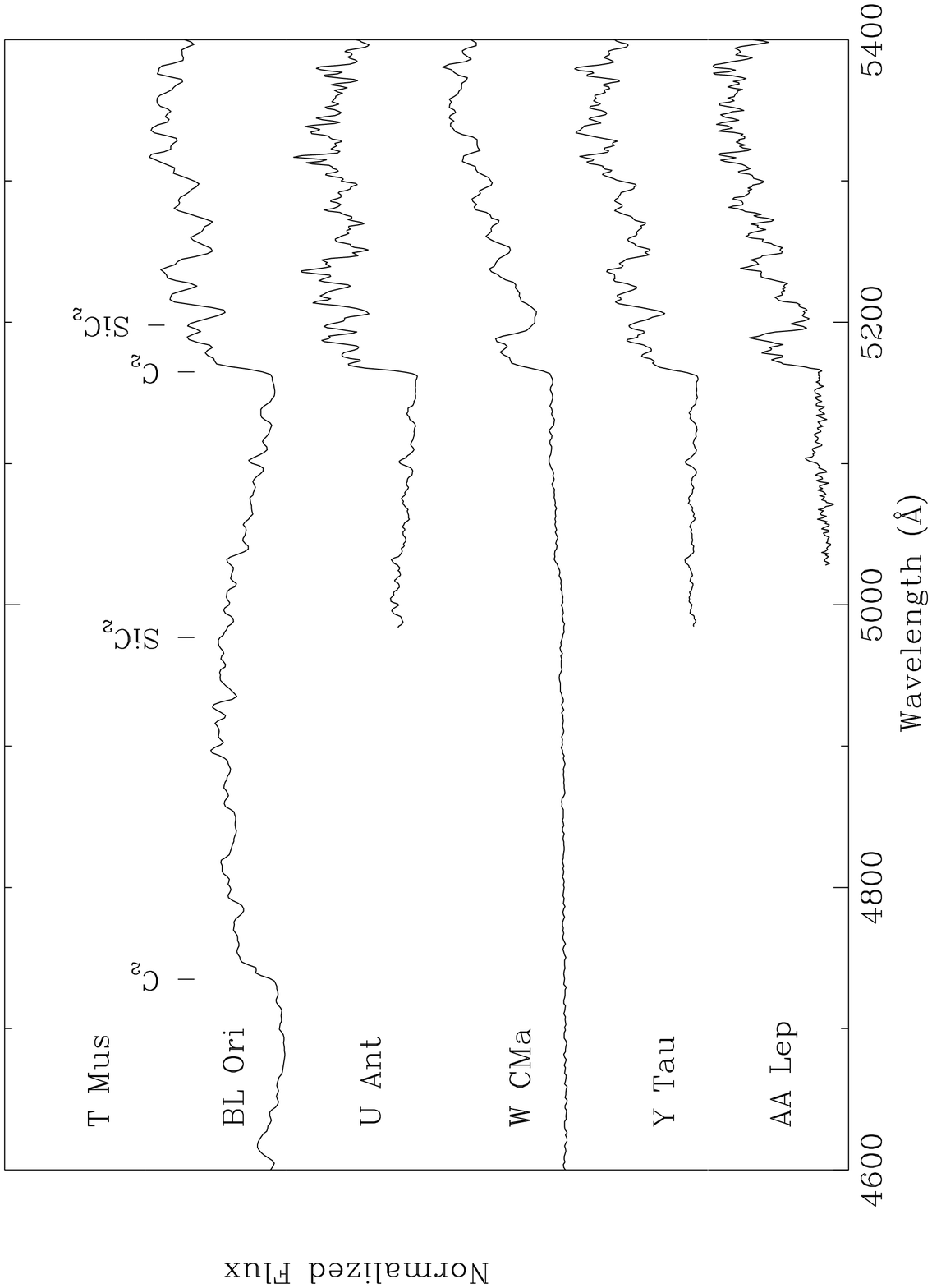}
\caption{C-N stars in a decreasing sequence of their temperature indices.} \label{specn2}
\end{center}
\end{figure}
\begin{figure}
\begin{center}
\plotone{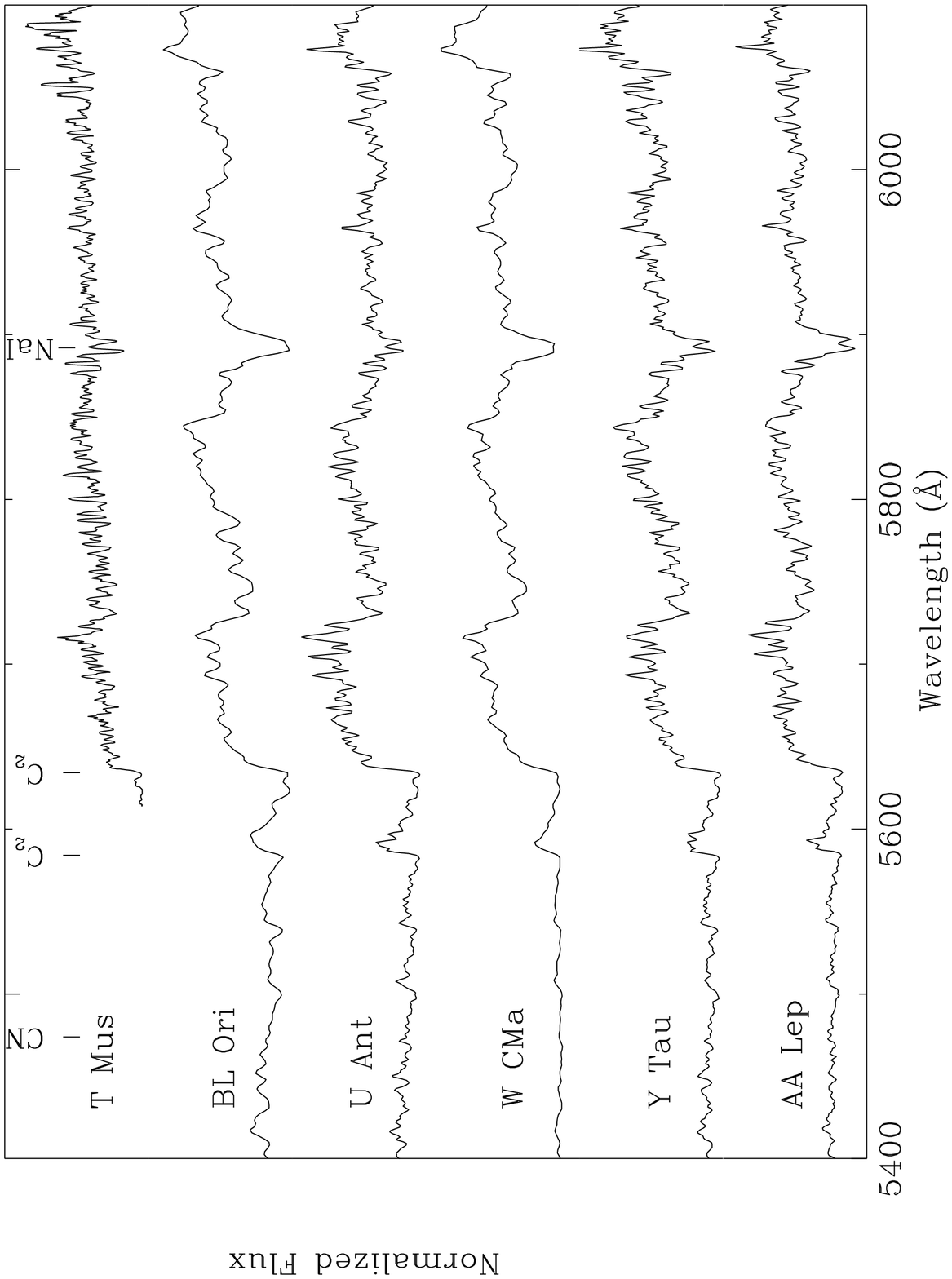}
\end{center}
Fig.~\ref{specn2}.--- continued
\end{figure}
\begin{figure}
\begin{center}
\plotone{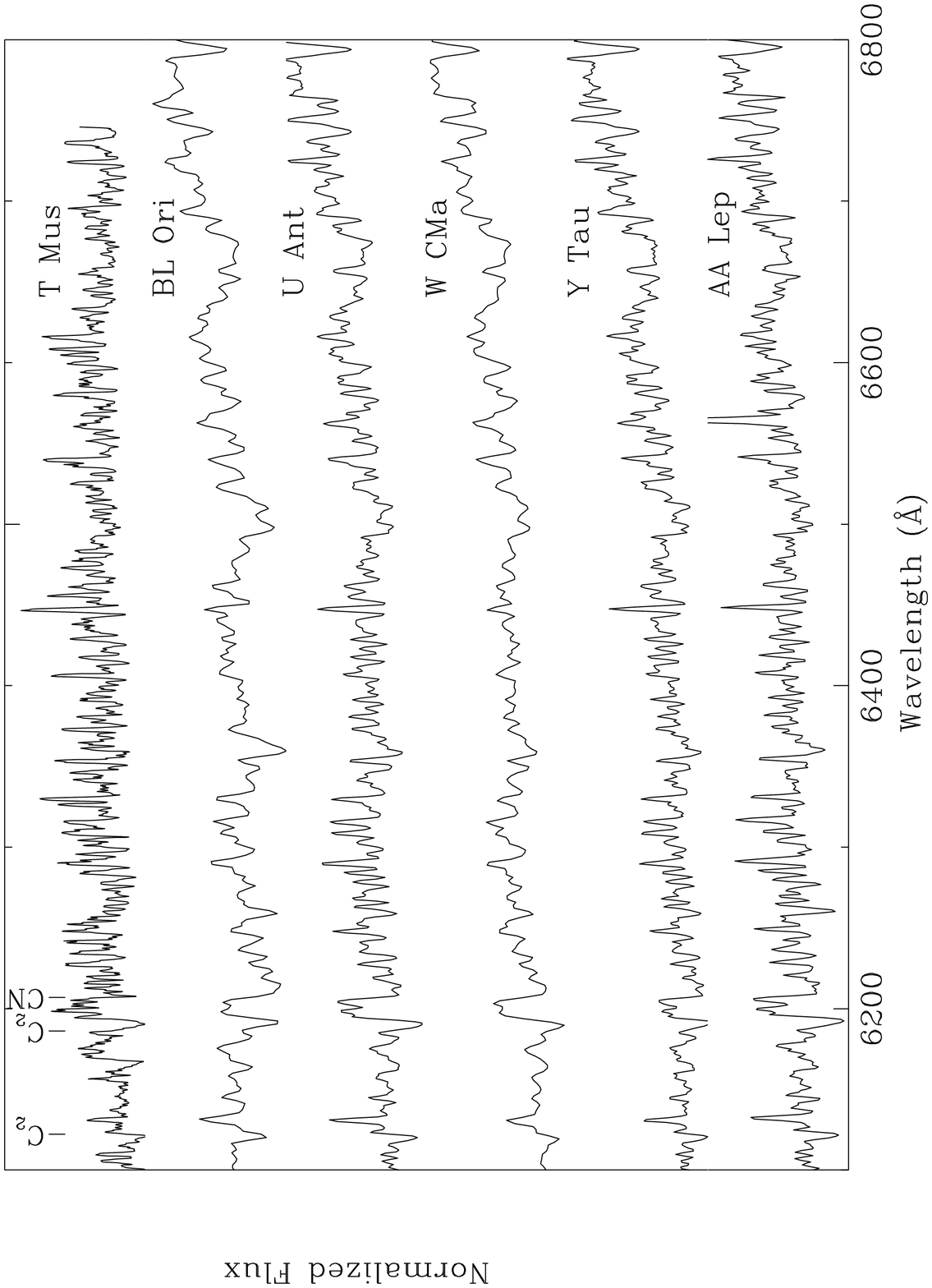}
\end{center}
Fig.~\ref{specn2}.--- continued
\end{figure}

\begin{figure}
\begin{center}
\plotone{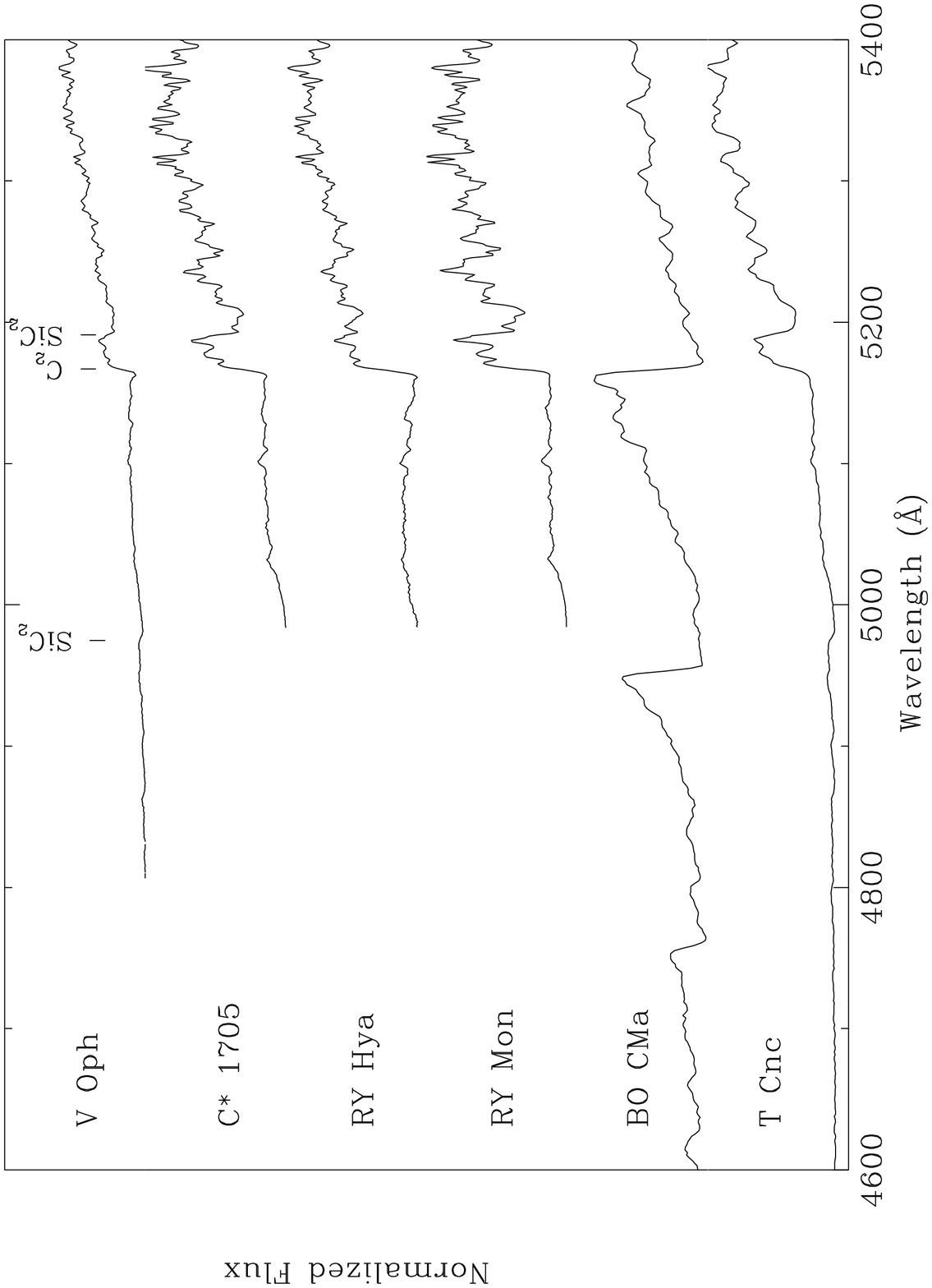}
\caption{C-N stars in a decreasing sequence of their temperature indices.} \label{specn3}
\end{center}
\end{figure}
\begin{figure}
\begin{center}
\plotone{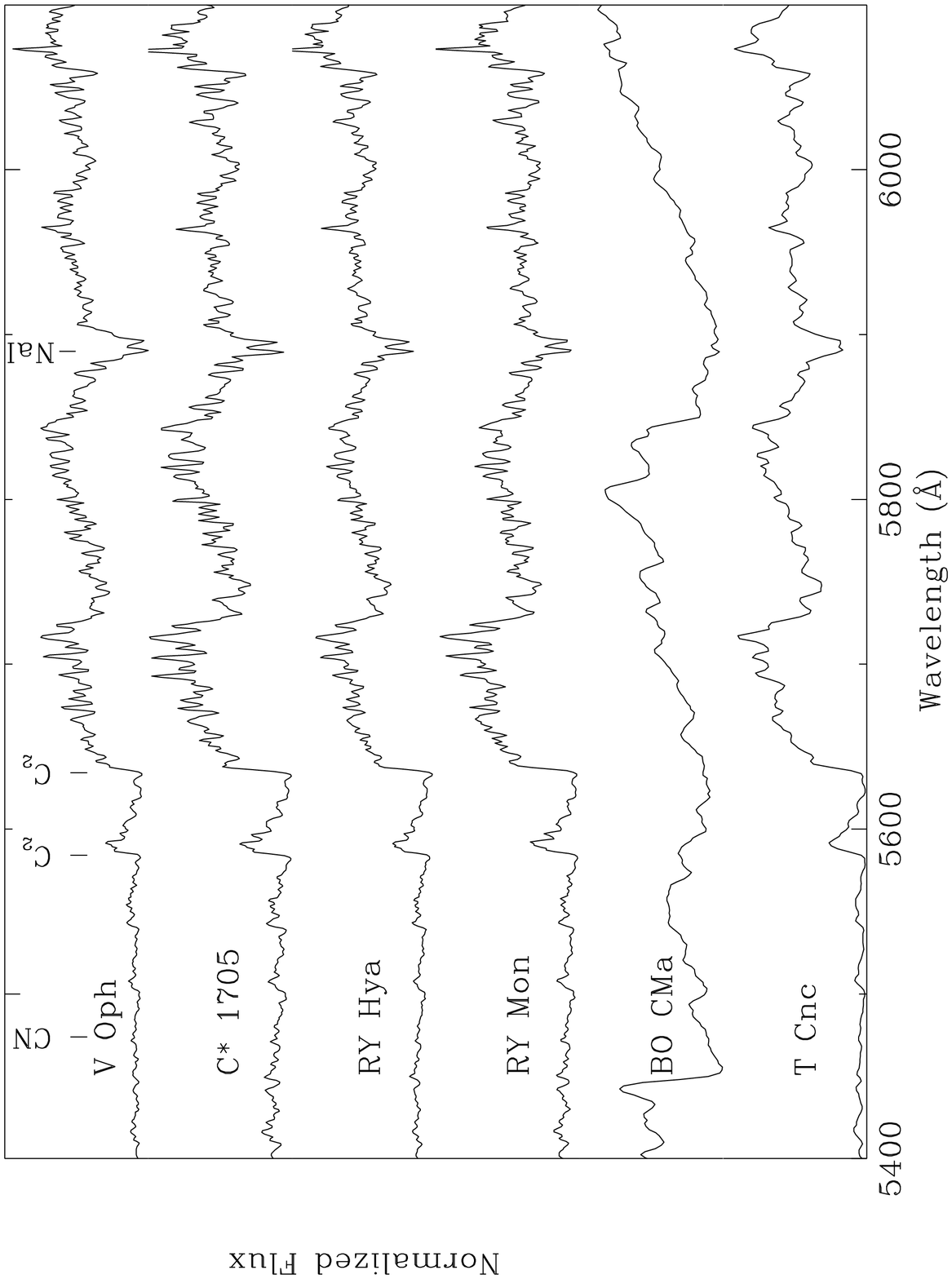}
\end{center}
Fig.~\ref{specn3}.--- continued
\end{figure}
\begin{figure}
\begin{center}
\plotone{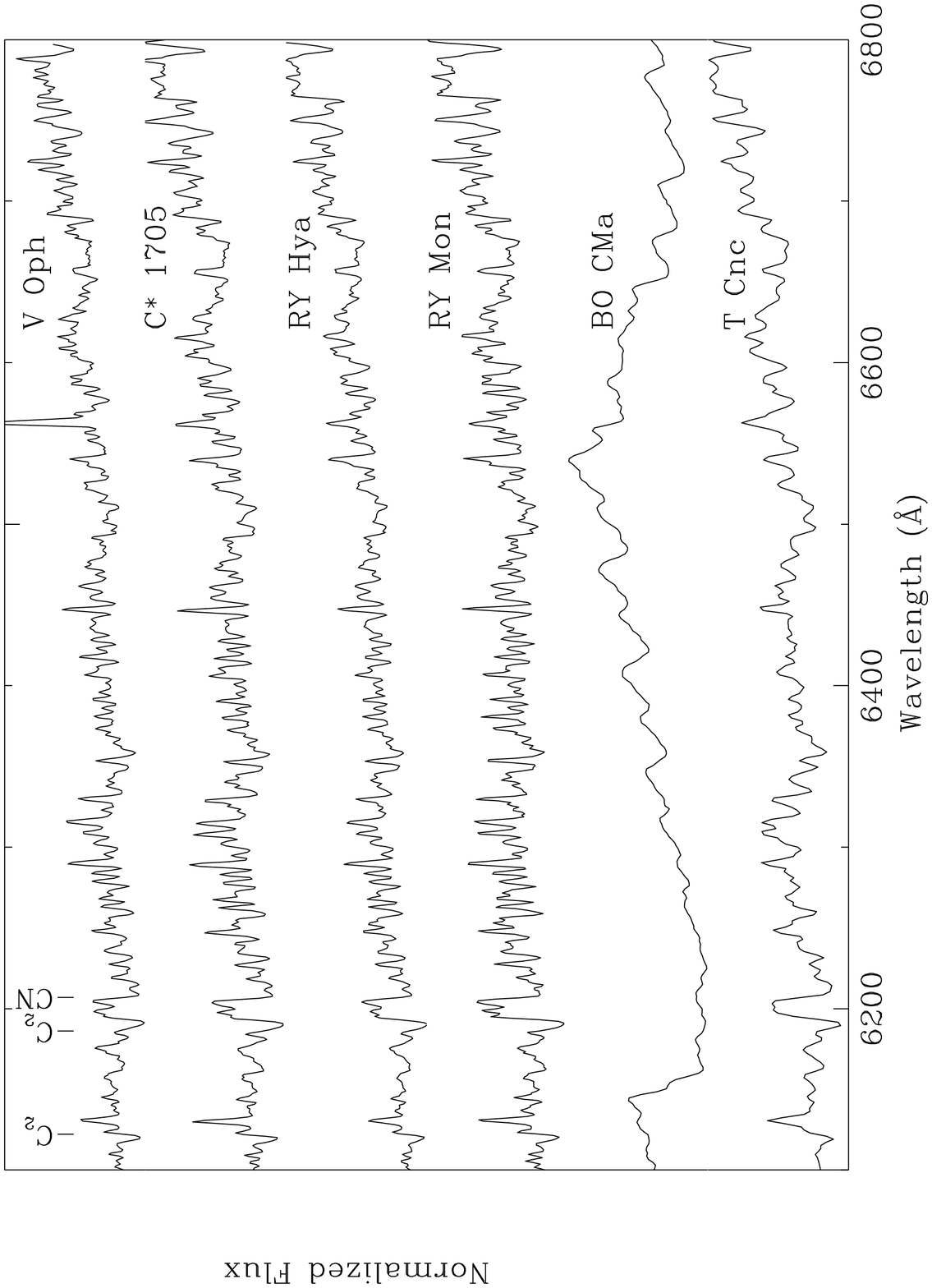}
\end{center}
Fig.~\ref{specn3}.--- continued
\end{figure}

\begin{figure}
\begin{center}
\plotone{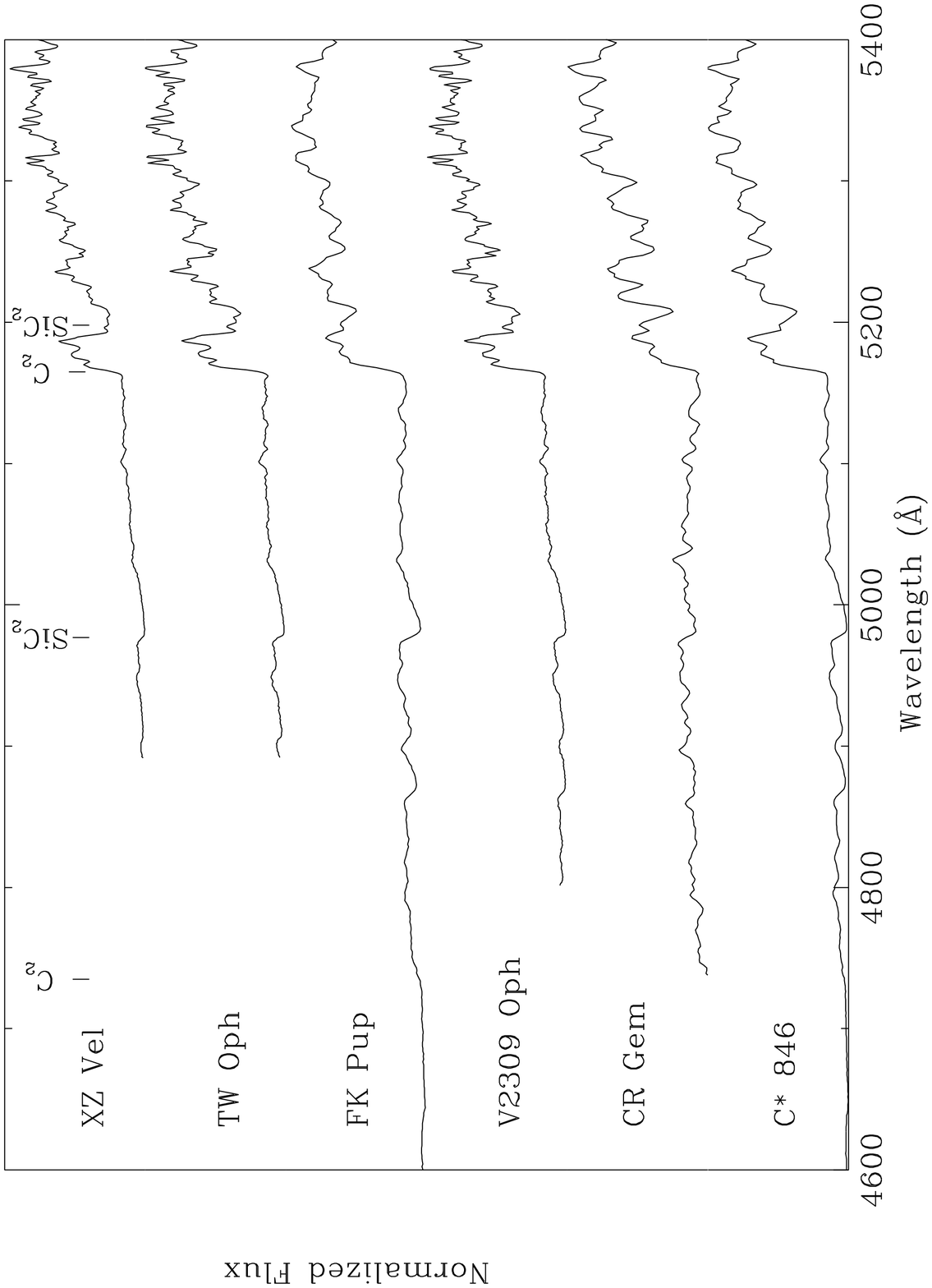}
\caption{C-N stars in a decreasing sequence of their temperature indices.} \label{specn4}
\end{center}
\end{figure}
\begin{figure}
\begin{center}
\plotone{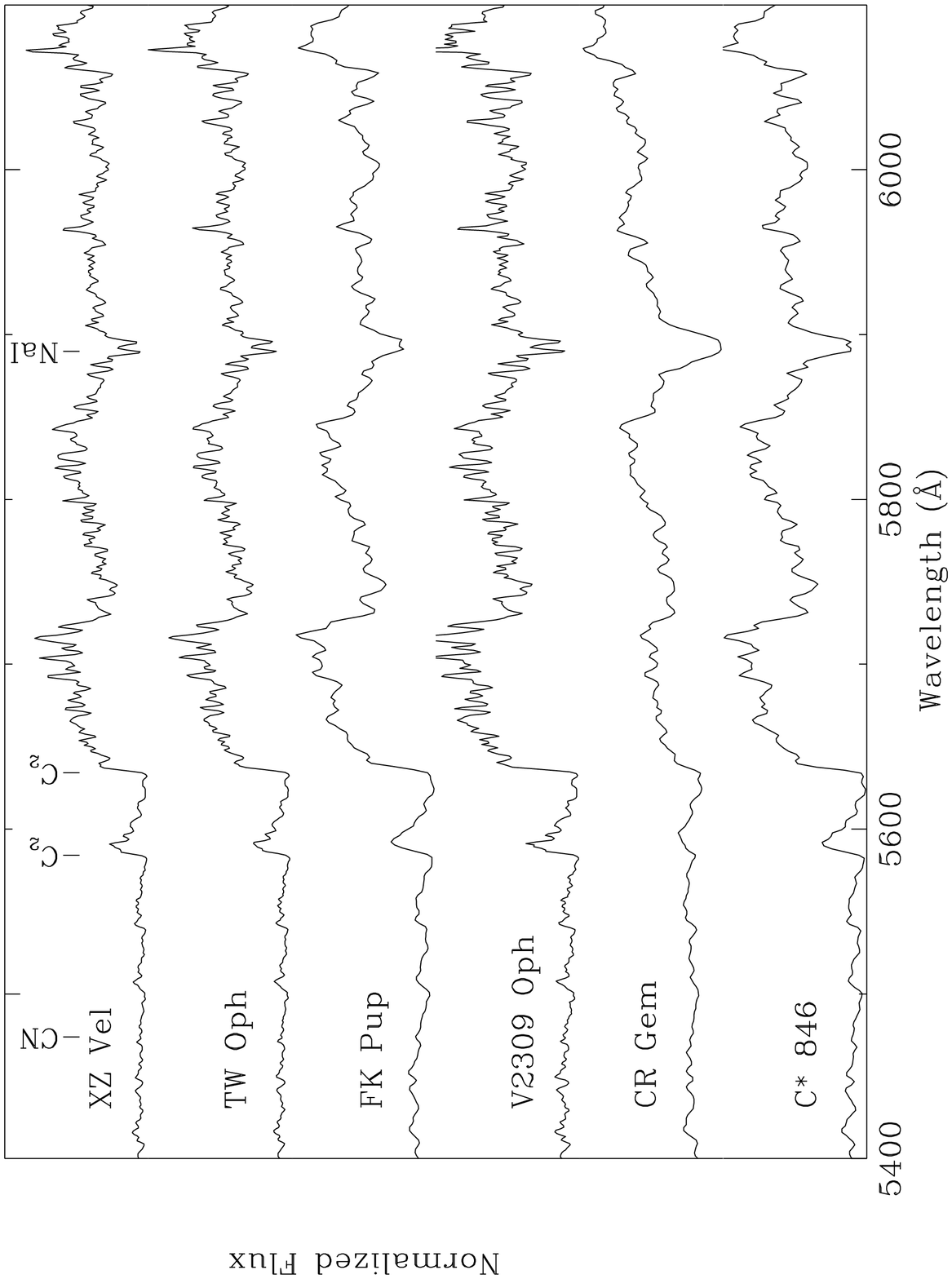}
\end{center}
Fig.~\ref{specn4}.--- continued
\end{figure}
\begin{figure}
\begin{center}
\plotone{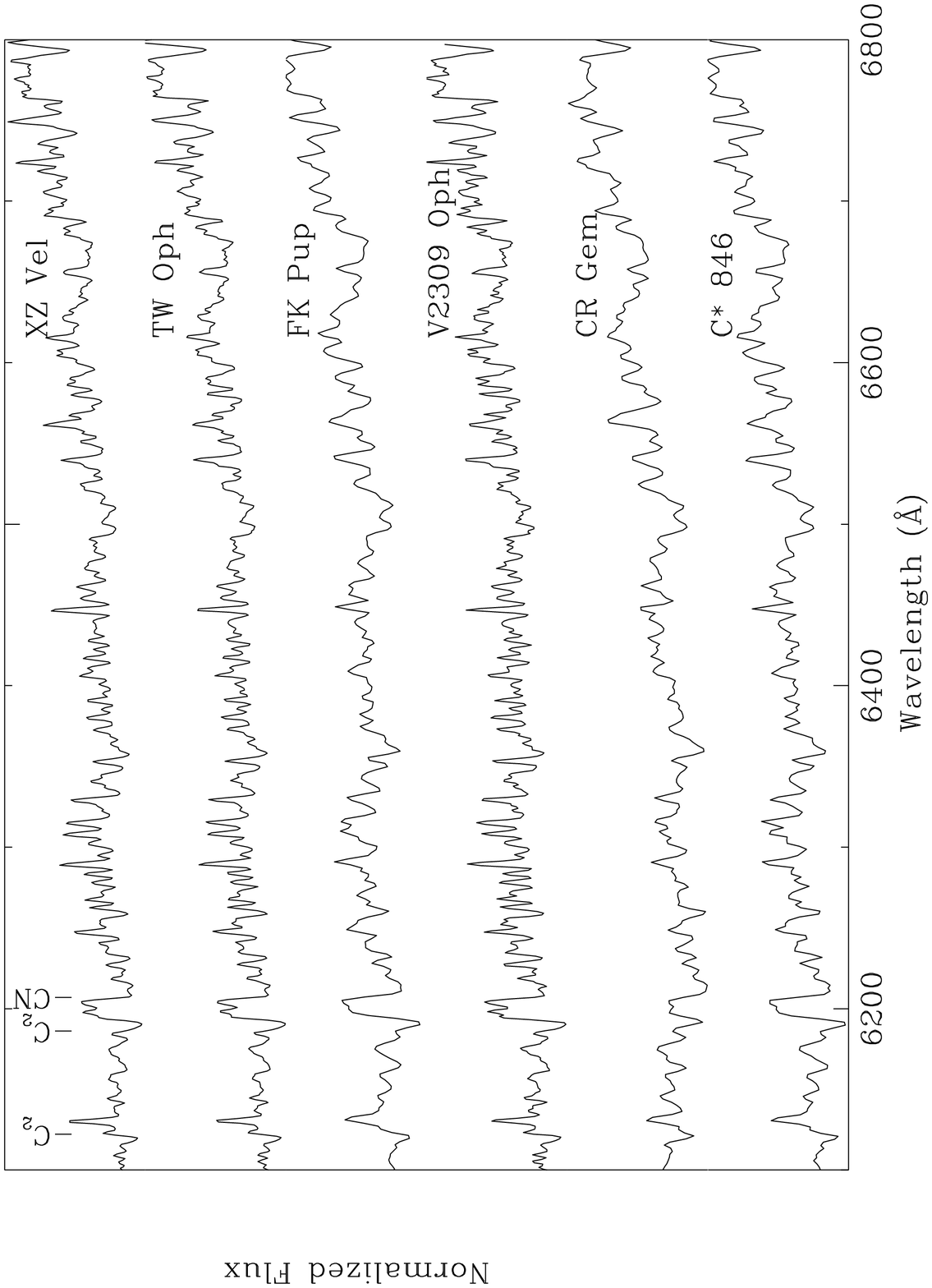}
\end{center}
Fig.~\ref{specn4}.--- continued
\end{figure}

\begin{figure}
\begin{center}
\plotone{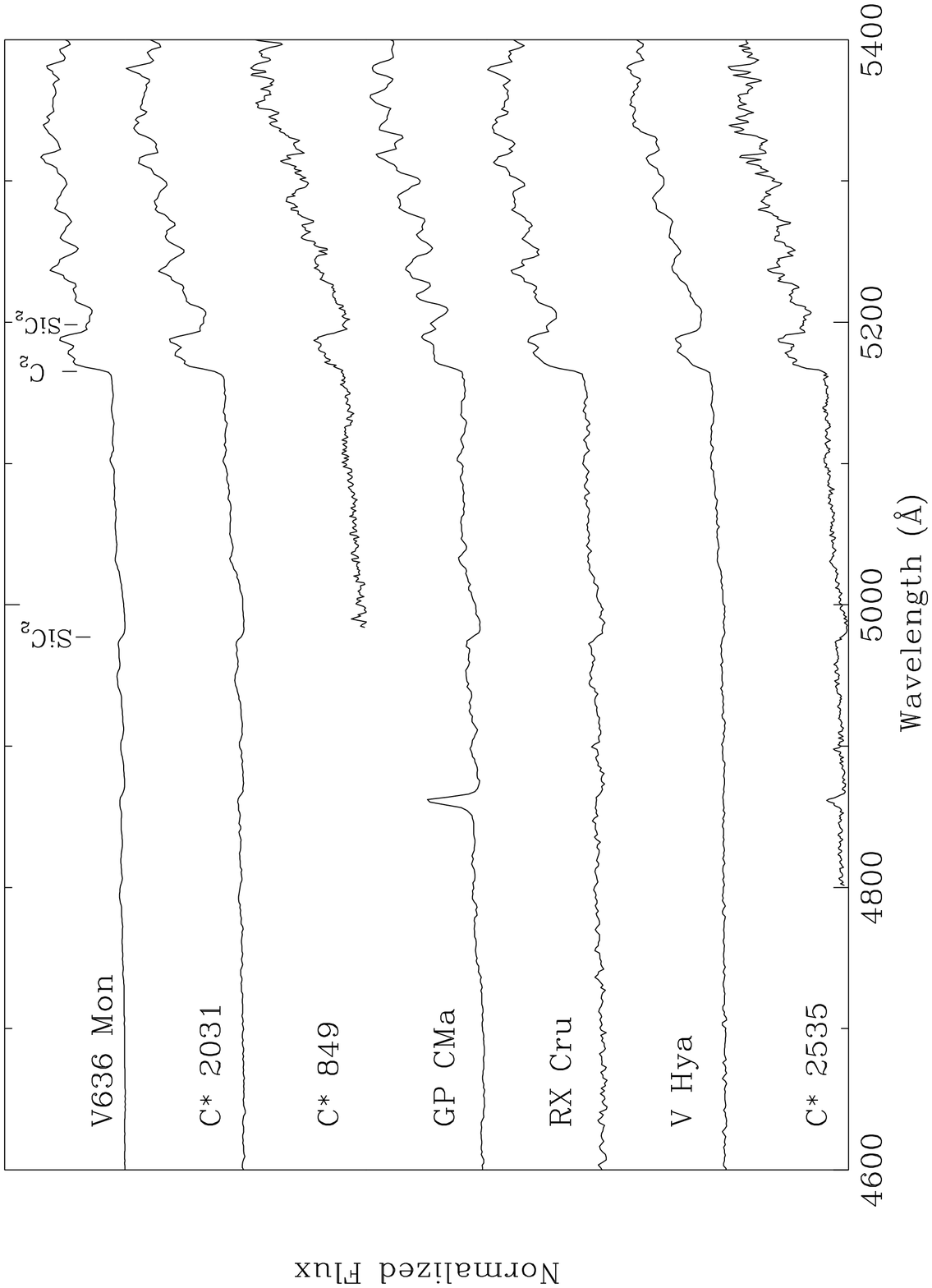}
\caption{C-N stars in a decreasing sequence of their temperature indices.} \label{specn5}
\end{center}
\end{figure}
\begin{figure}
\begin{center}
\plotone{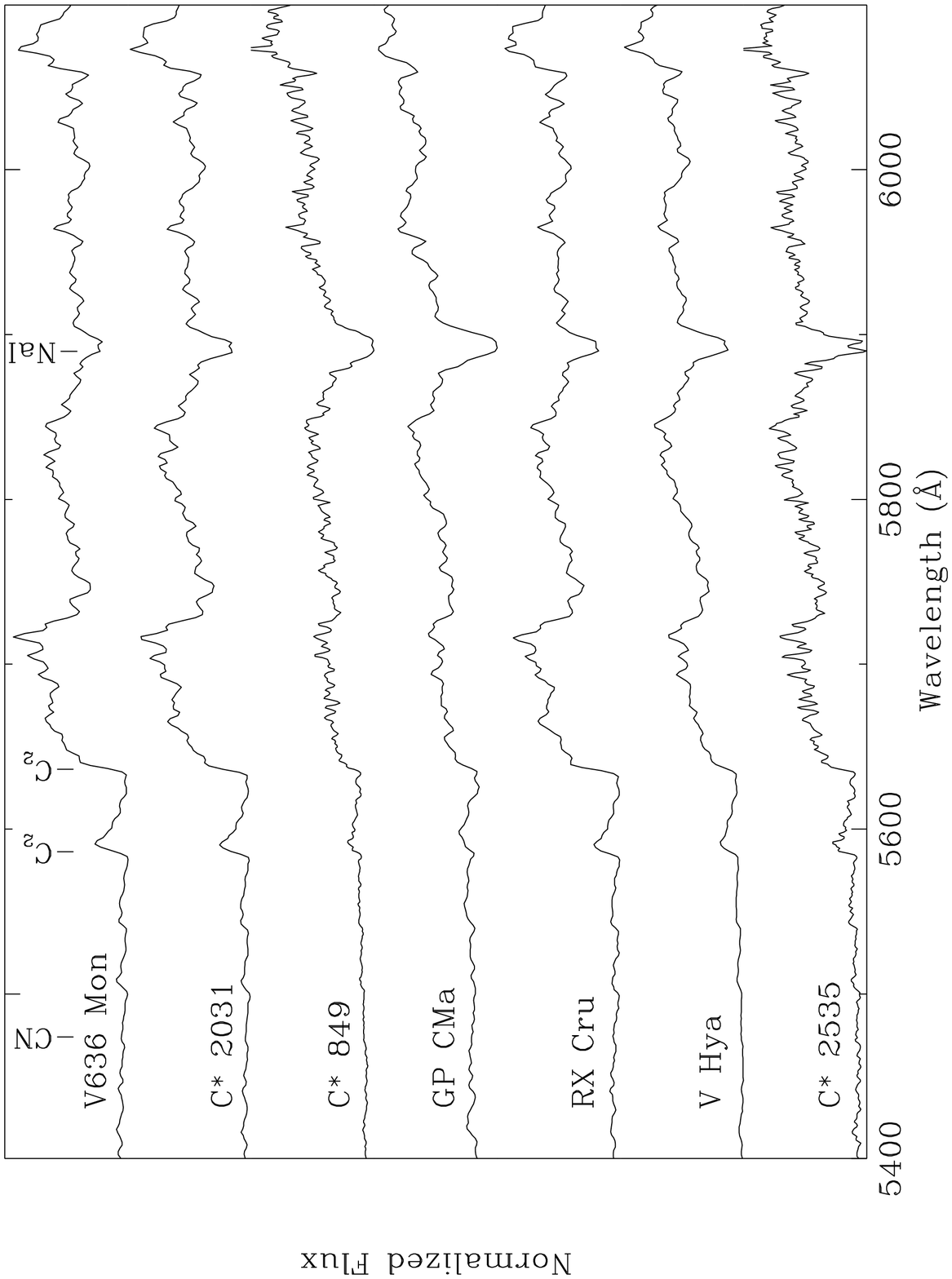}
\end{center}
Fig.~\ref{specn5}.--- continued
\end{figure}
\begin{figure}
\begin{center}
\plotone{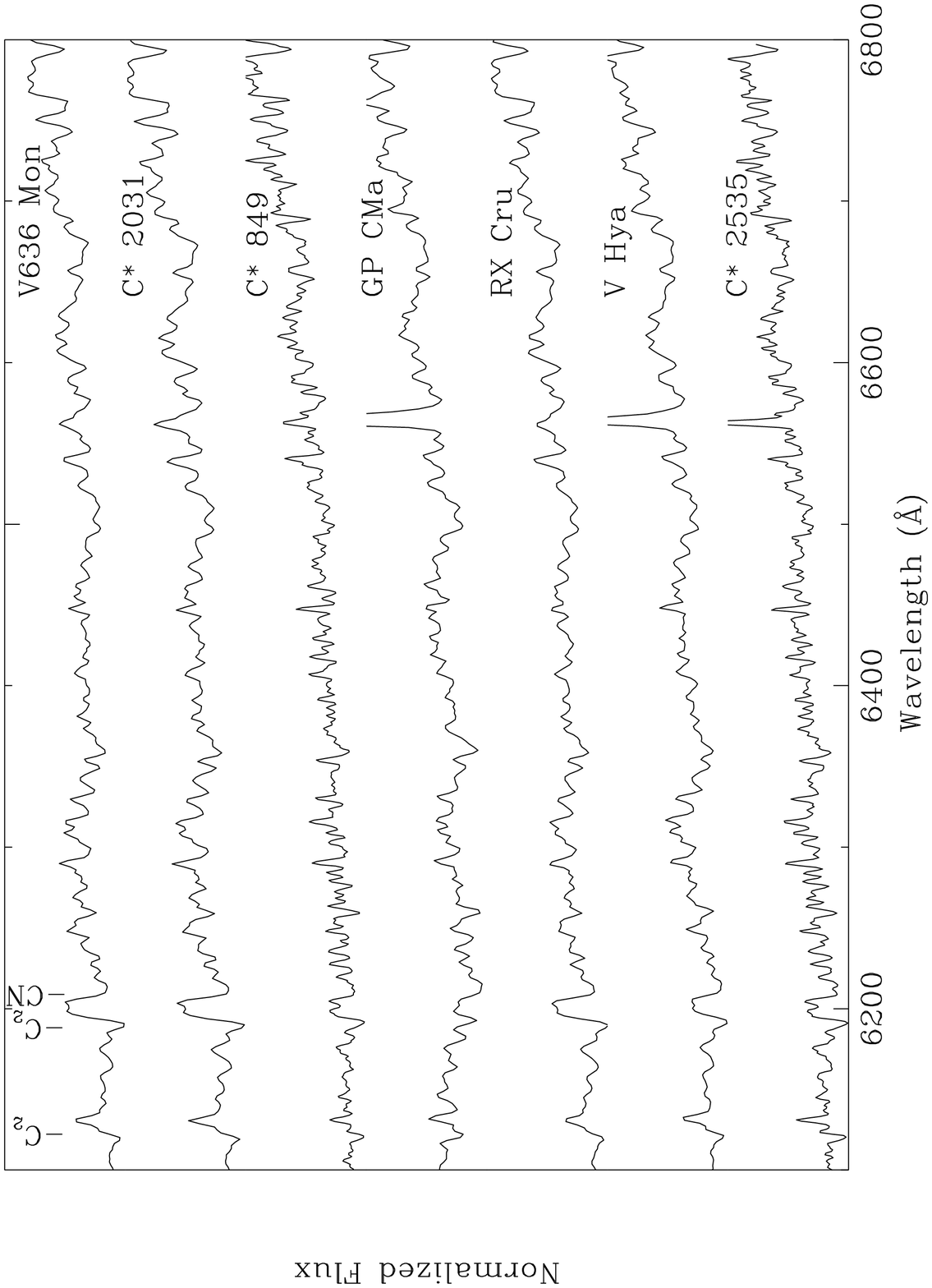}
\end{center}
Fig.~\ref{specn5}.--- continued
\end{figure}
\clearpage

\begin{figure}
\begin{center}
\plotone{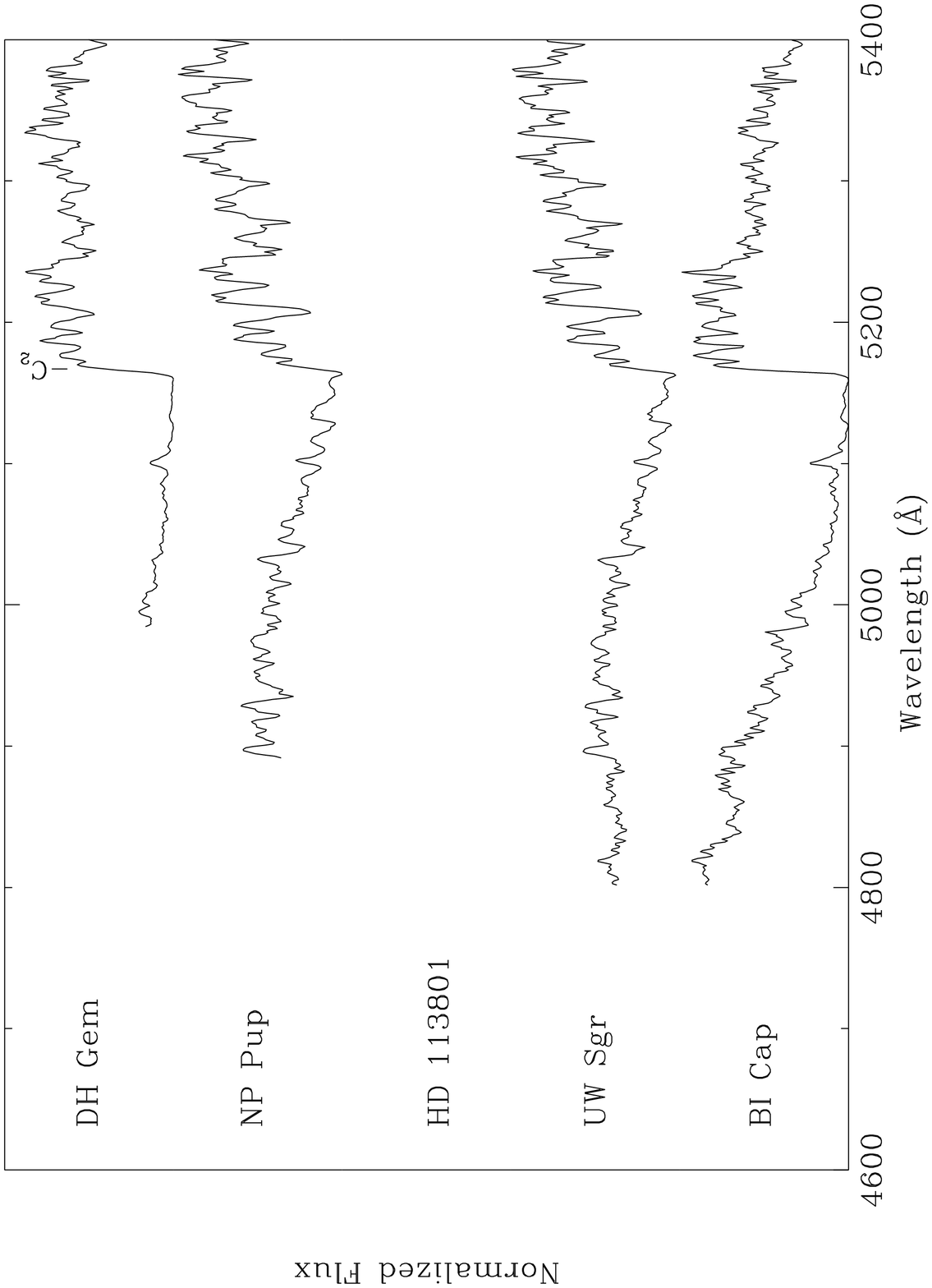}
\caption{C-R stars in a decreasing sequence of their temperature indices.} \label{specn6}
\end{center}
\end{figure}
\begin{figure}
\begin{center}
\plotone{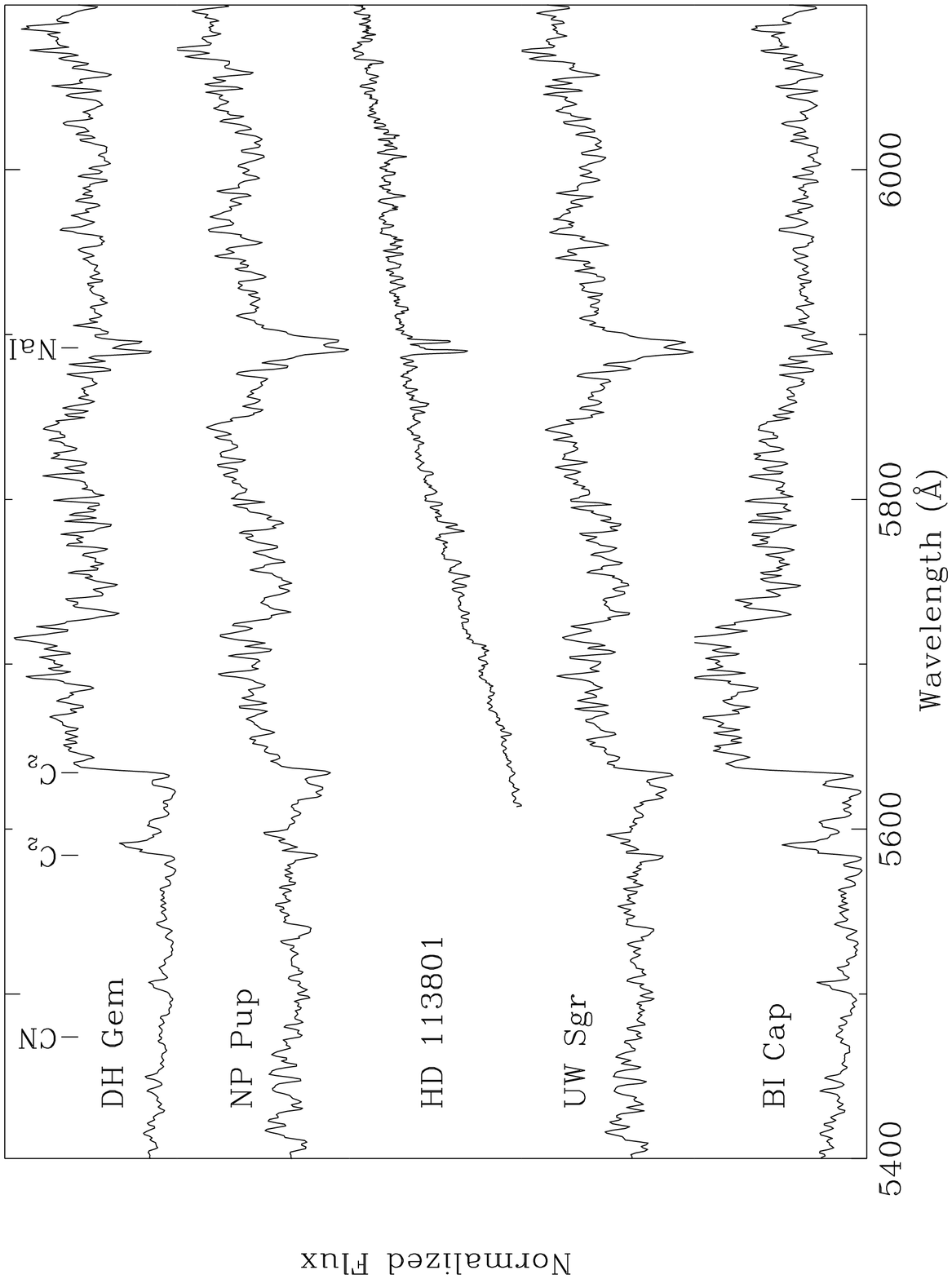}
\end{center}
Fig.~\ref{specn6}.--- continued
\end{figure}
\begin{figure}
\begin{center}
\plotone{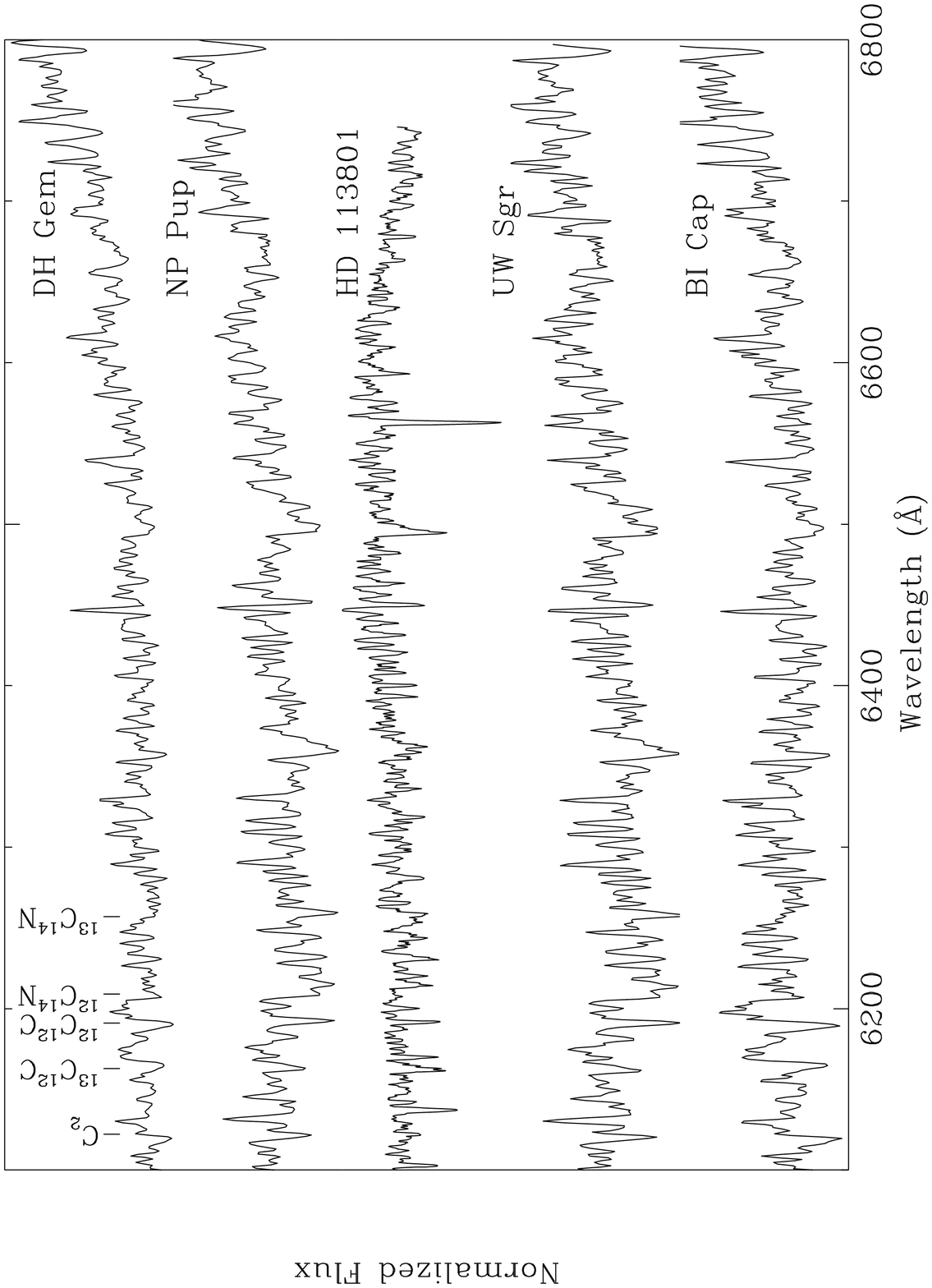}
\end{center}
Fig.~\ref{specn6}.--- continued
\end{figure}
\clearpage

\begin{figure}
\begin{center}
\plotone{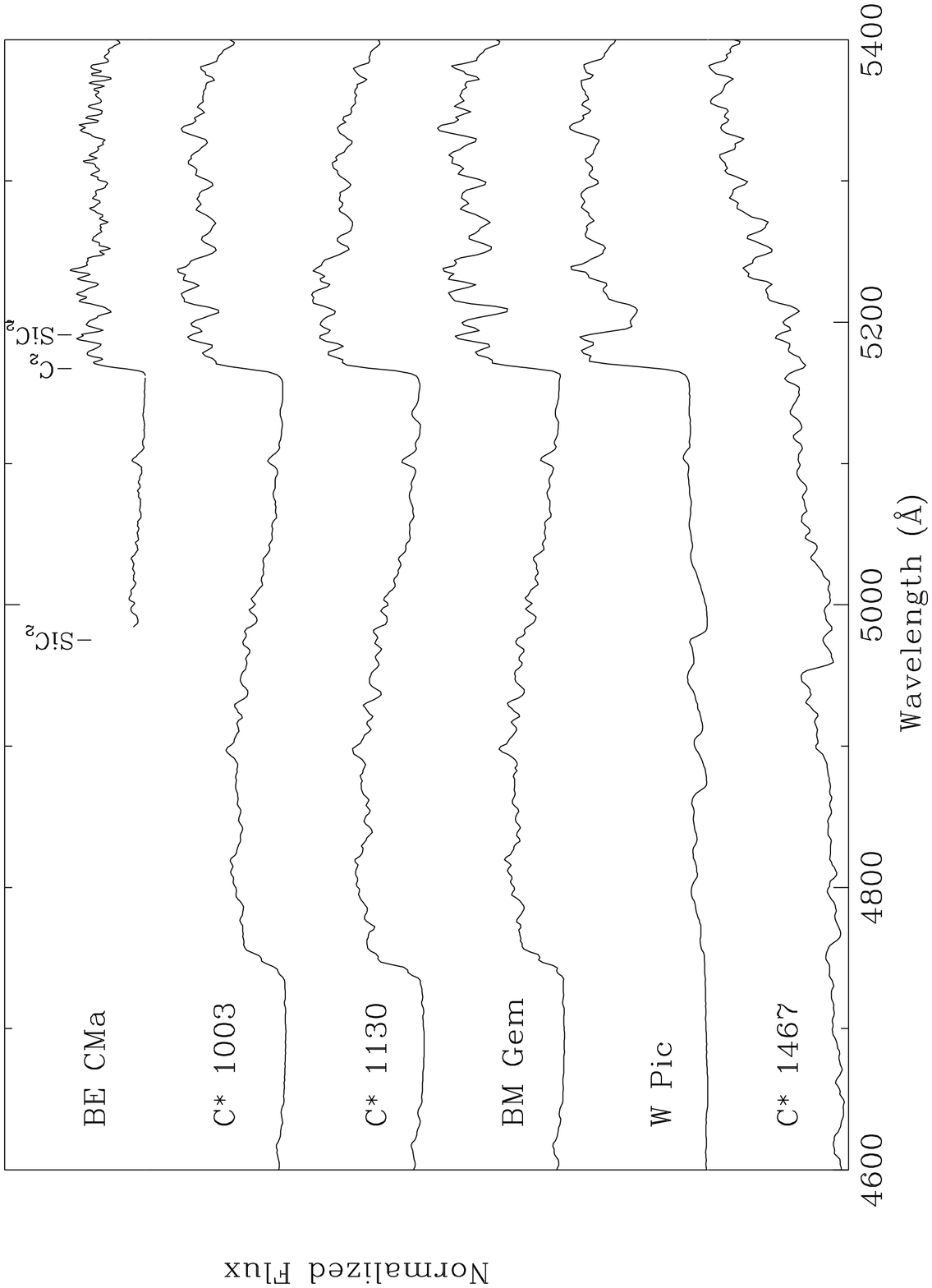}
\caption{C-J stars in a decreasing sequence of their temperature indices.} \label{specn7}
\end{center}
\end{figure}
\begin{figure}
\begin{center}
\plotone{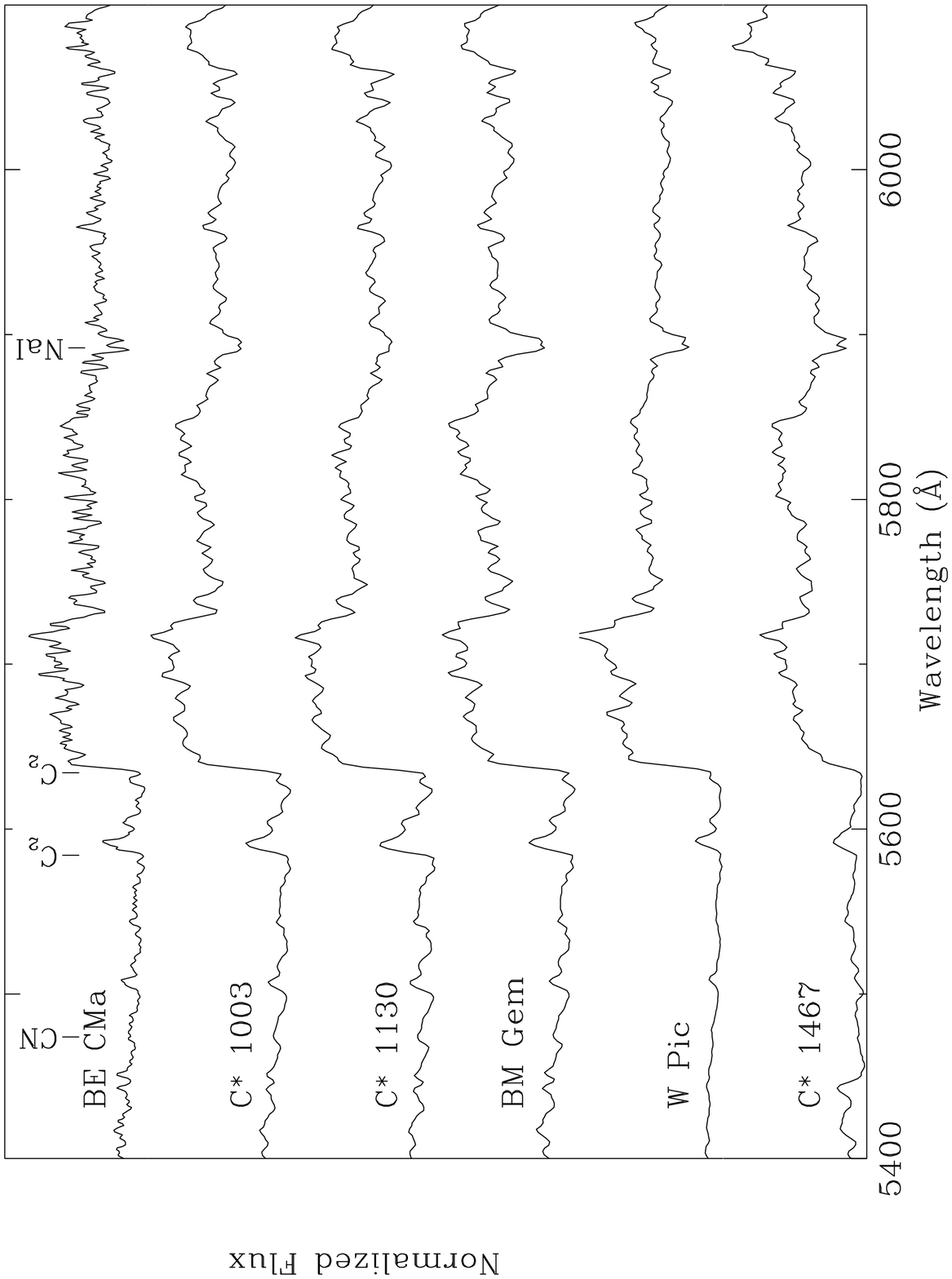}
\end{center}
Fig.~\ref{specn7}.--- continued
\end{figure}
\begin{figure}
\begin{center}
\plotone{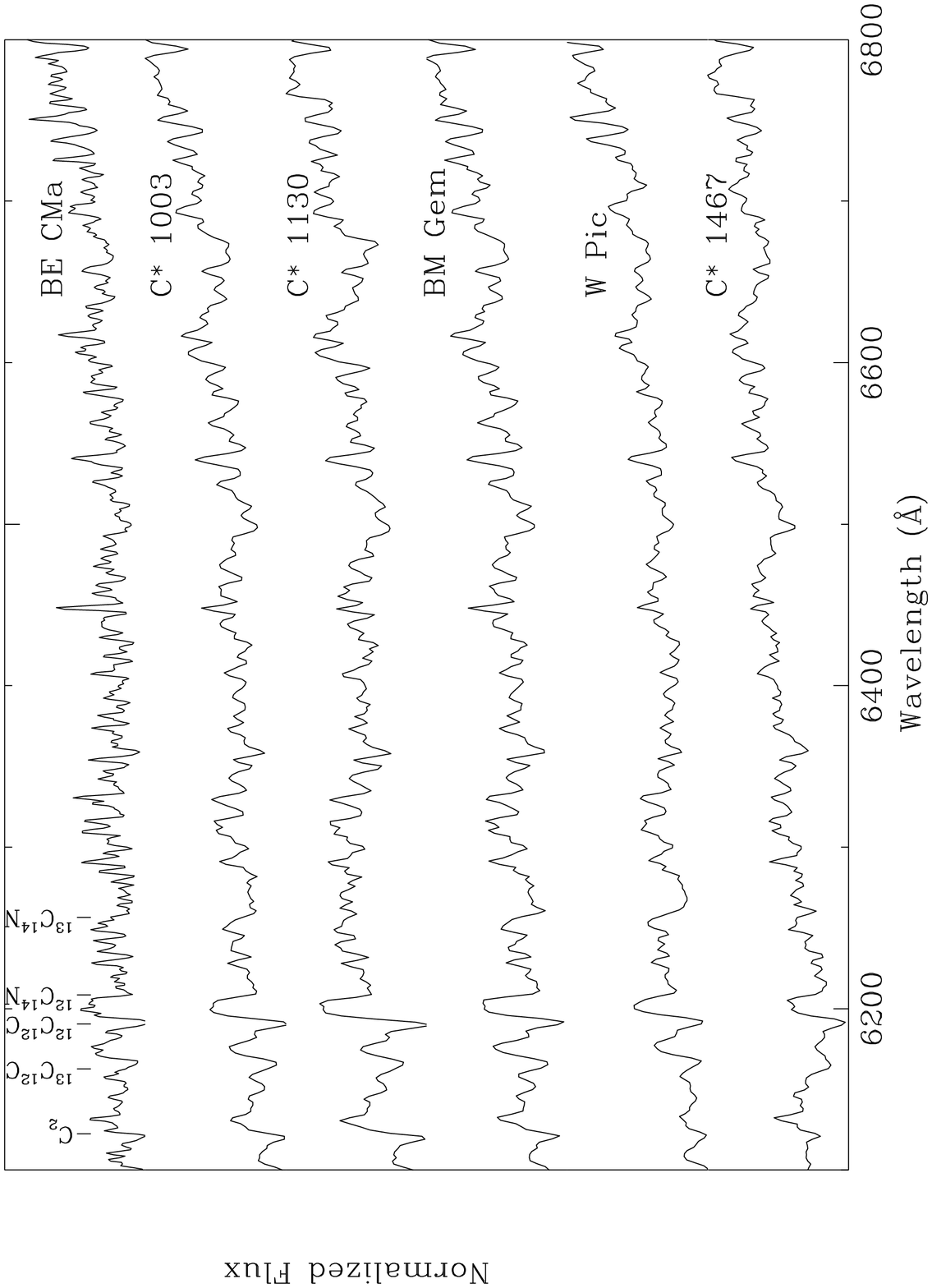}
\end{center}
Fig.~\ref{specn7}.--- continued
\end{figure}
\clearpage

\begin{figure}
\begin{center}
\plotone{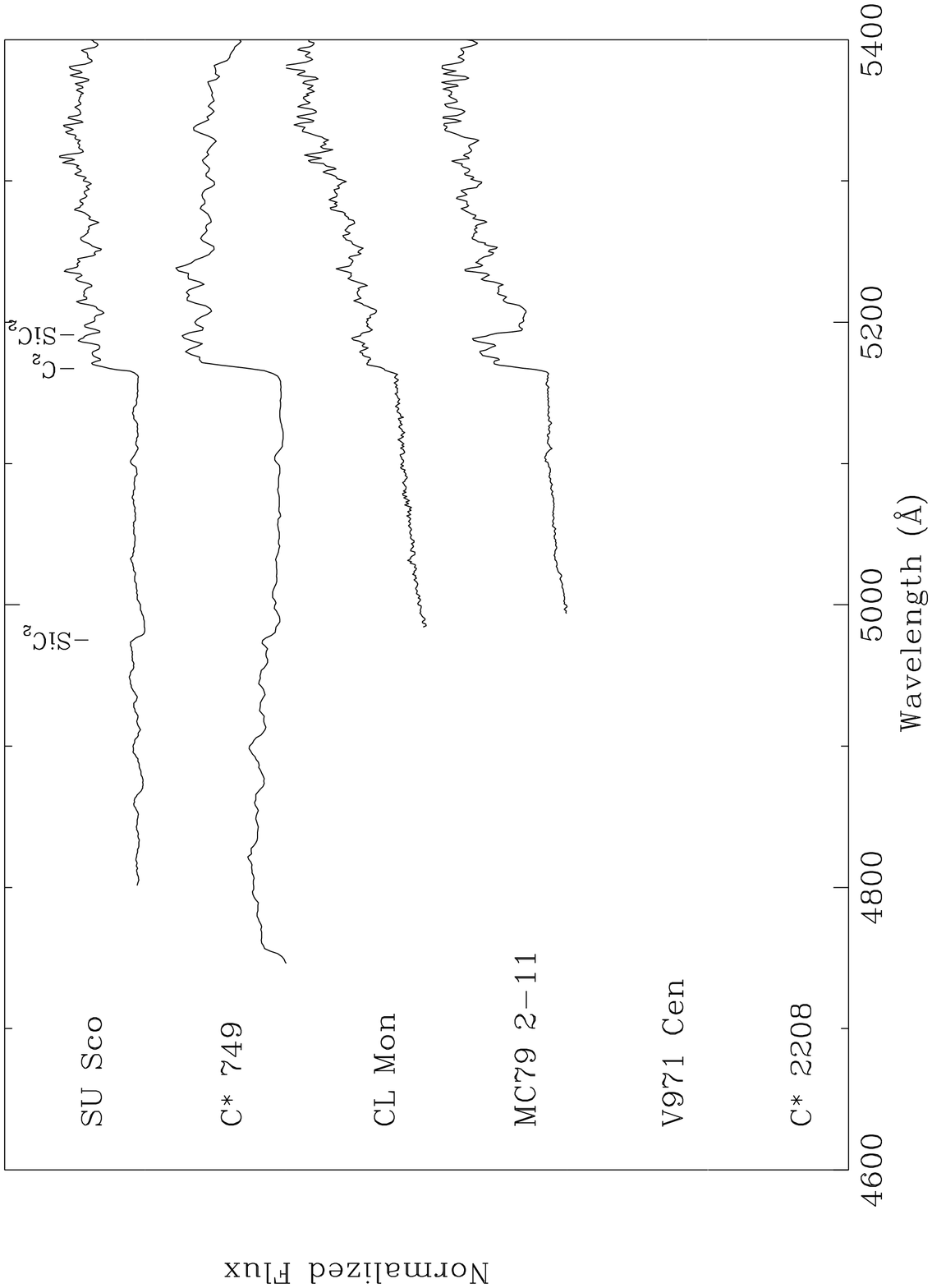}
\caption{C-J stars in a decreasing sequence of their temperature indices.} \label{specn8}
\end{center}
\end{figure}
\begin{figure}
\begin{center}
\plotone{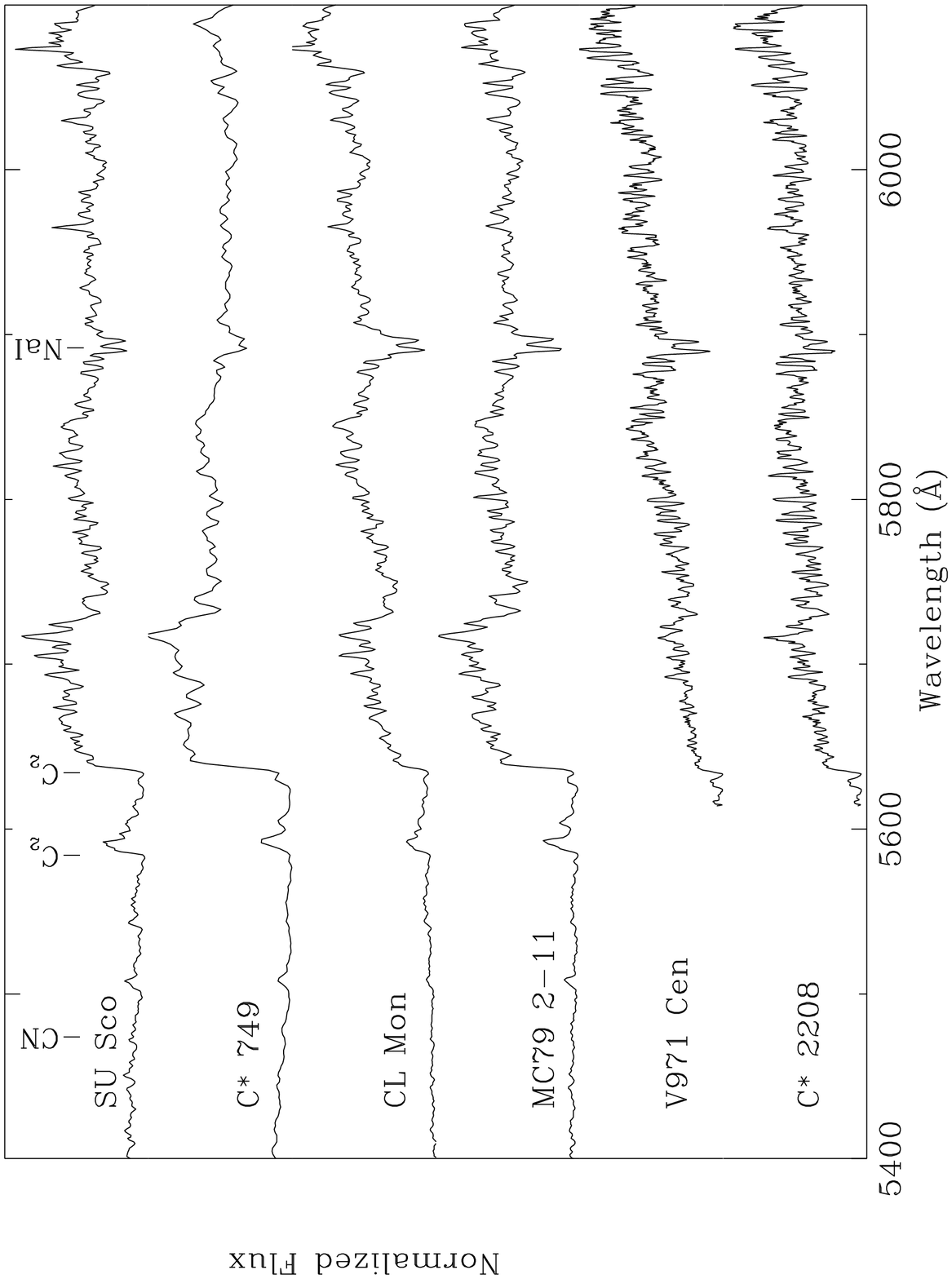}
\end{center}
Fig.~\ref{specn8}.--- continued
\end{figure}
\begin{figure}
\begin{center}
\plotone{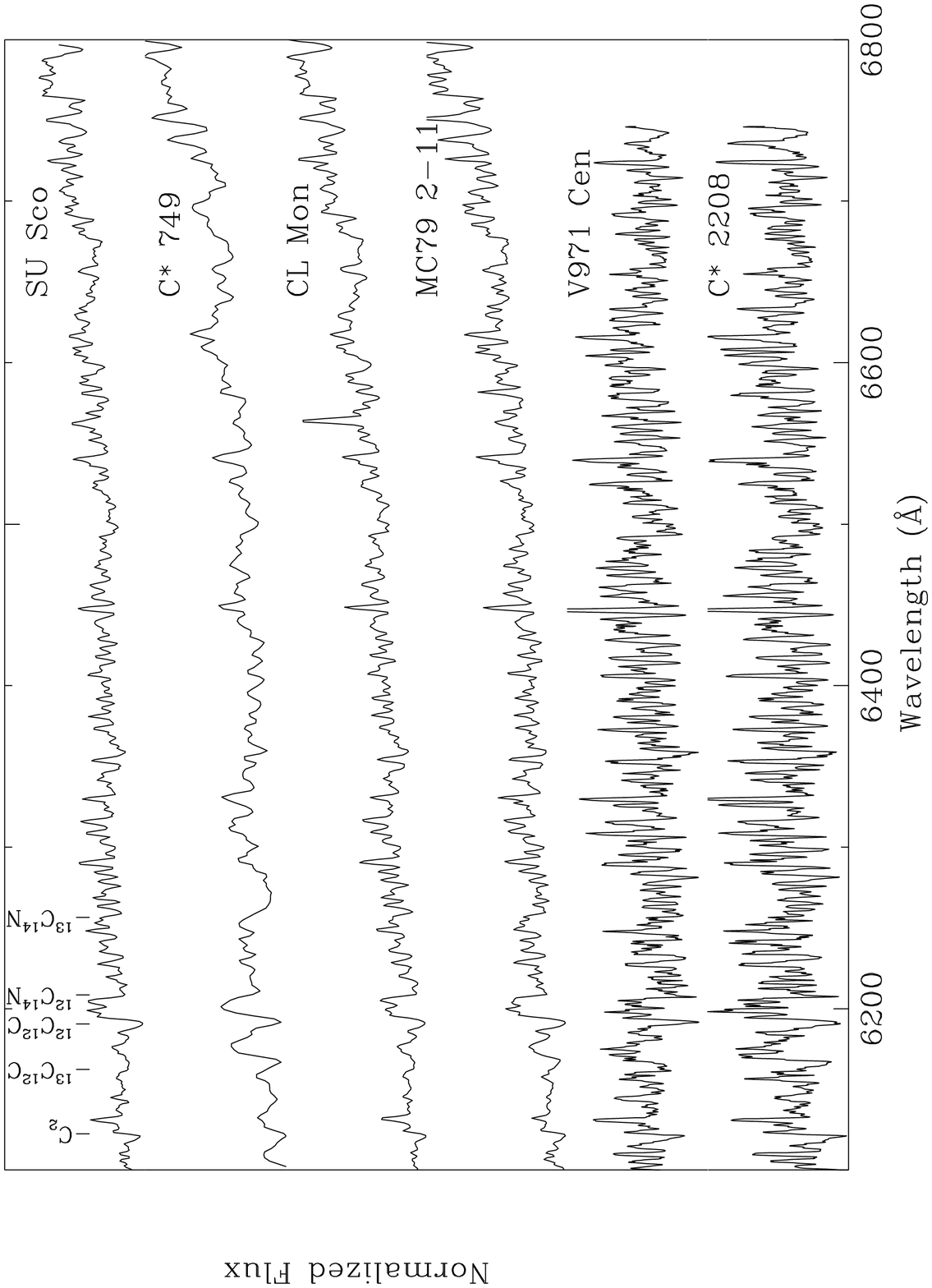}
\end{center}
Fig.~\ref{specn8}.--- continued
\end{figure}
\clearpage

\begin{figure}
\epsscale{0.8}
\begin{center}
\plottwo{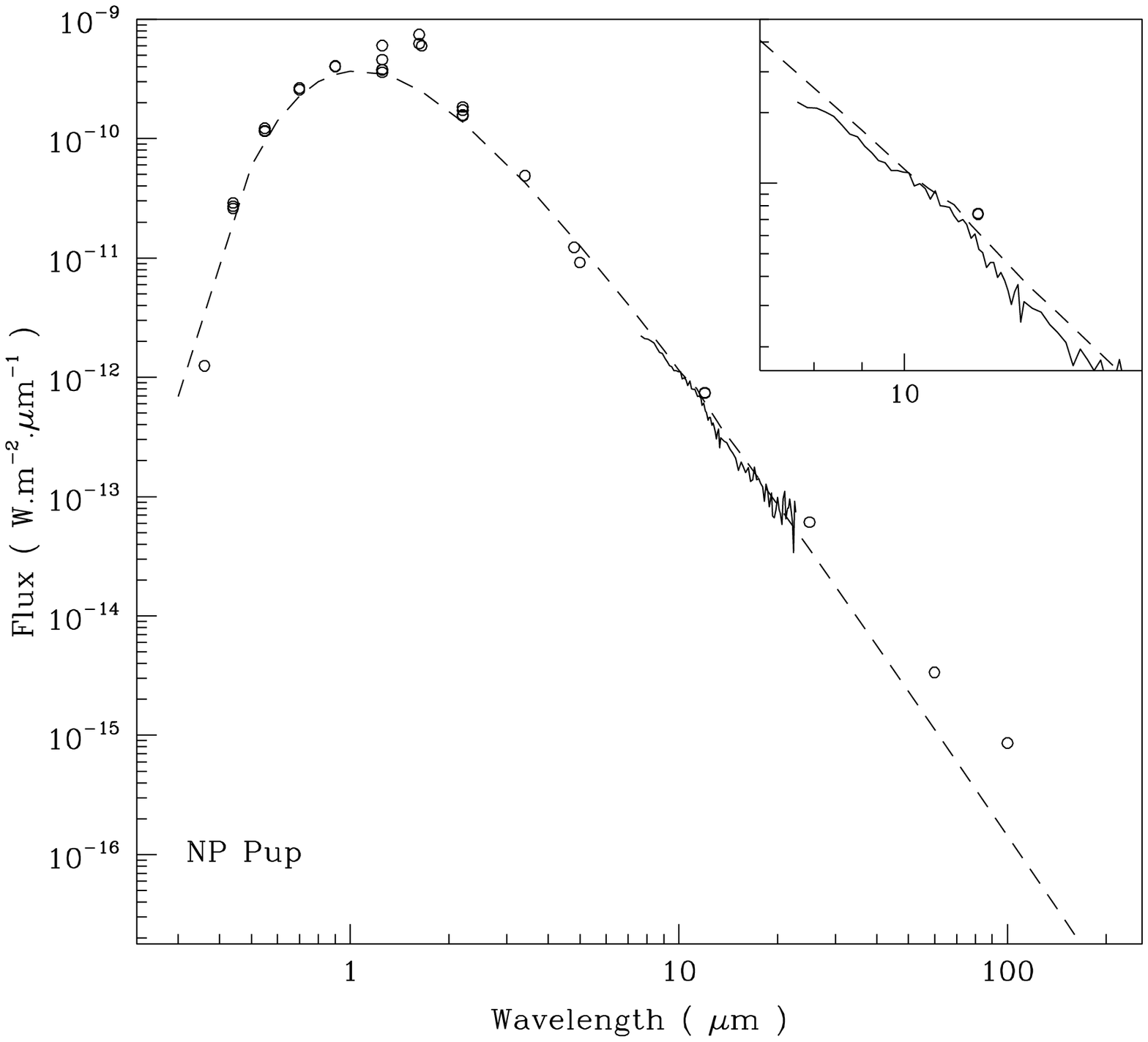}{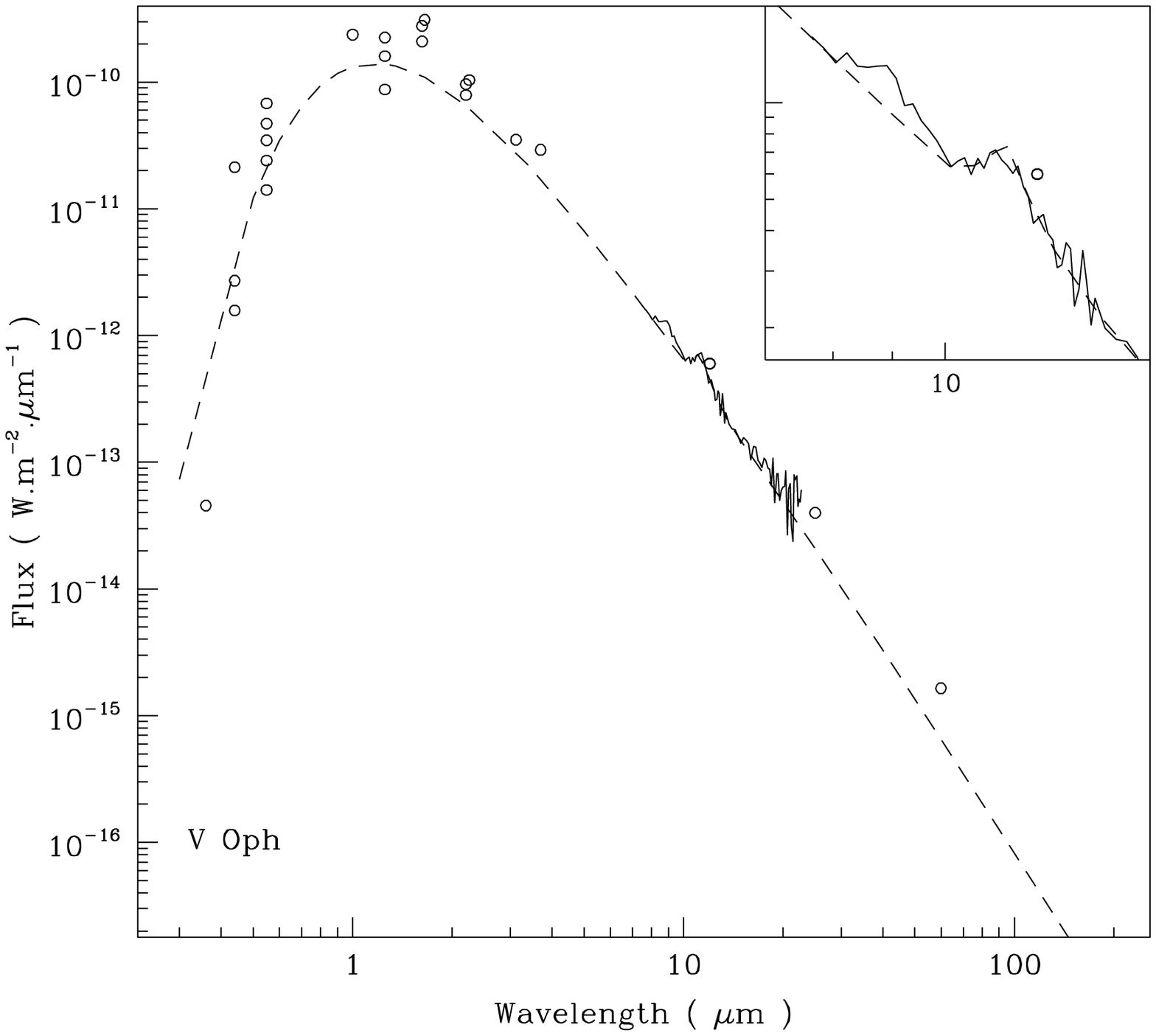}
\plottwo{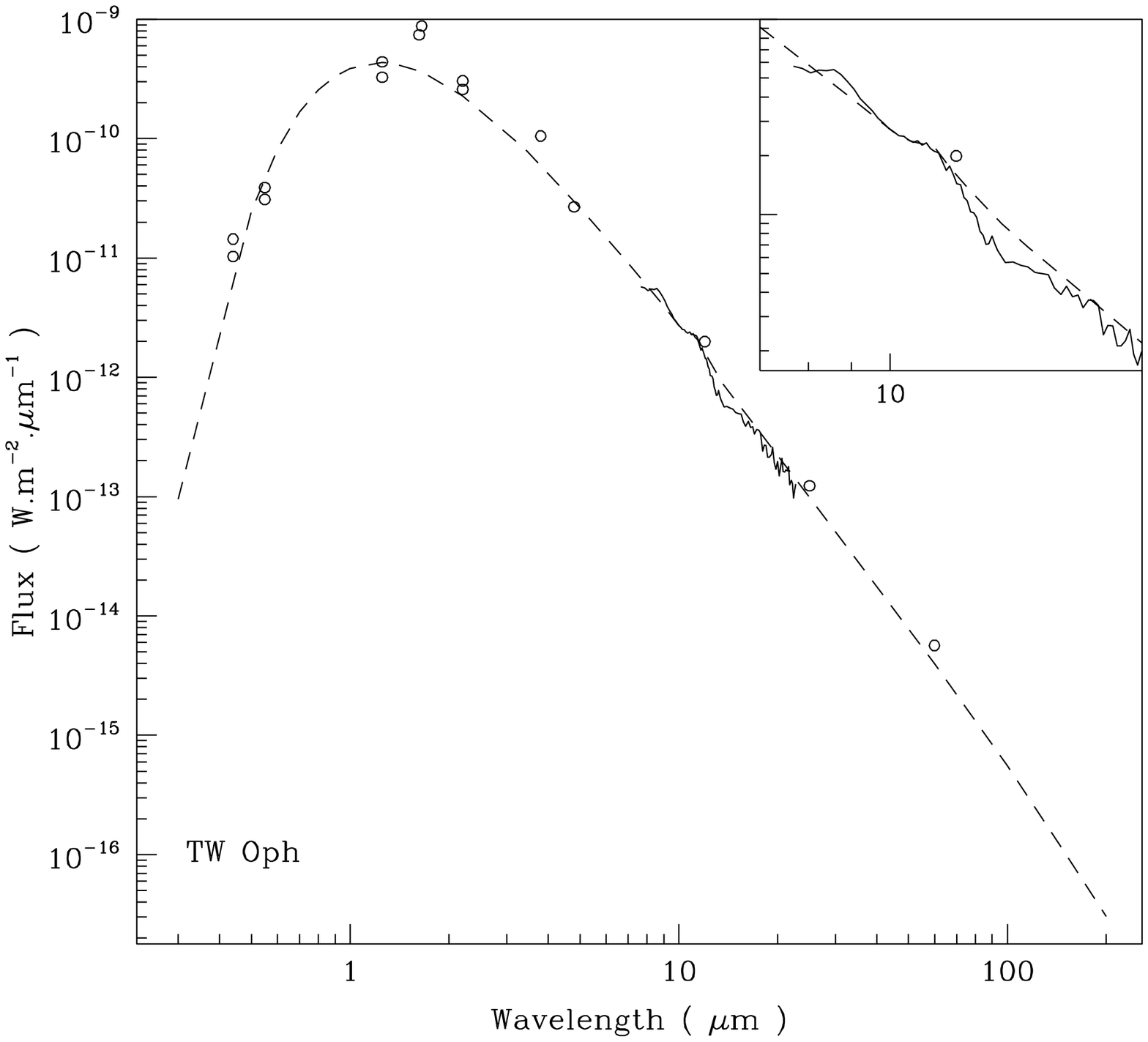}{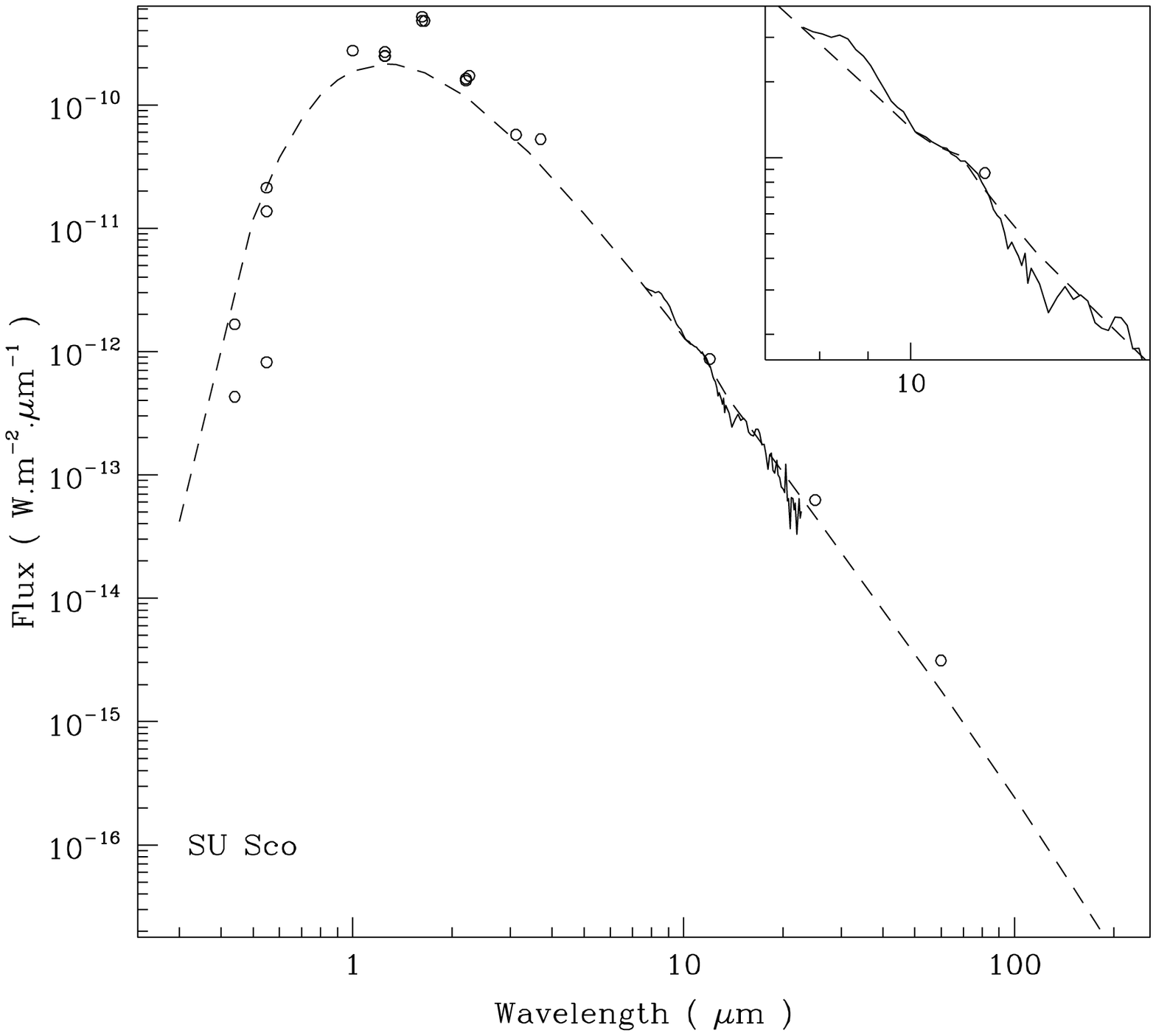}
\plottwo{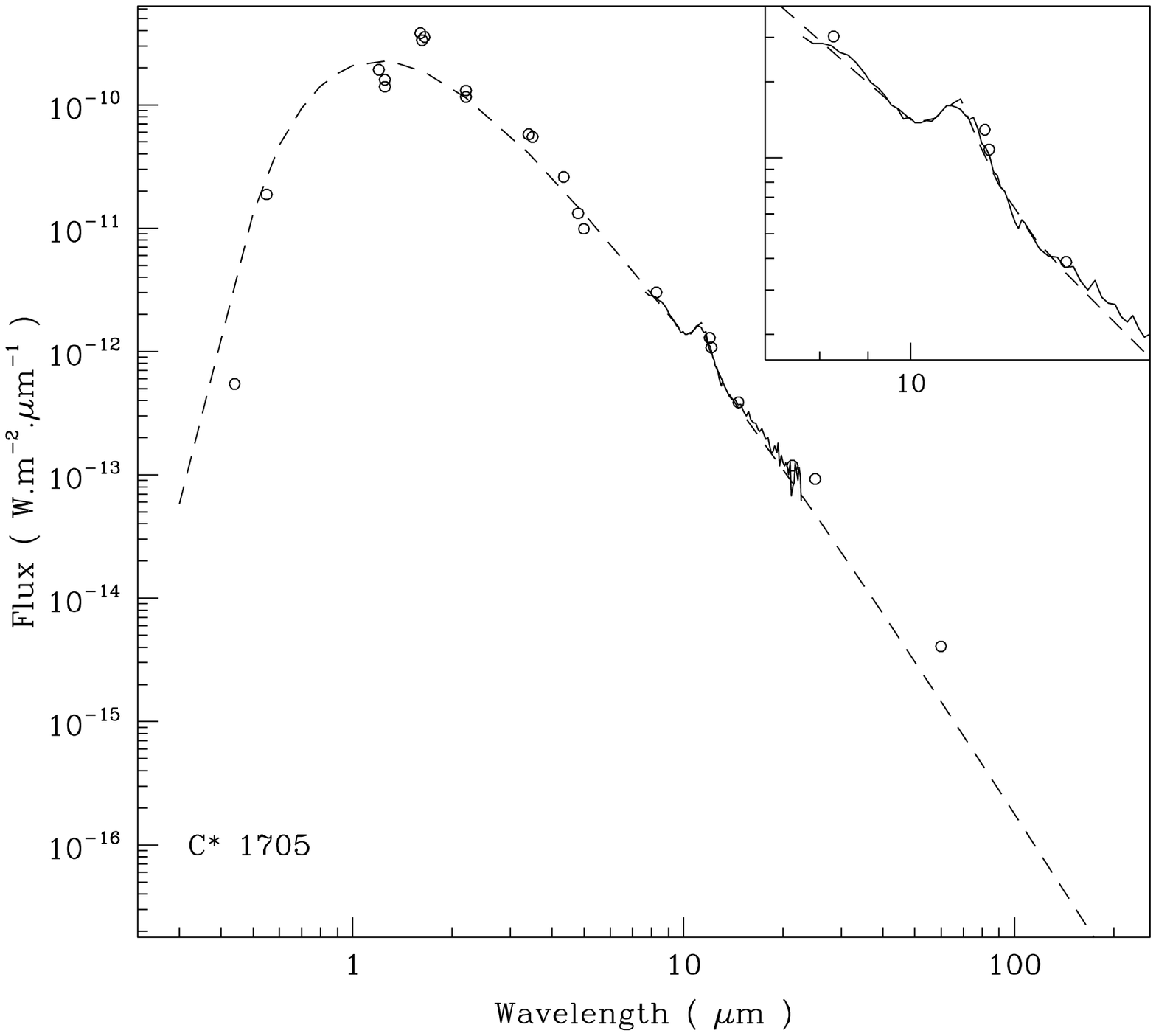}{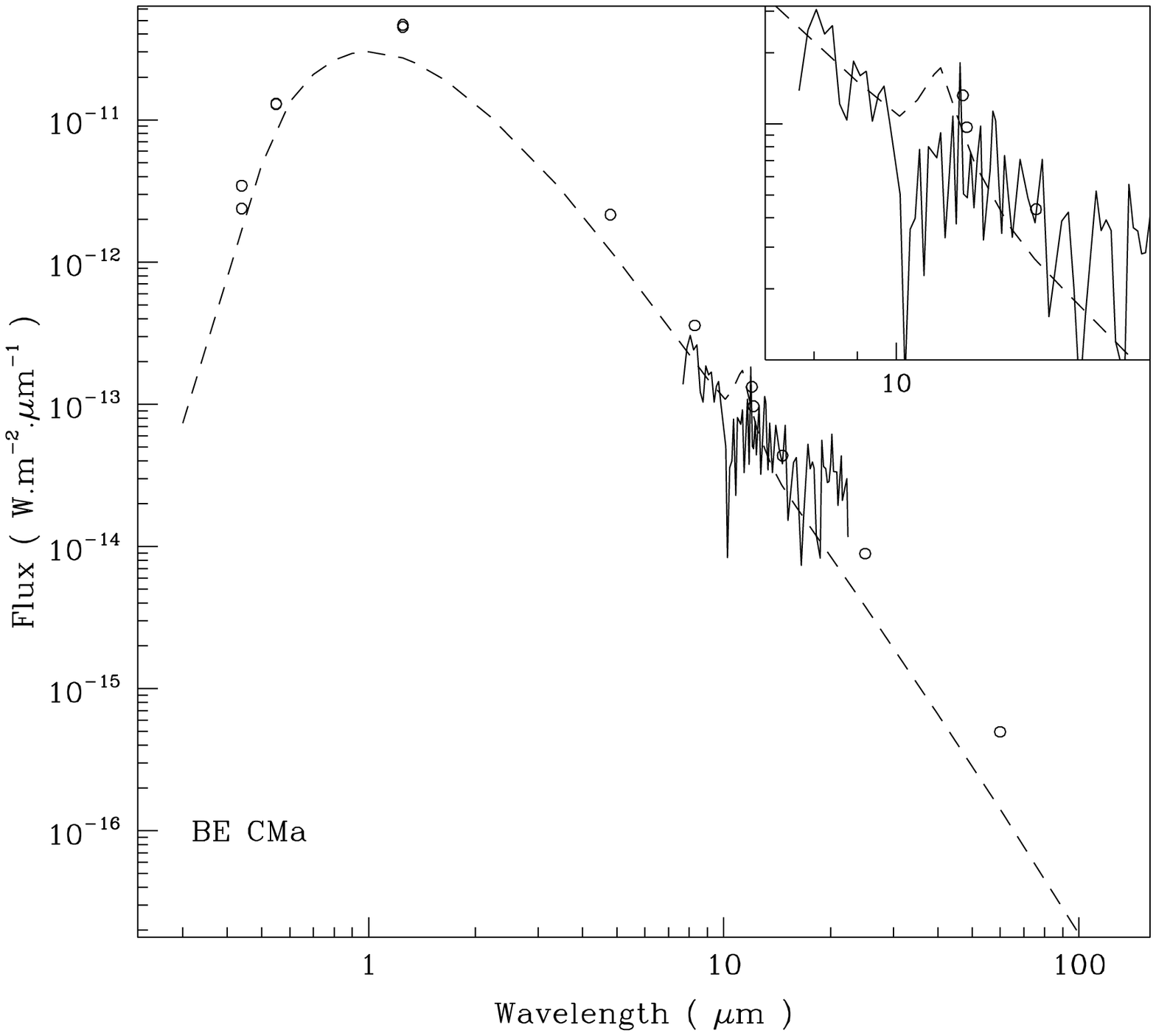}
\caption{Best fit model to the 11.3 $\mu$m feature for the sample} \label{modnormal}
\end{center}
\end{figure}
\clearpage
\begin{figure}
\begin{center}
\plottwo{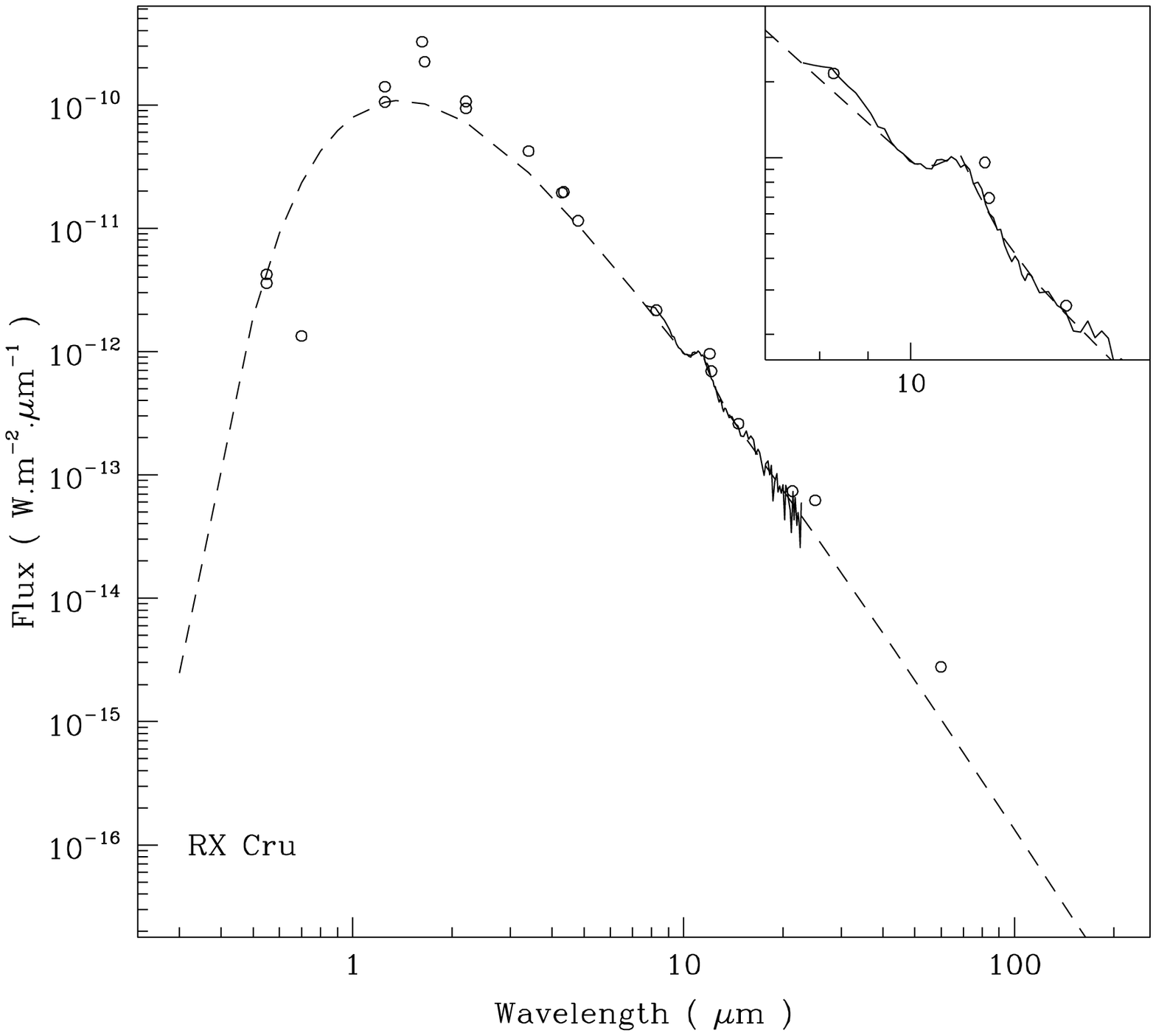}{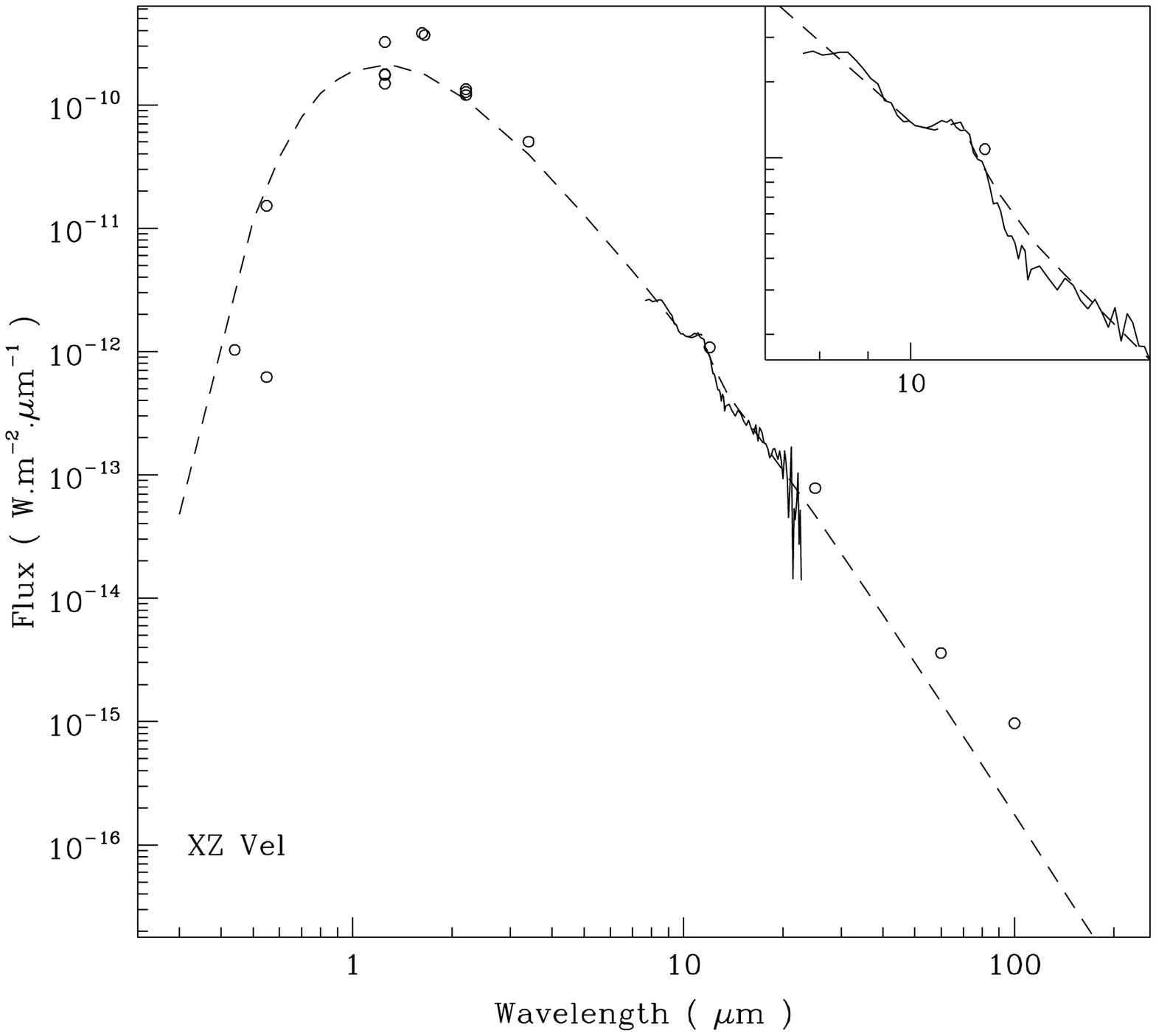}
\plottwo{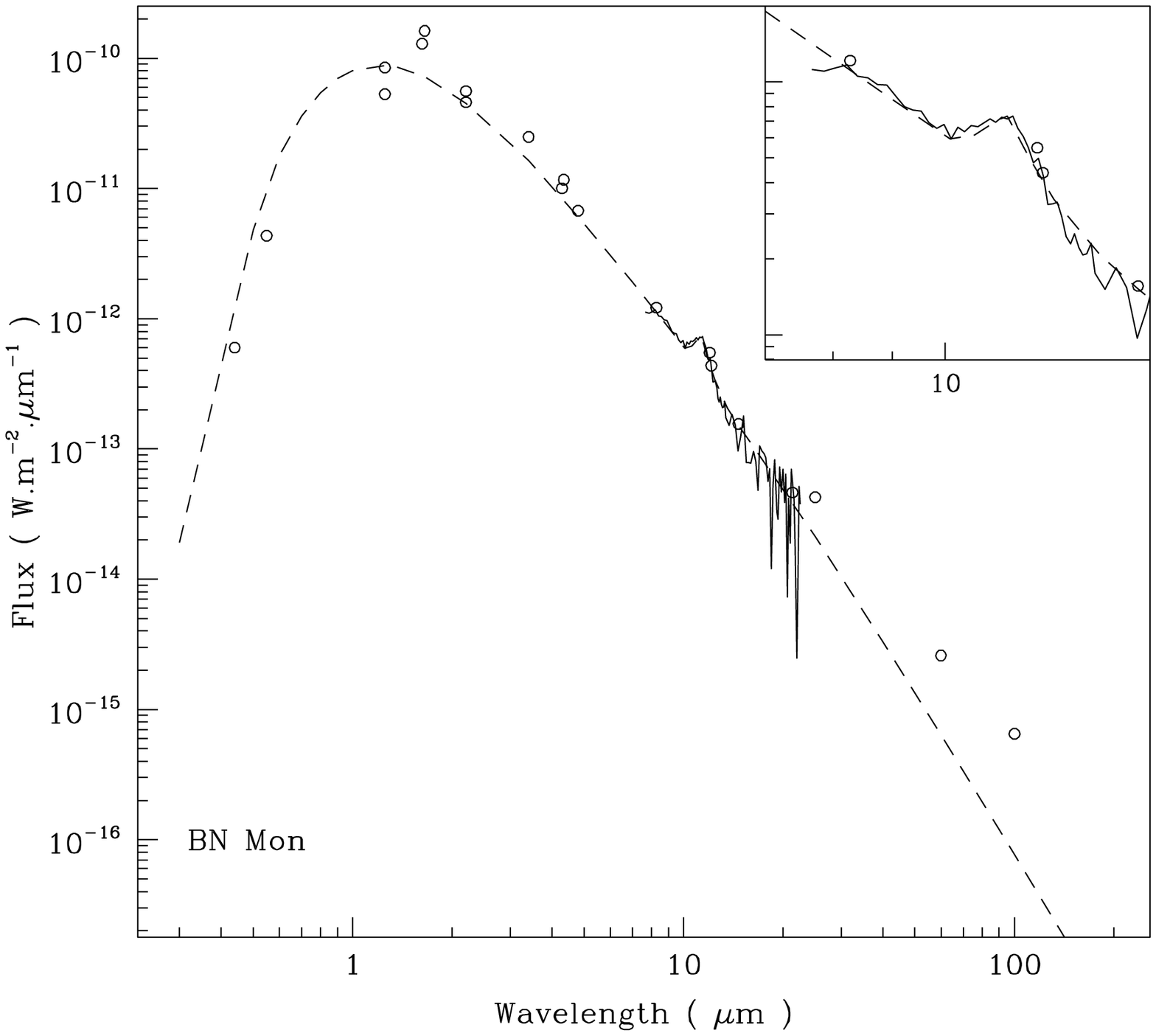}{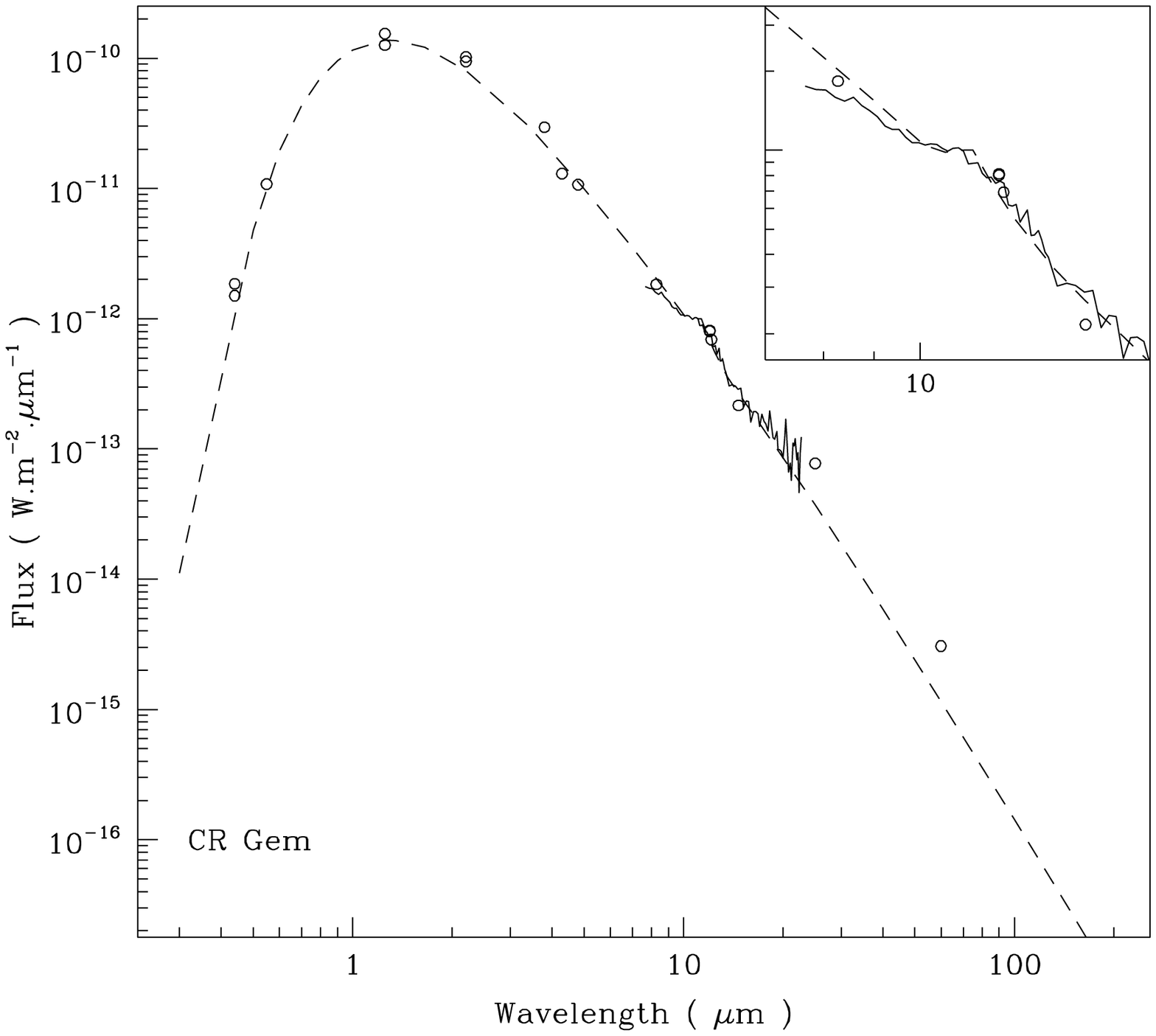}
\plottwo{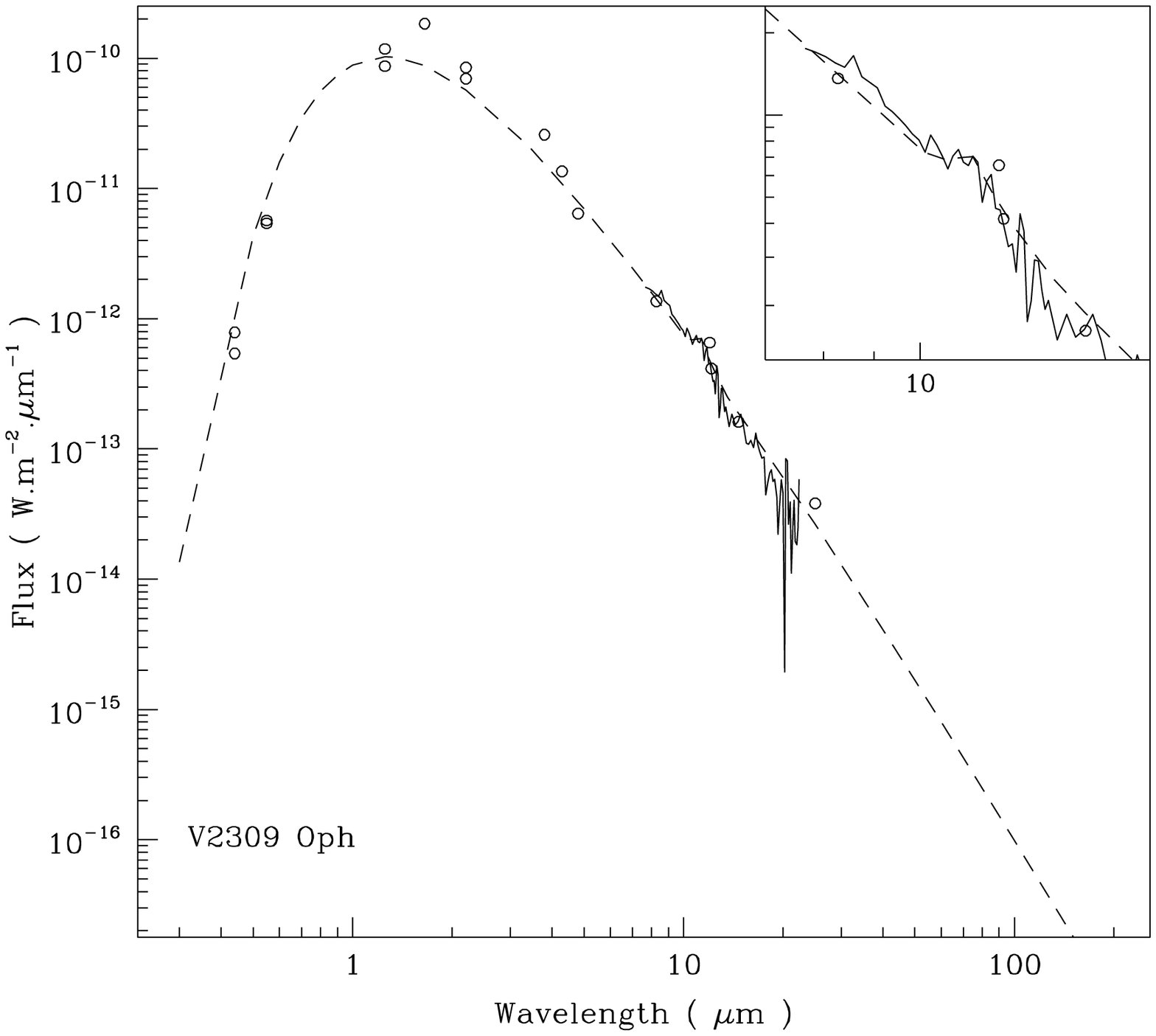}{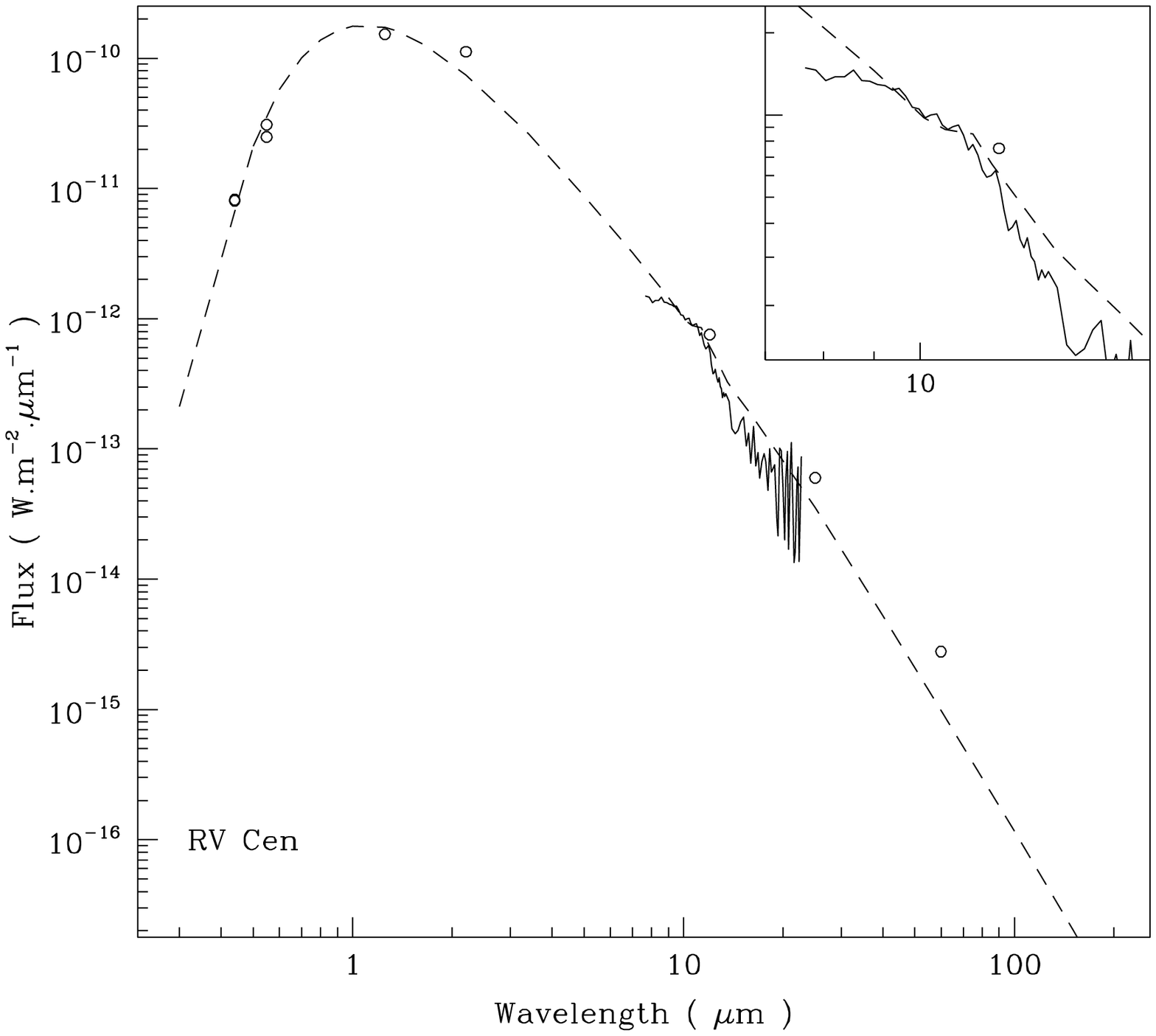}
\end{center}

Fig.~\ref{modnormal}.--- continued
\end{figure}
\clearpage
\begin{figure}
\begin{center}
\plottwo{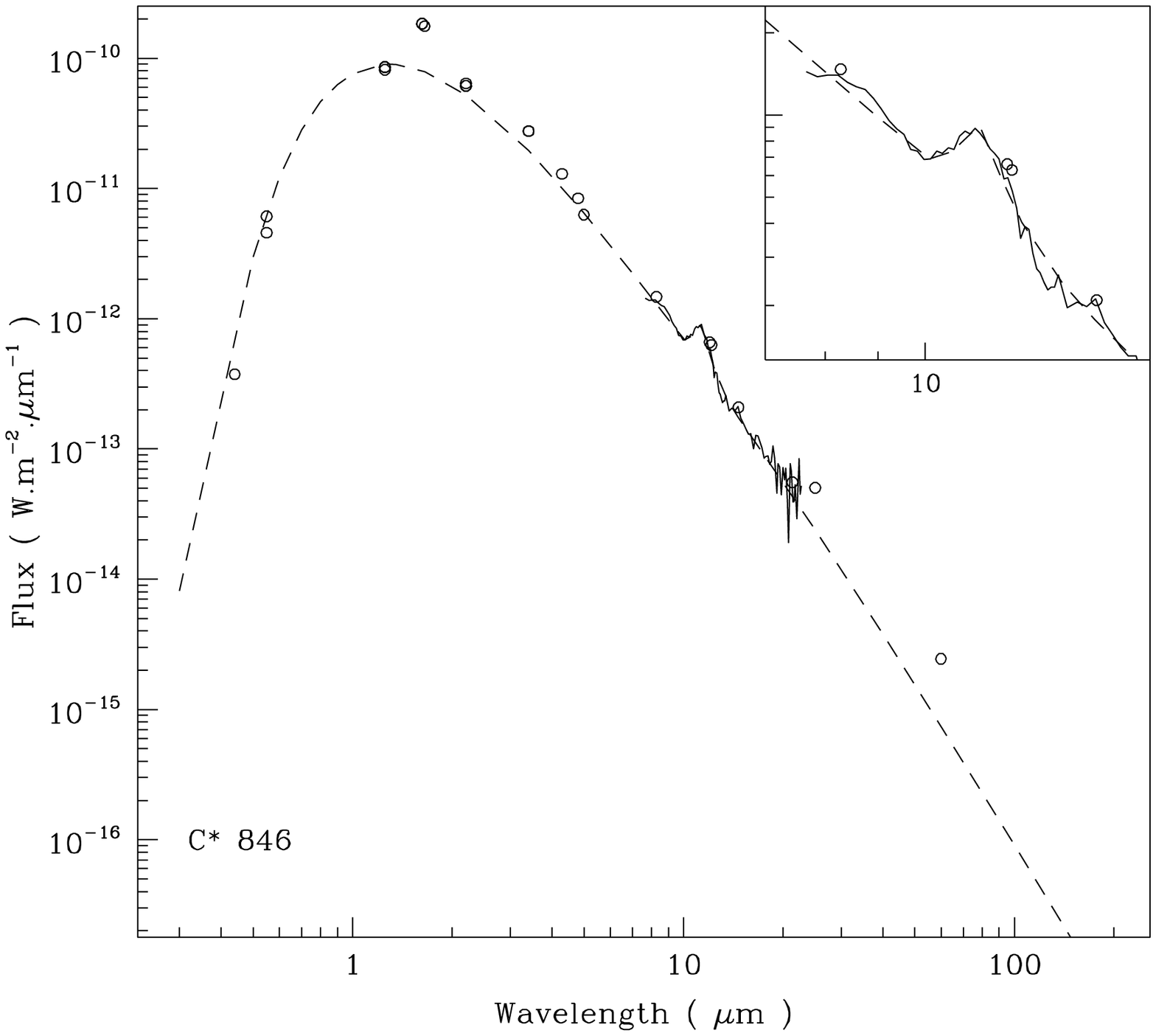}{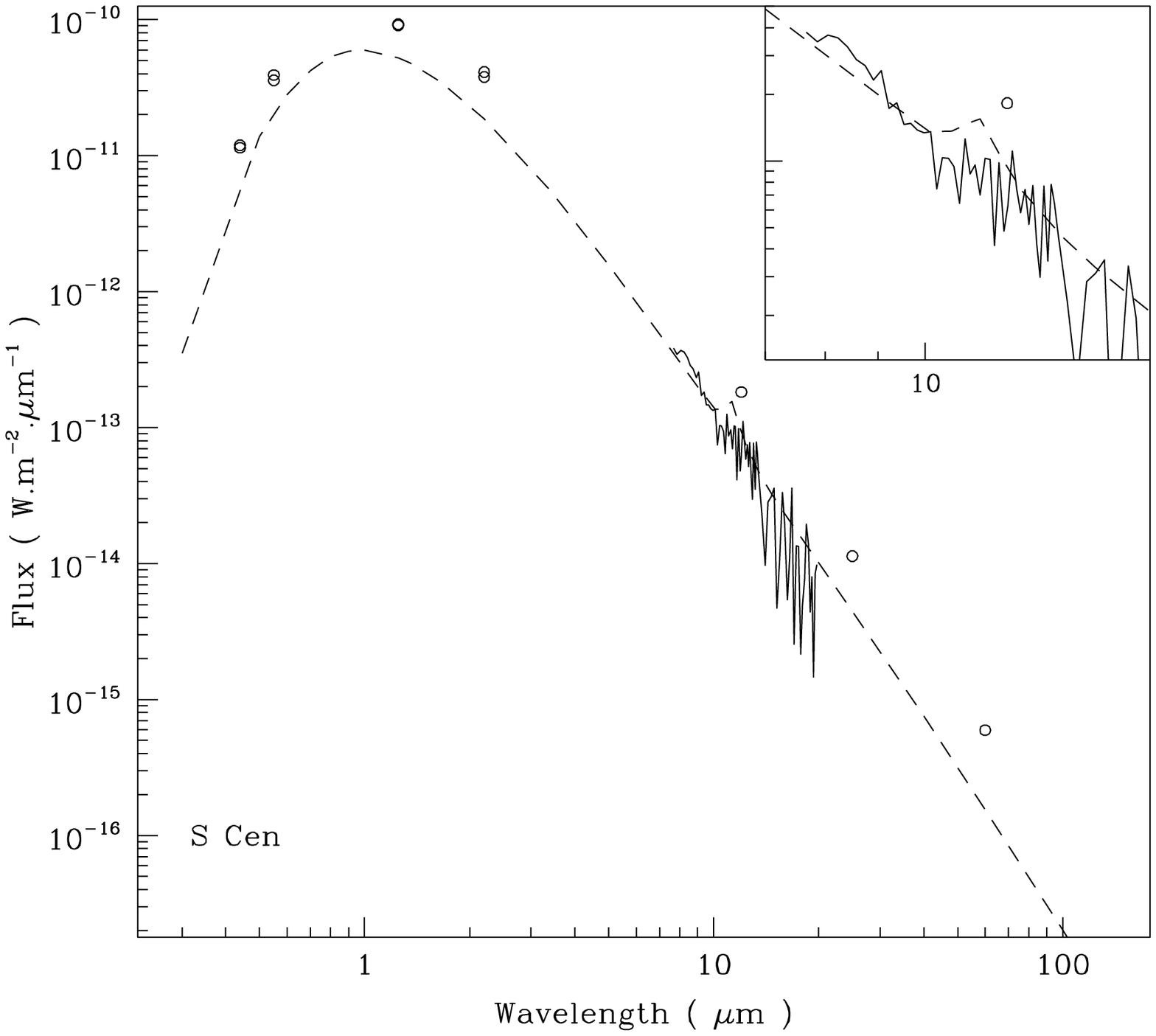}
\plottwo{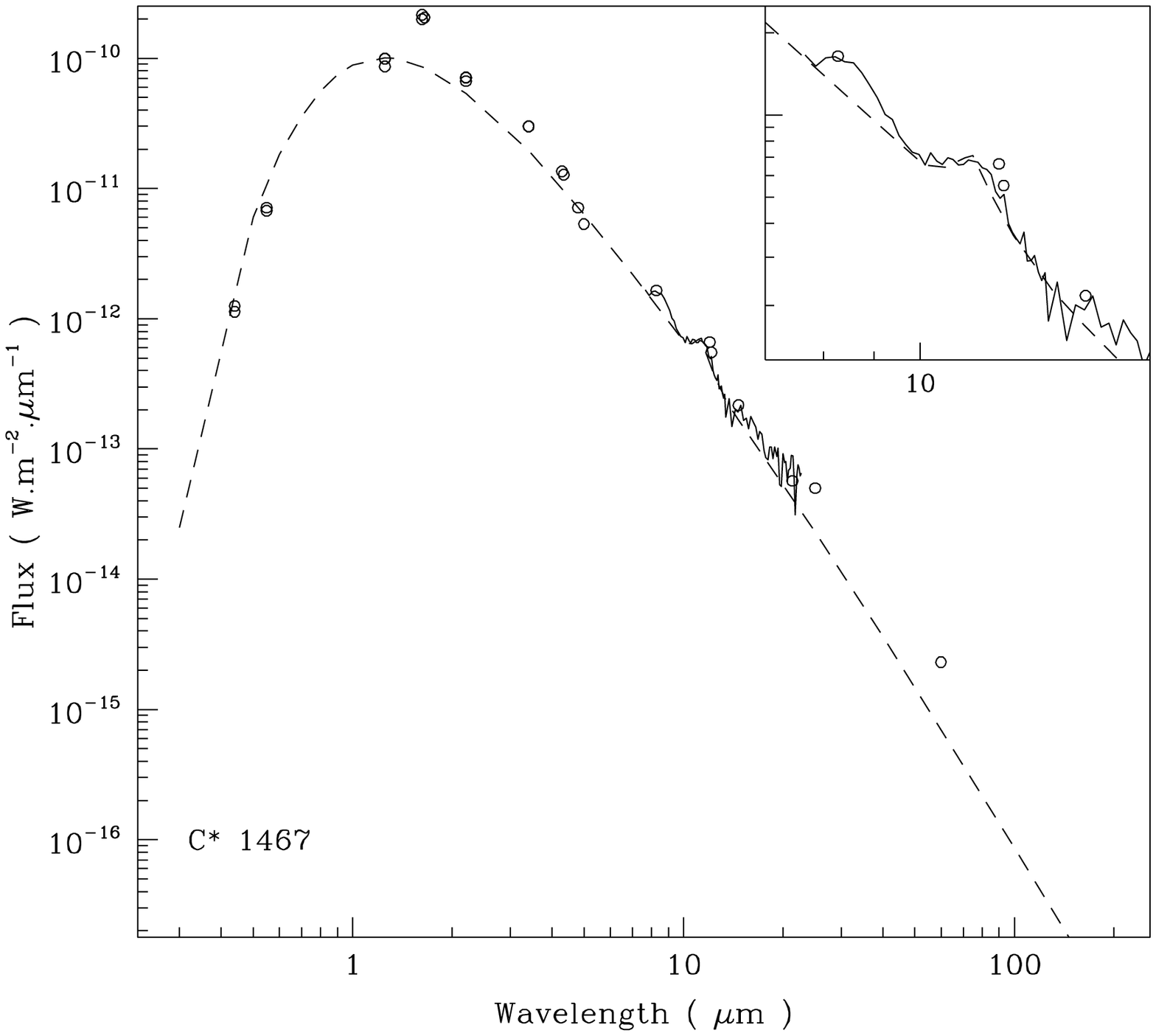}{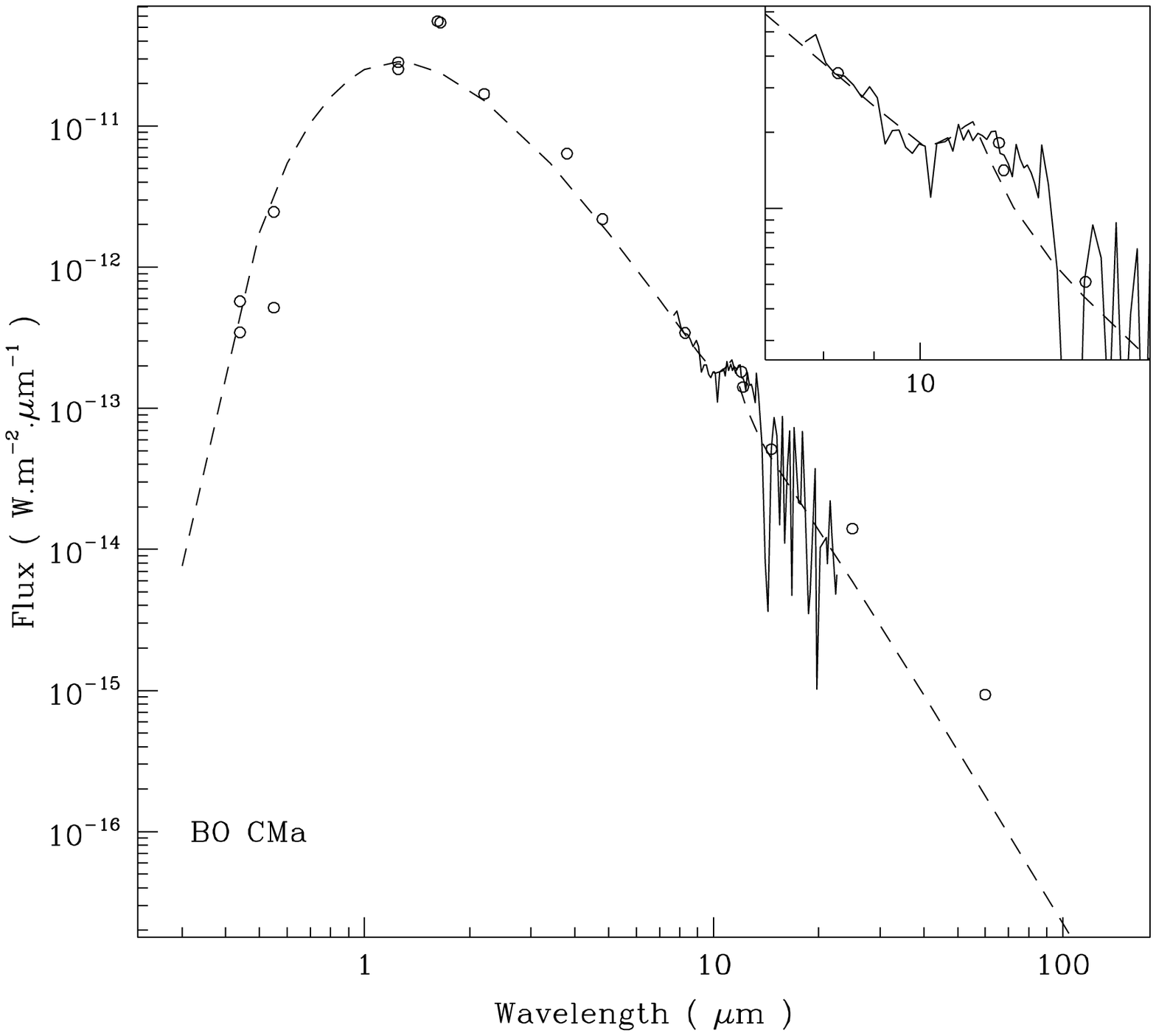}
\plottwo{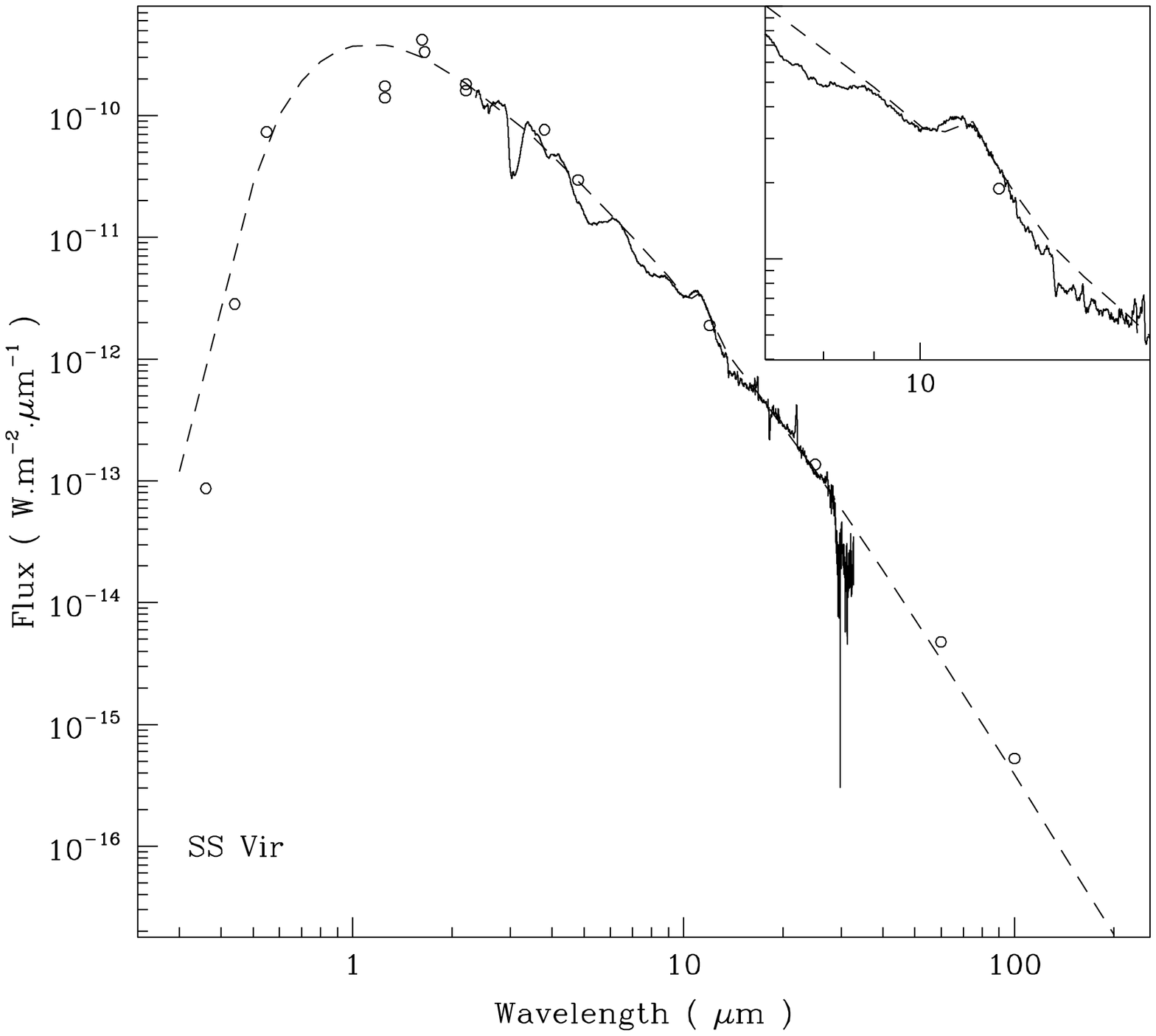}{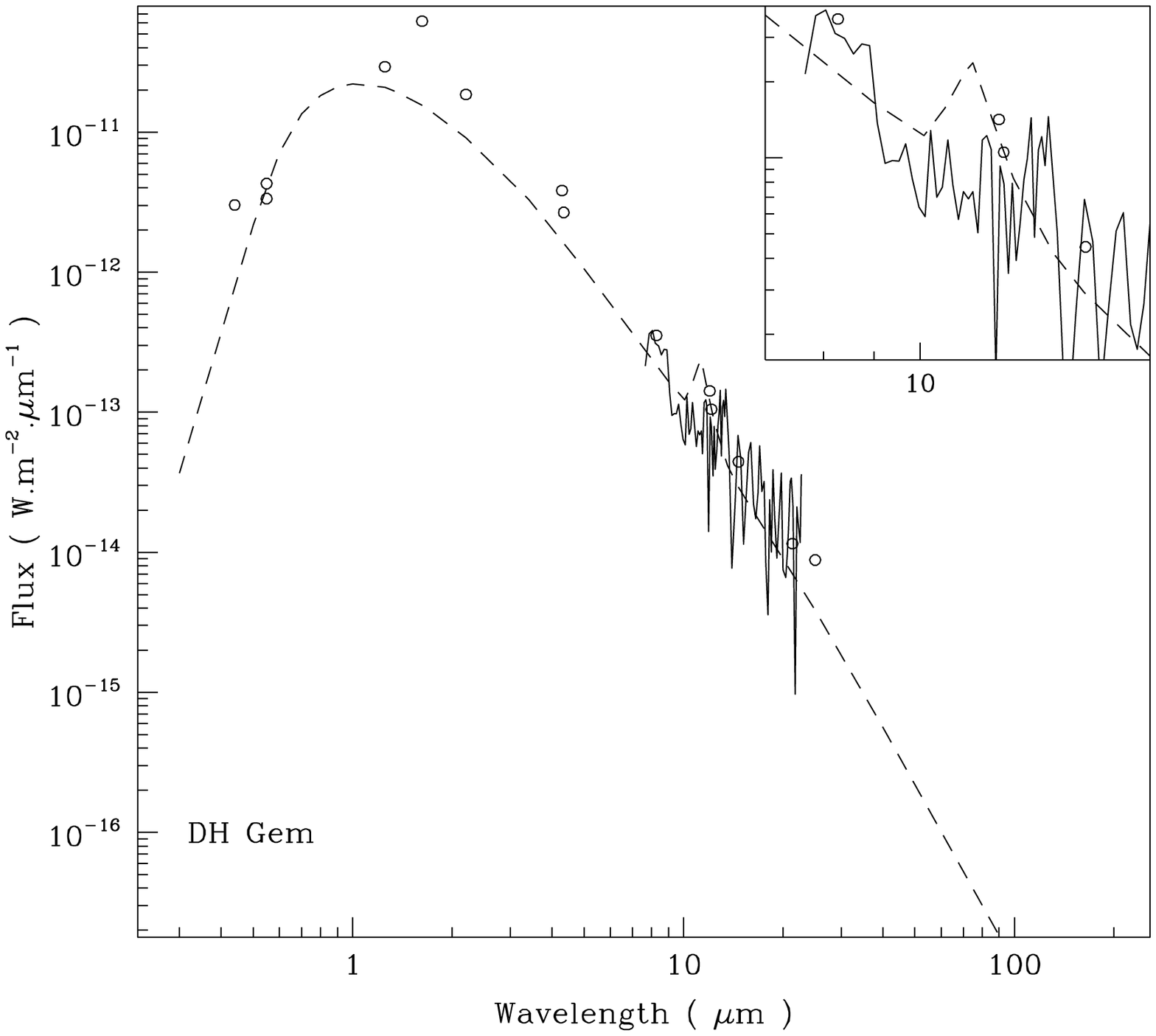}

\end{center}
Fig.~\ref{modnormal}.--- continued
\end{figure}
\clearpage
\begin{figure}
\begin{center}
\plottwo{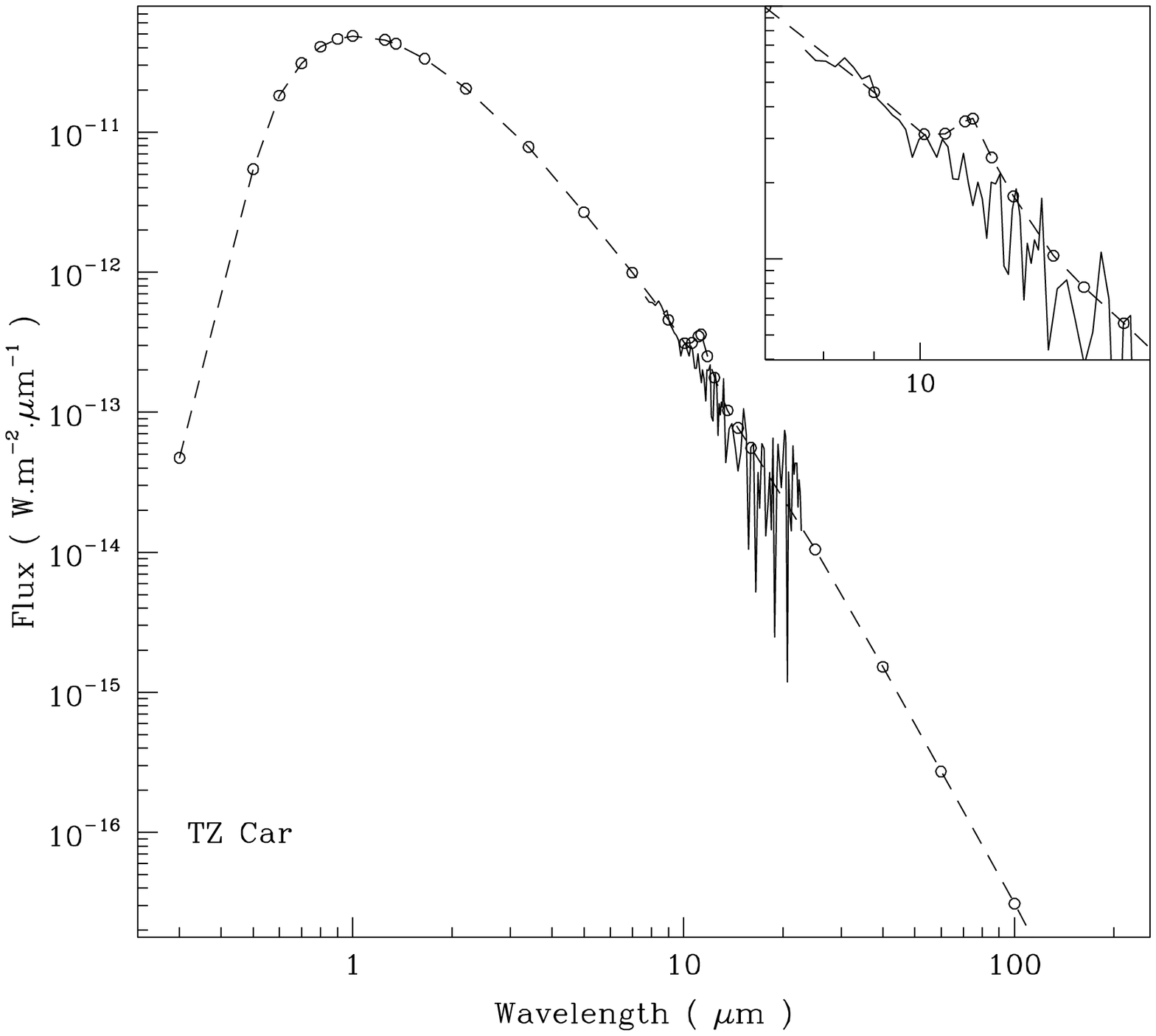}{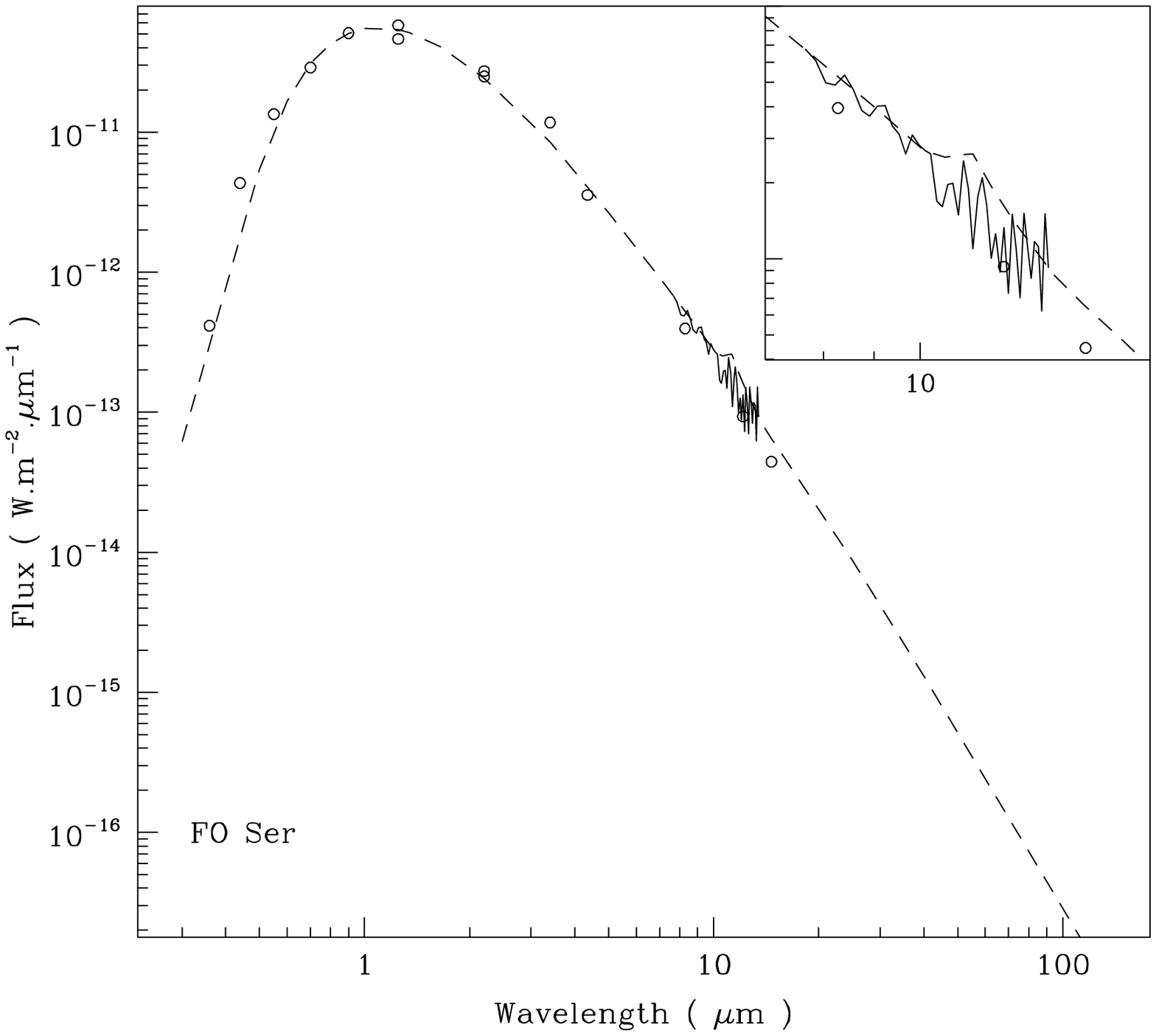}
\plottwo{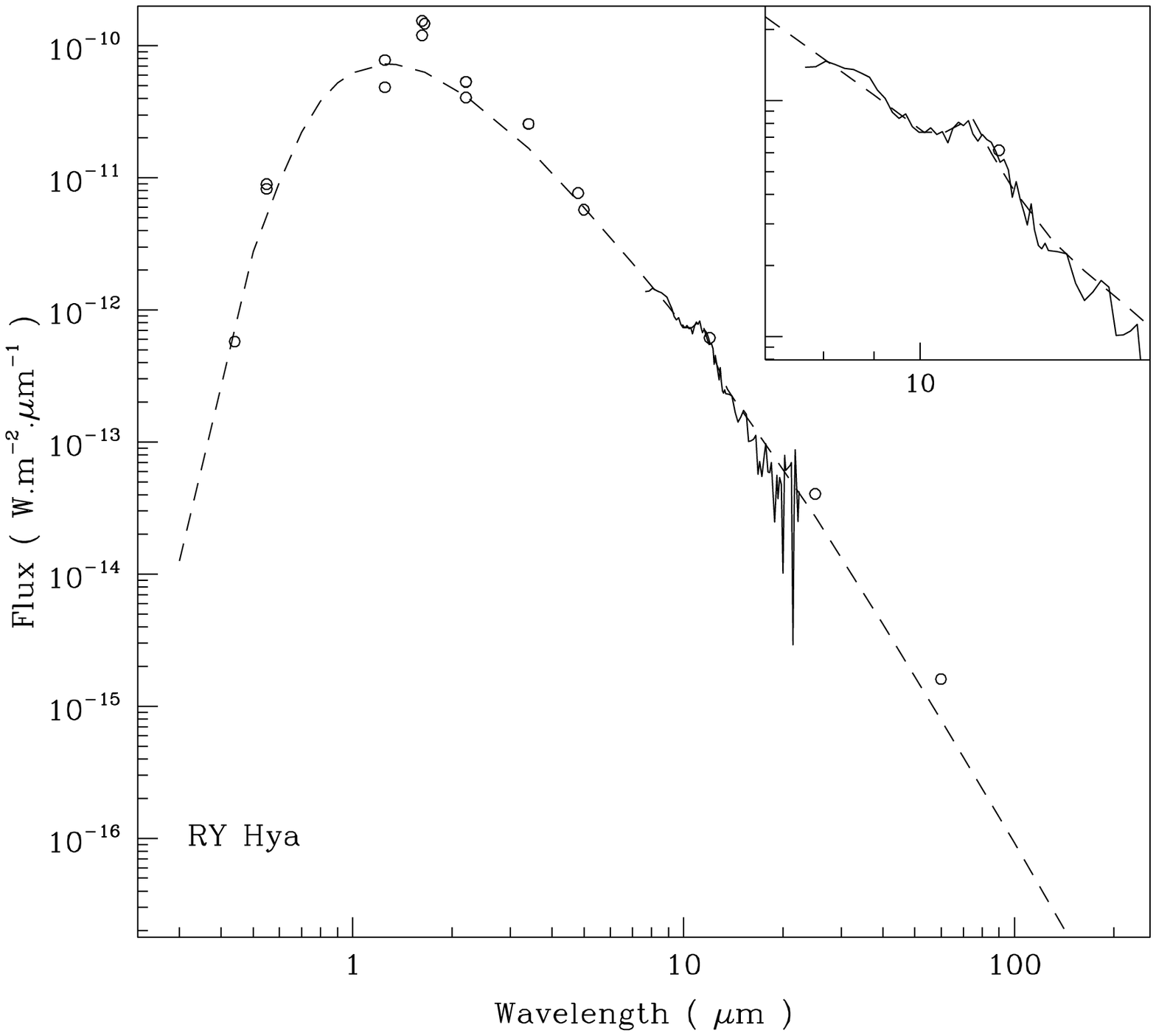}{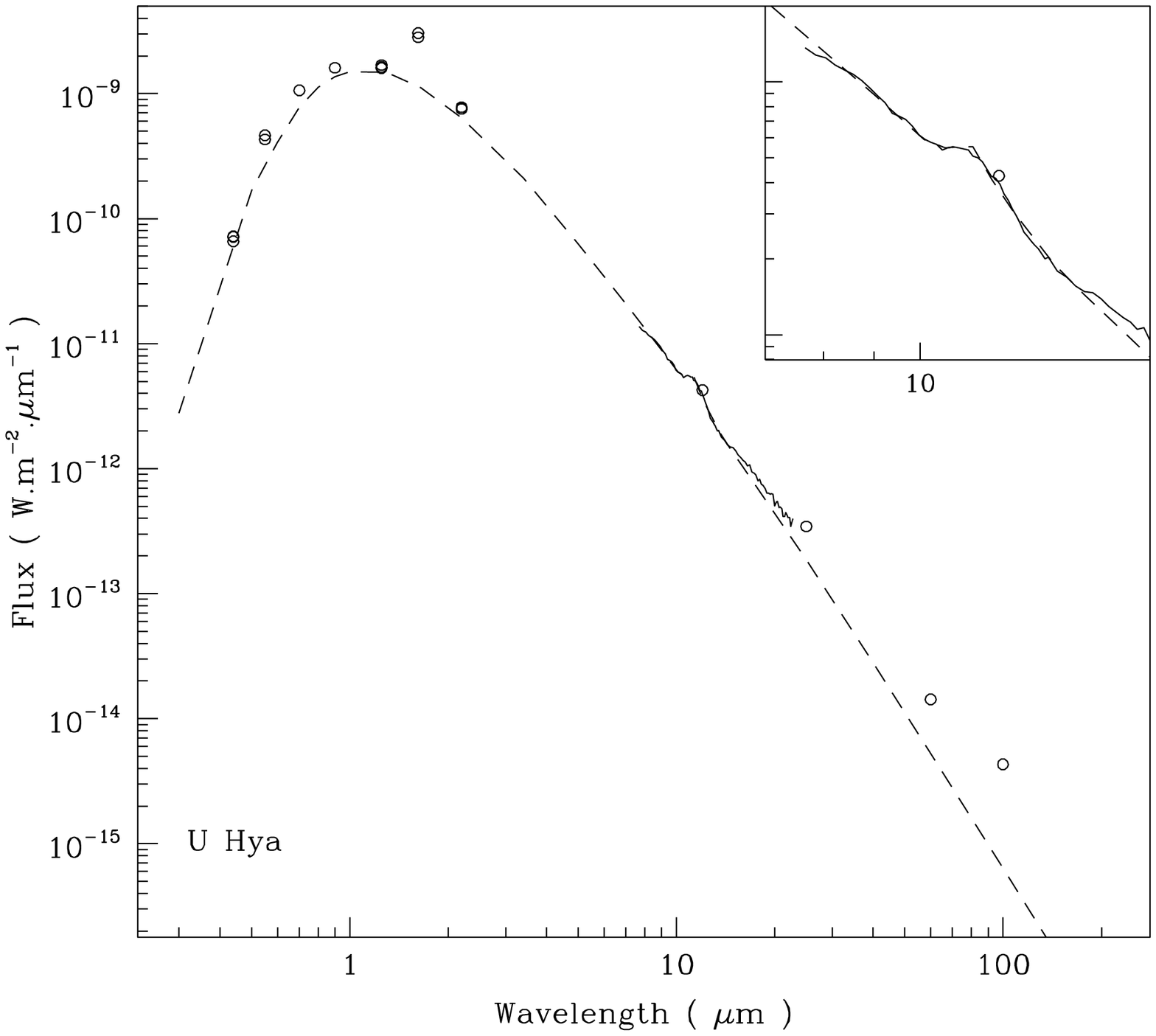}
\plottwo{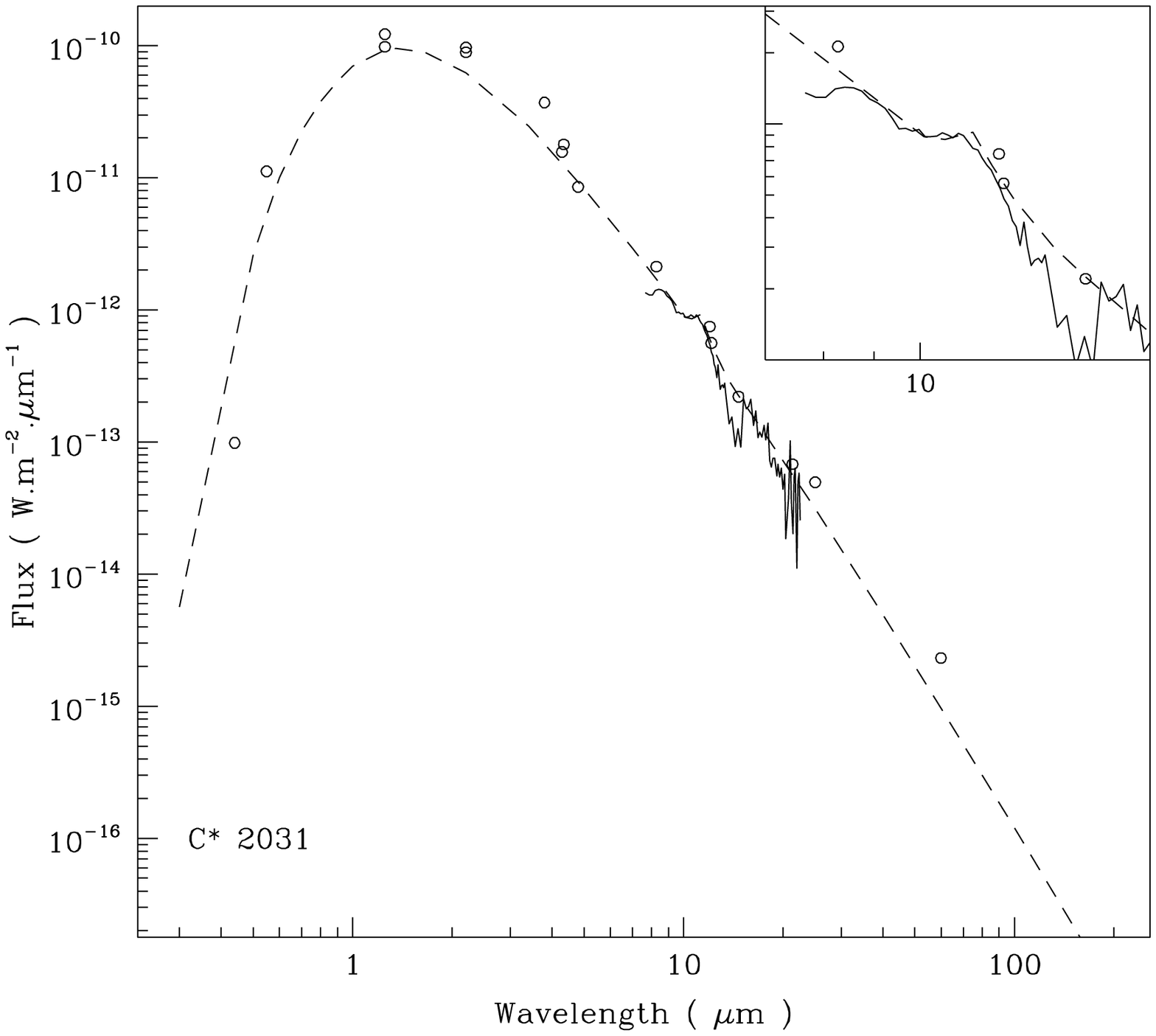}{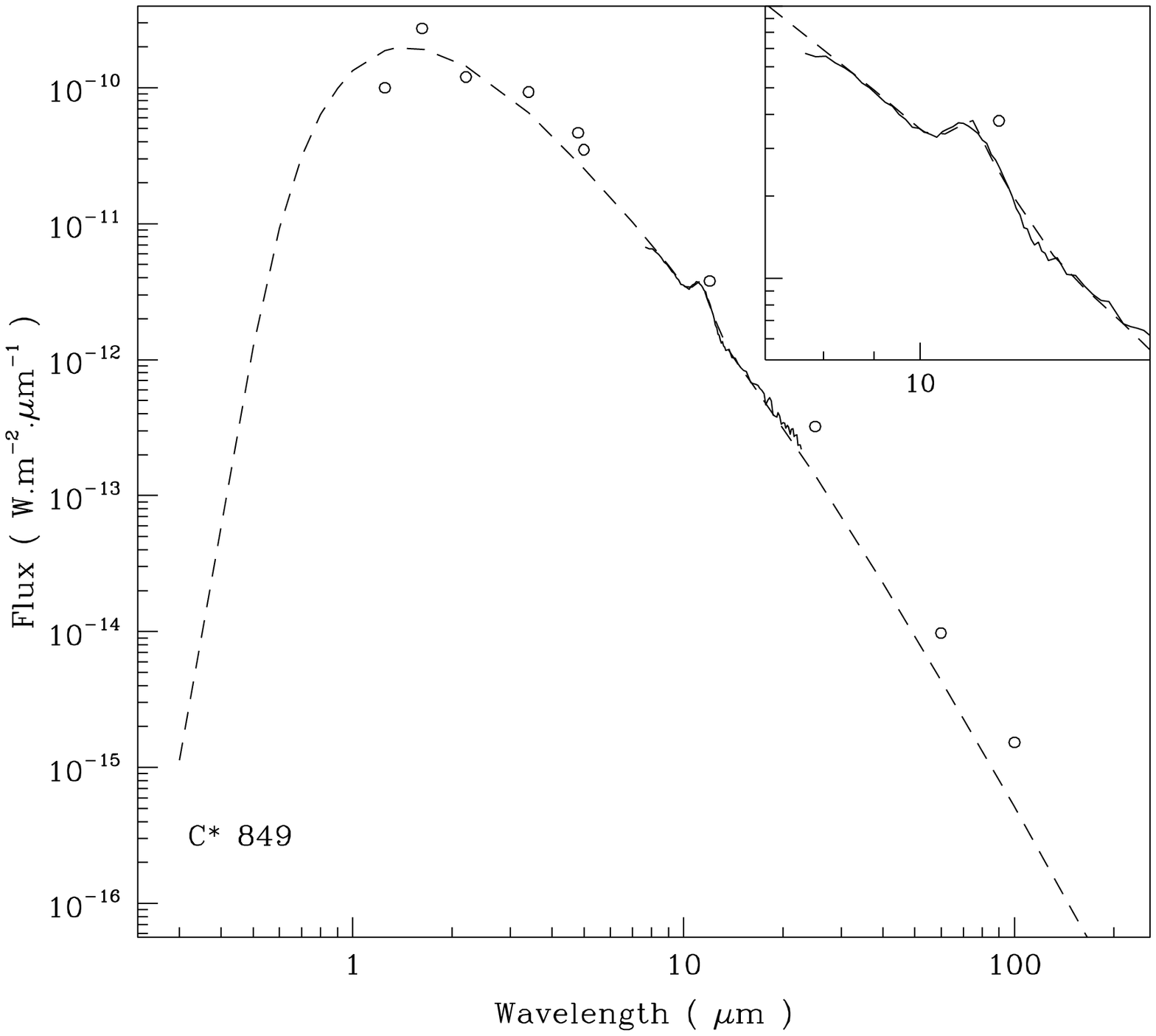}
\end{center}

Fig.~\ref{modnormal}.--- continued
\end{figure}
\newpage
\clearpage
\begin{figure}
\begin{center}
\plottwo{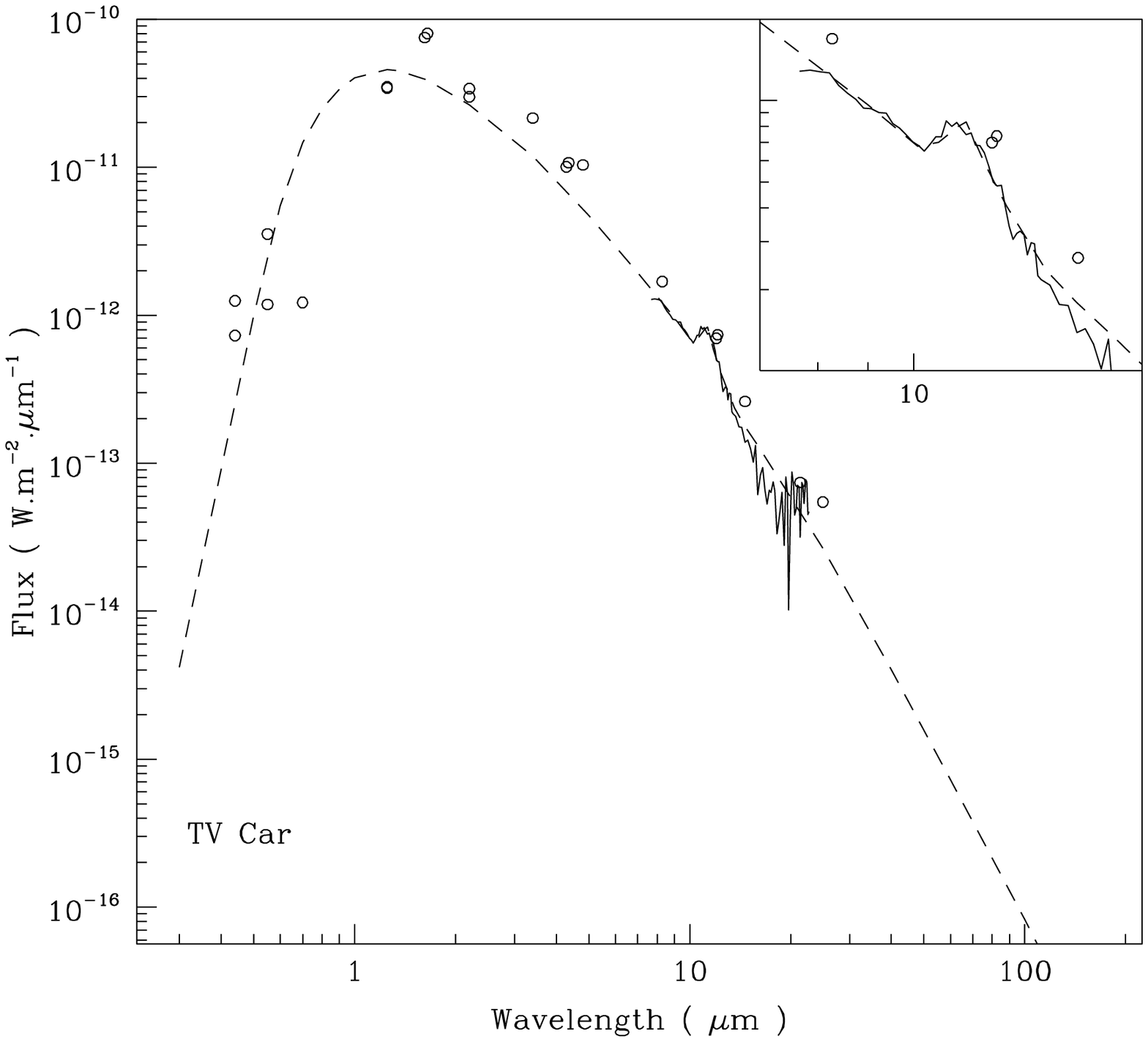}{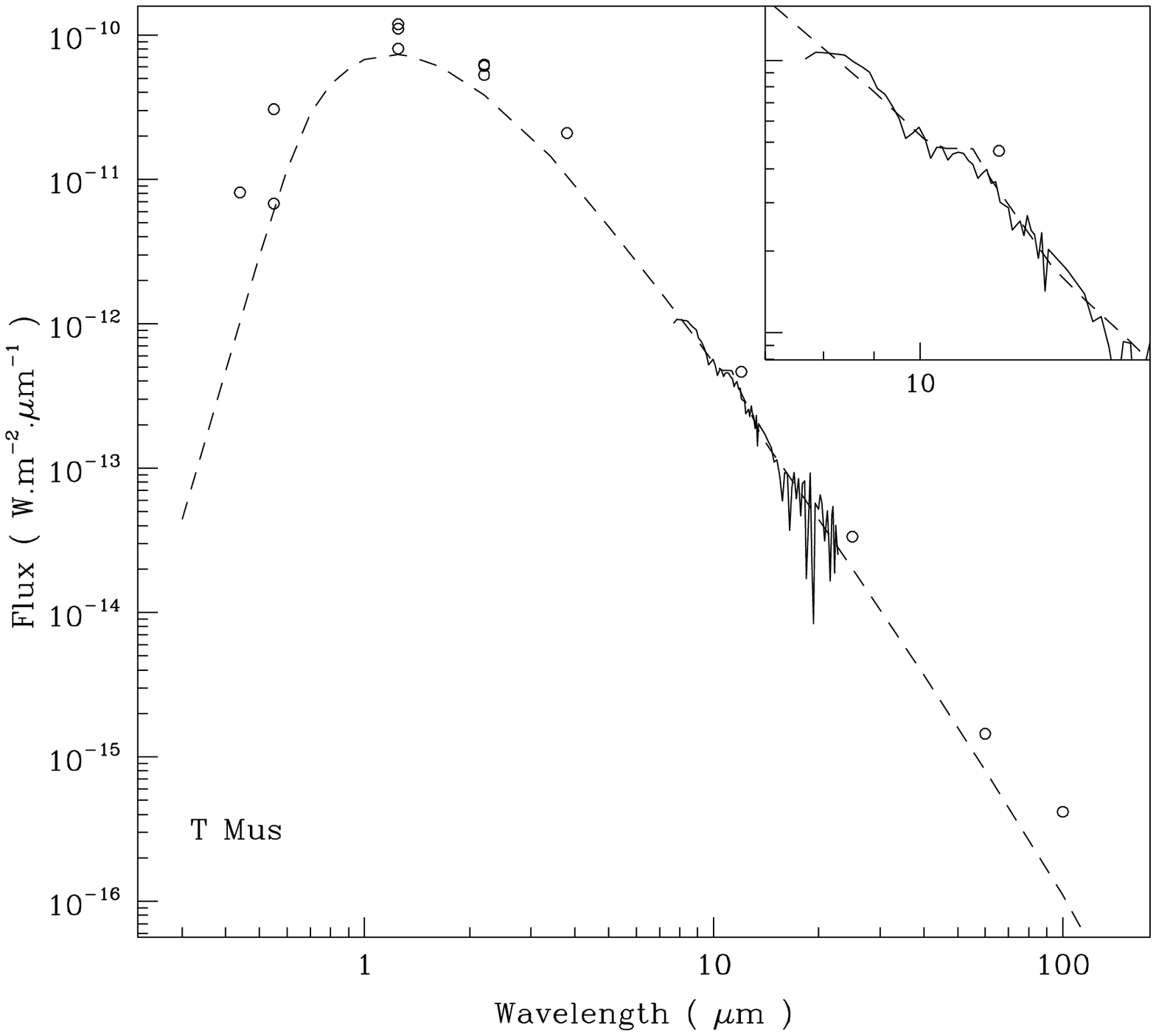}
\plottwo{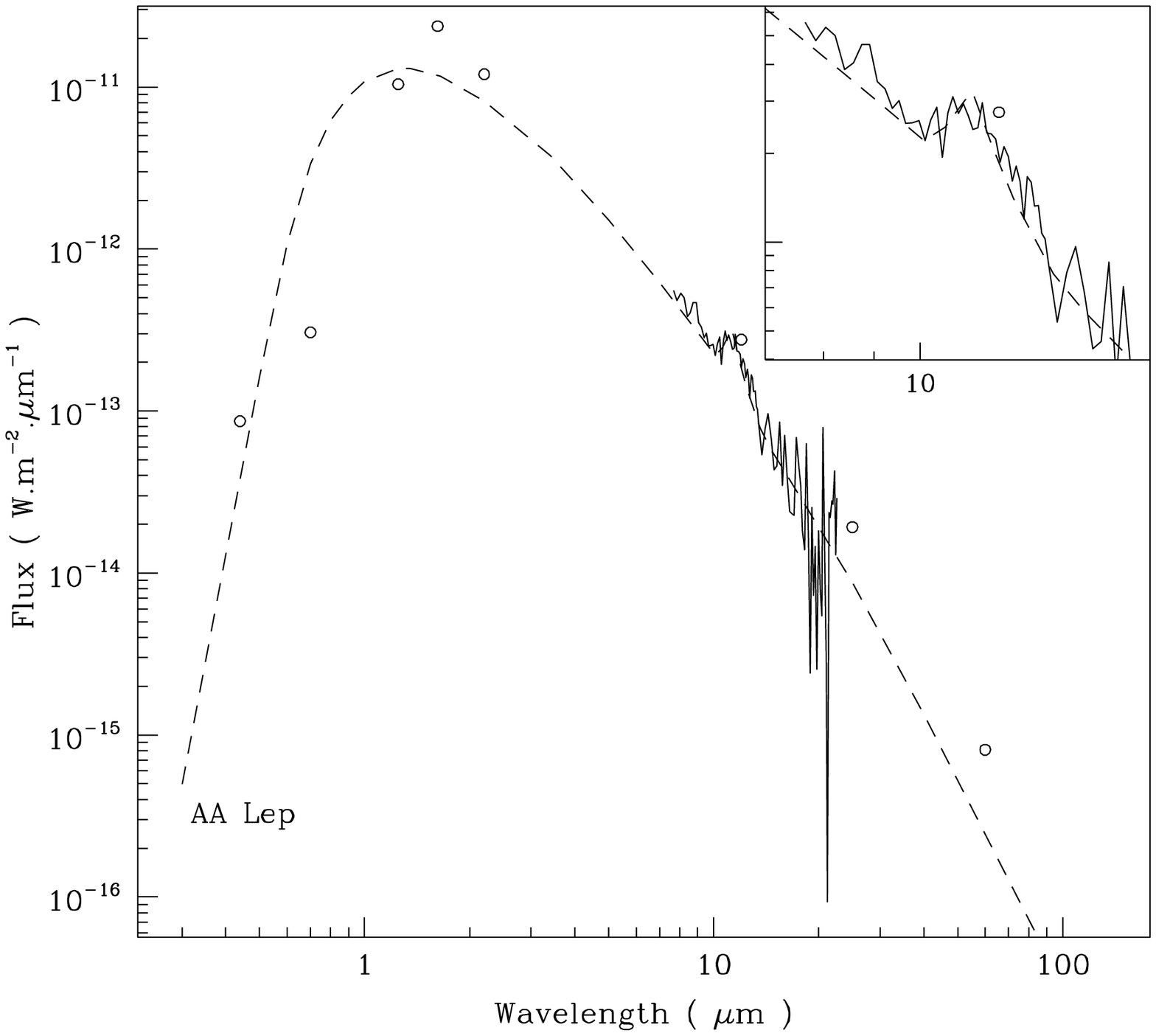}{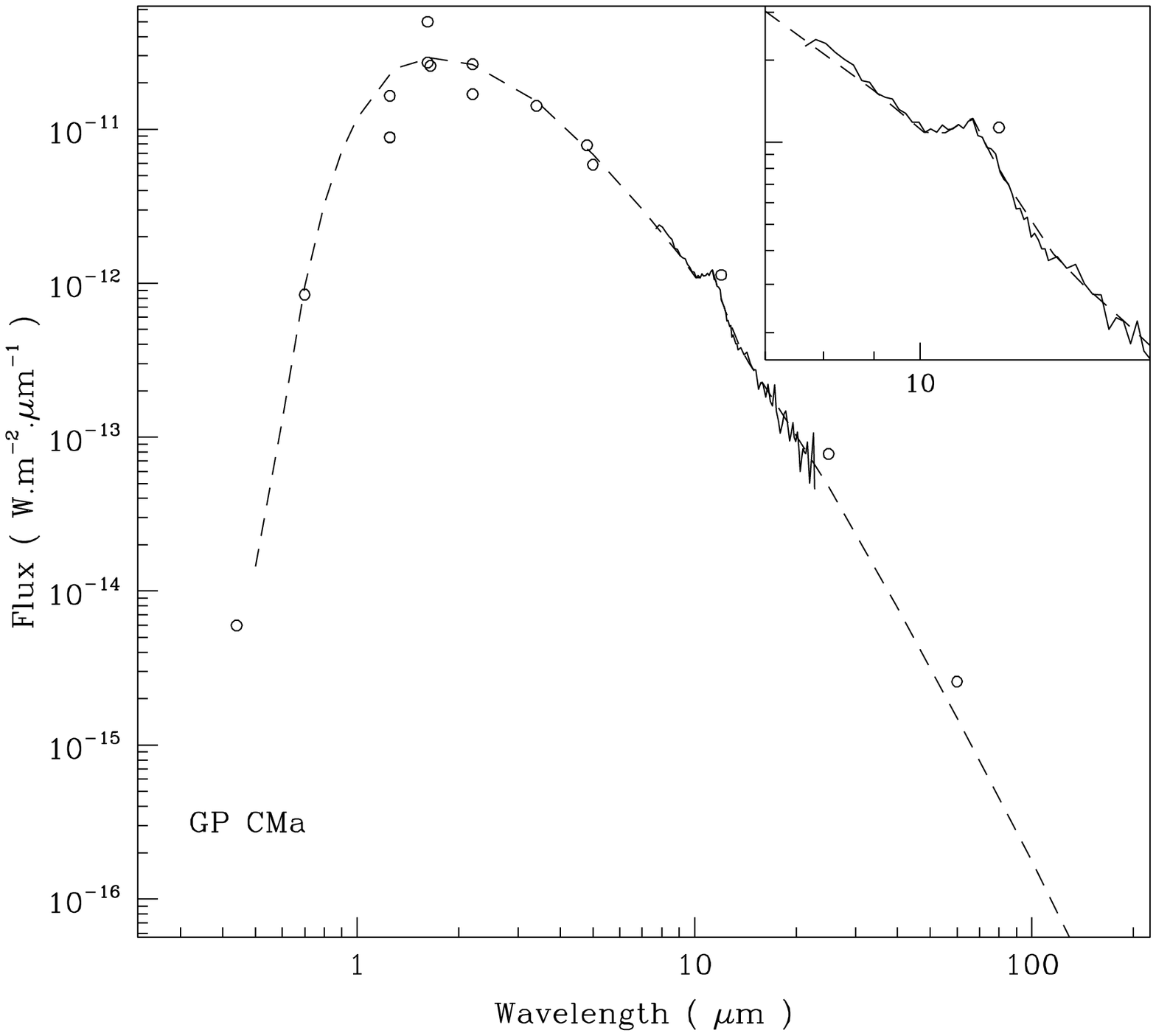}
\plottwo{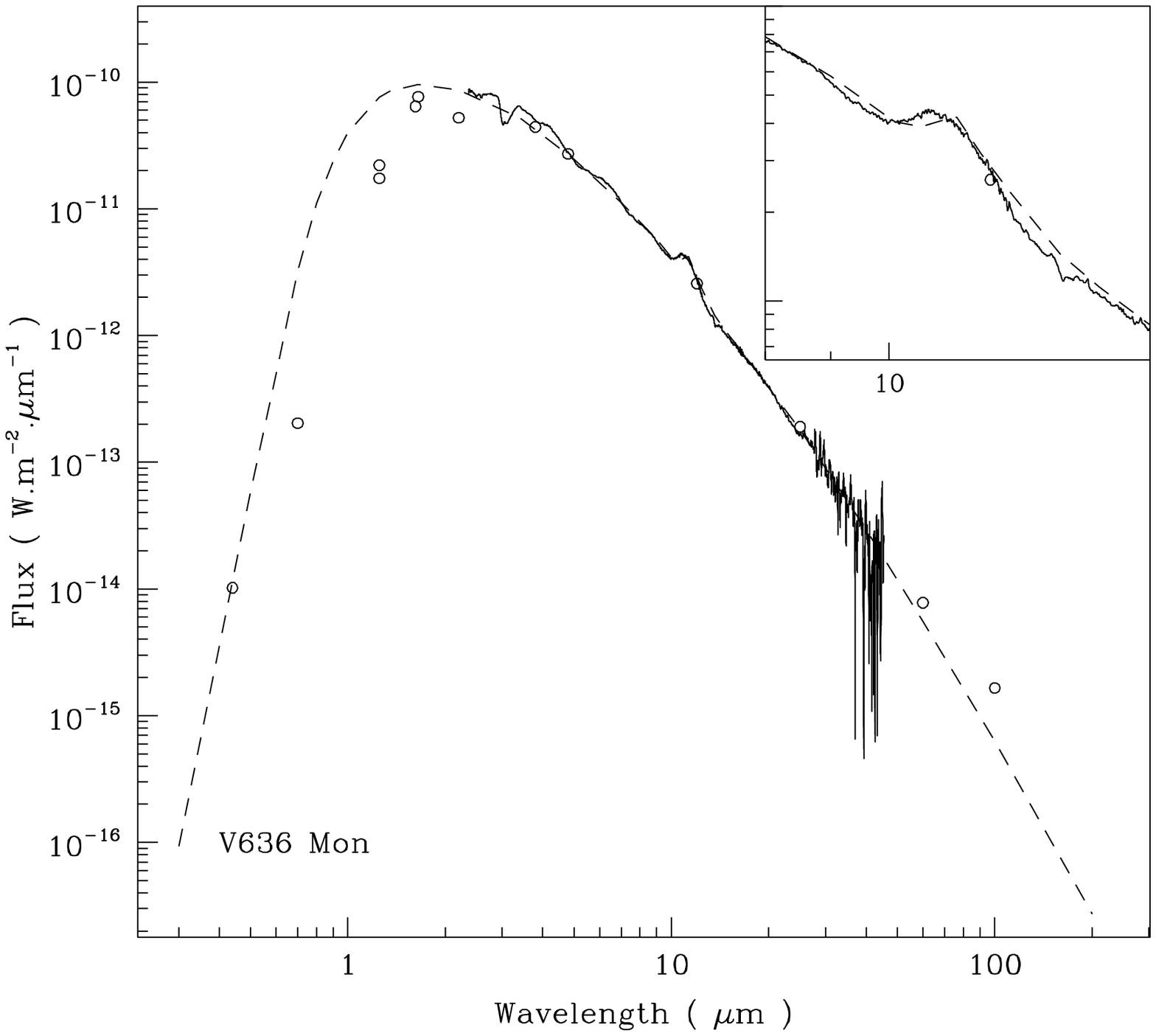}{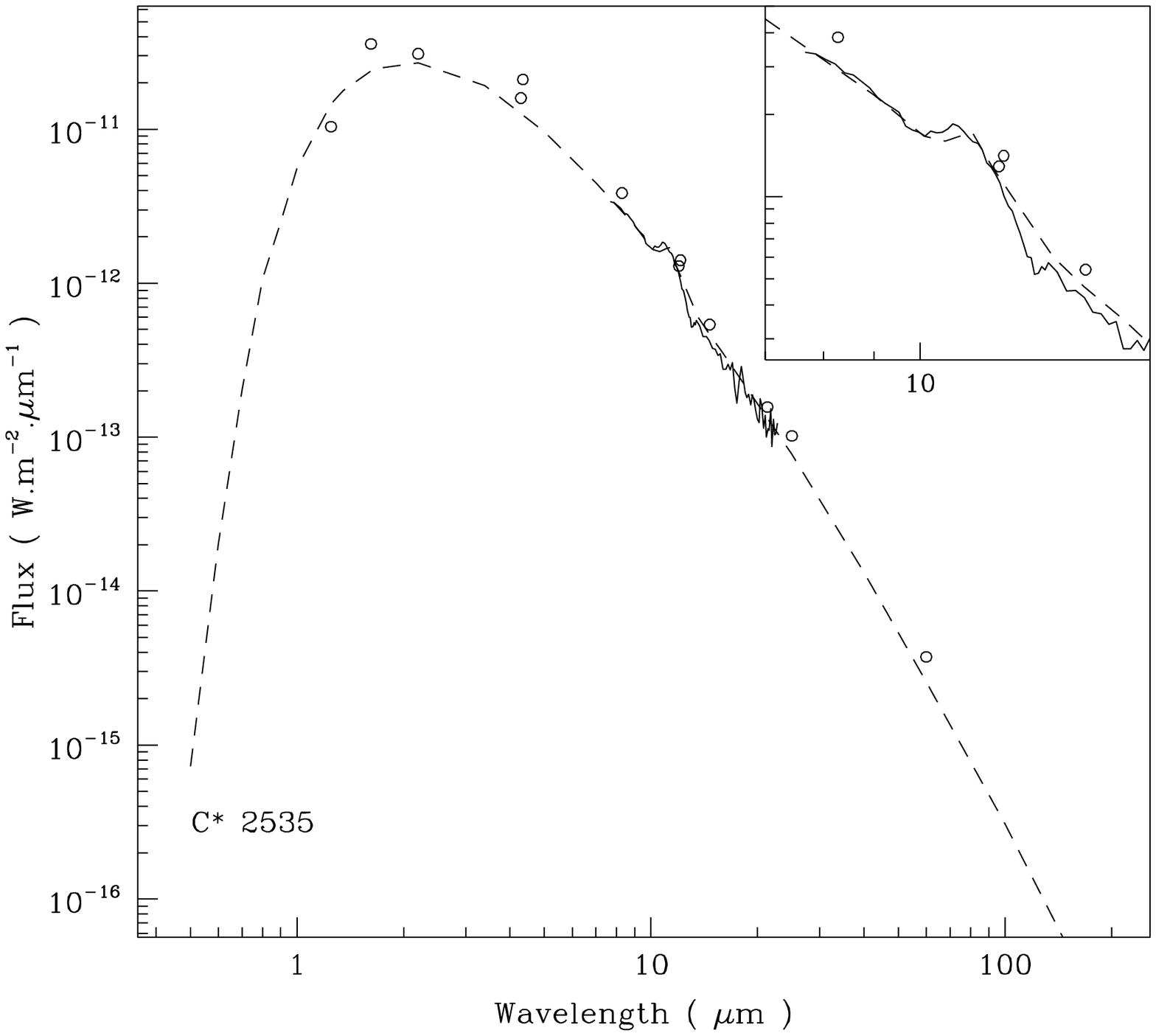} 
\epsscale{0.37}
\plotone{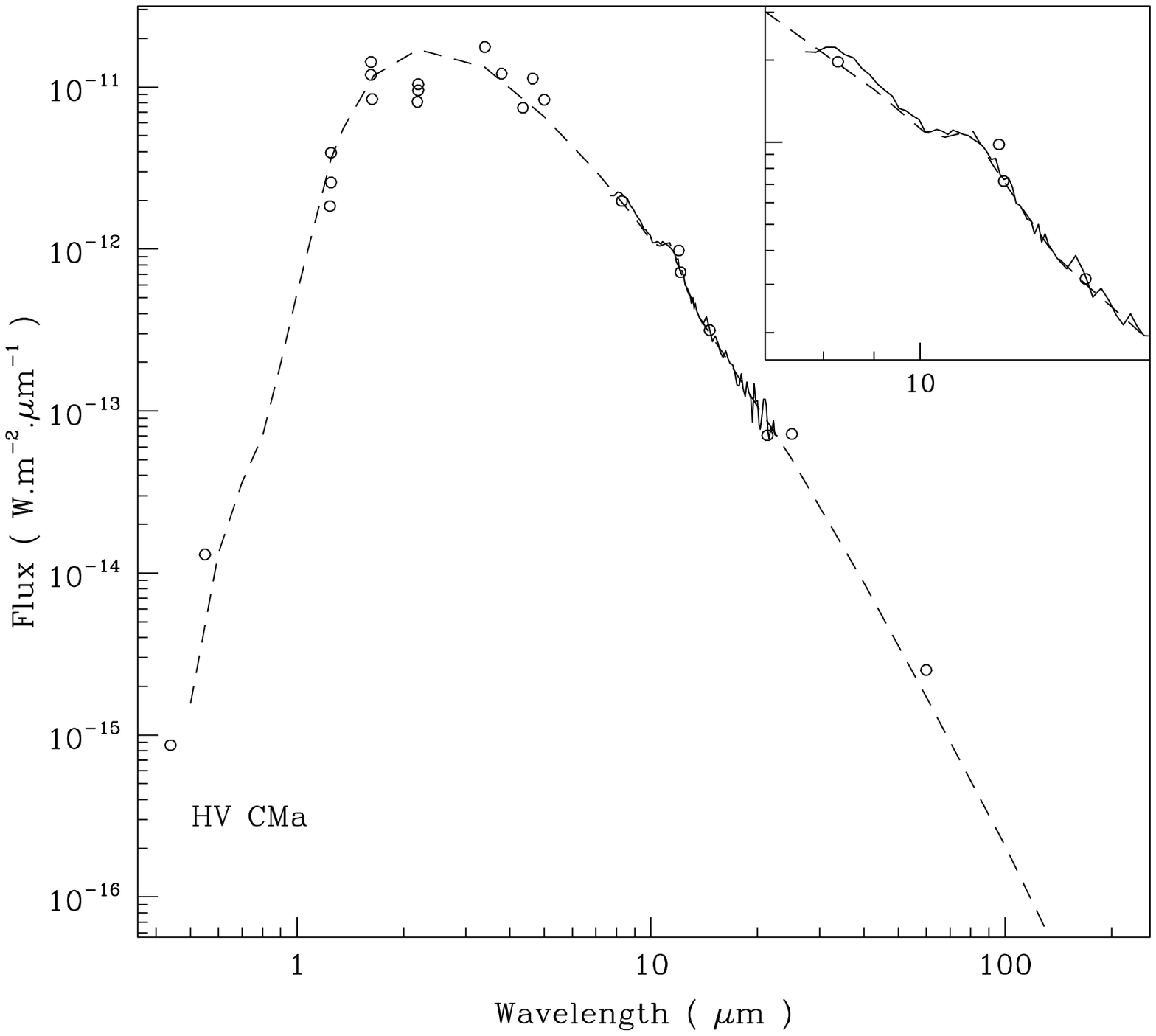}
\end{center}

Fig.~\ref{modnormal}.--- continued
\end{figure}
\newpage
\clearpage

\begin{figure}
\epsscale{0.8}
\begin{center}
\plottwo{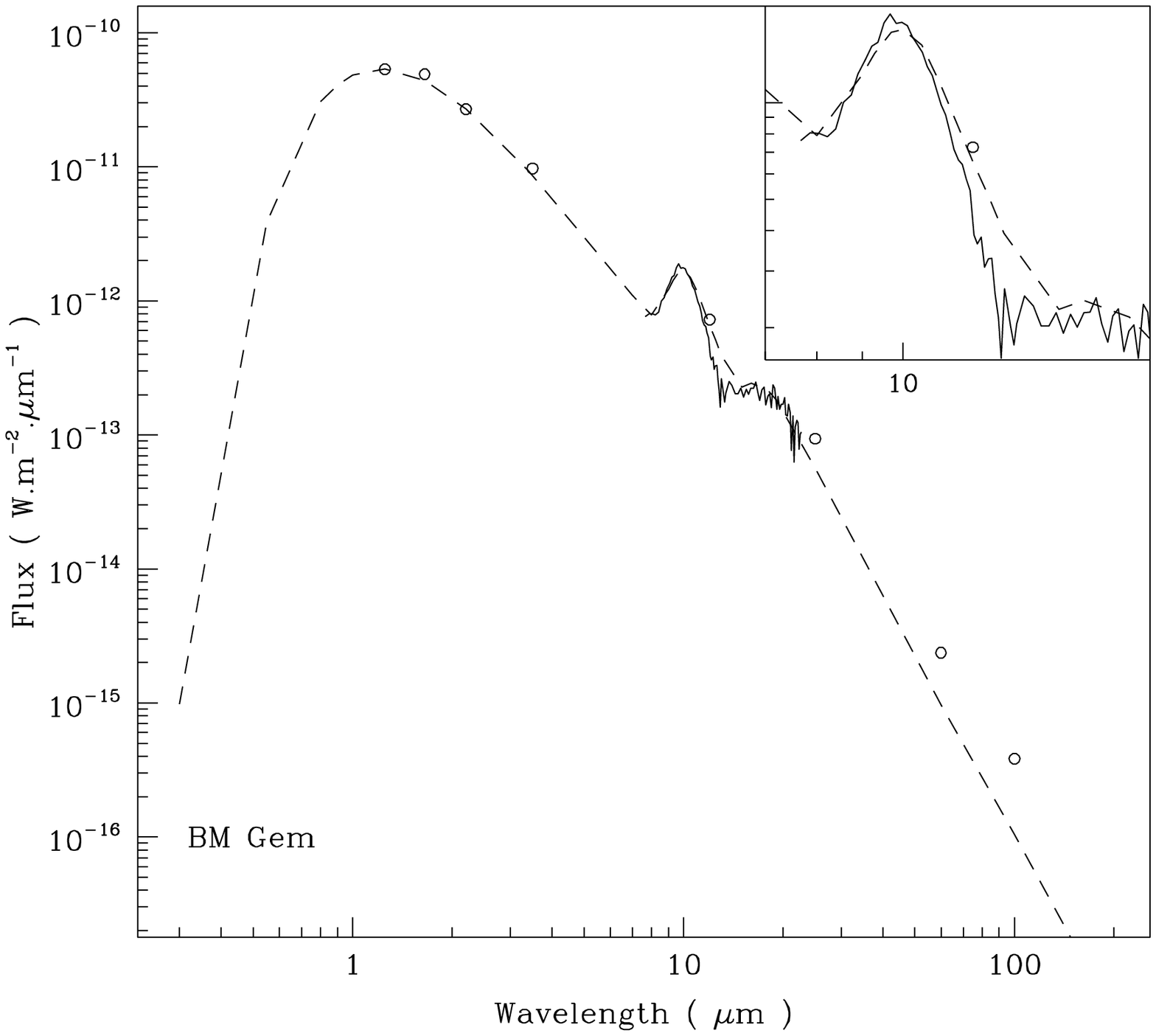}{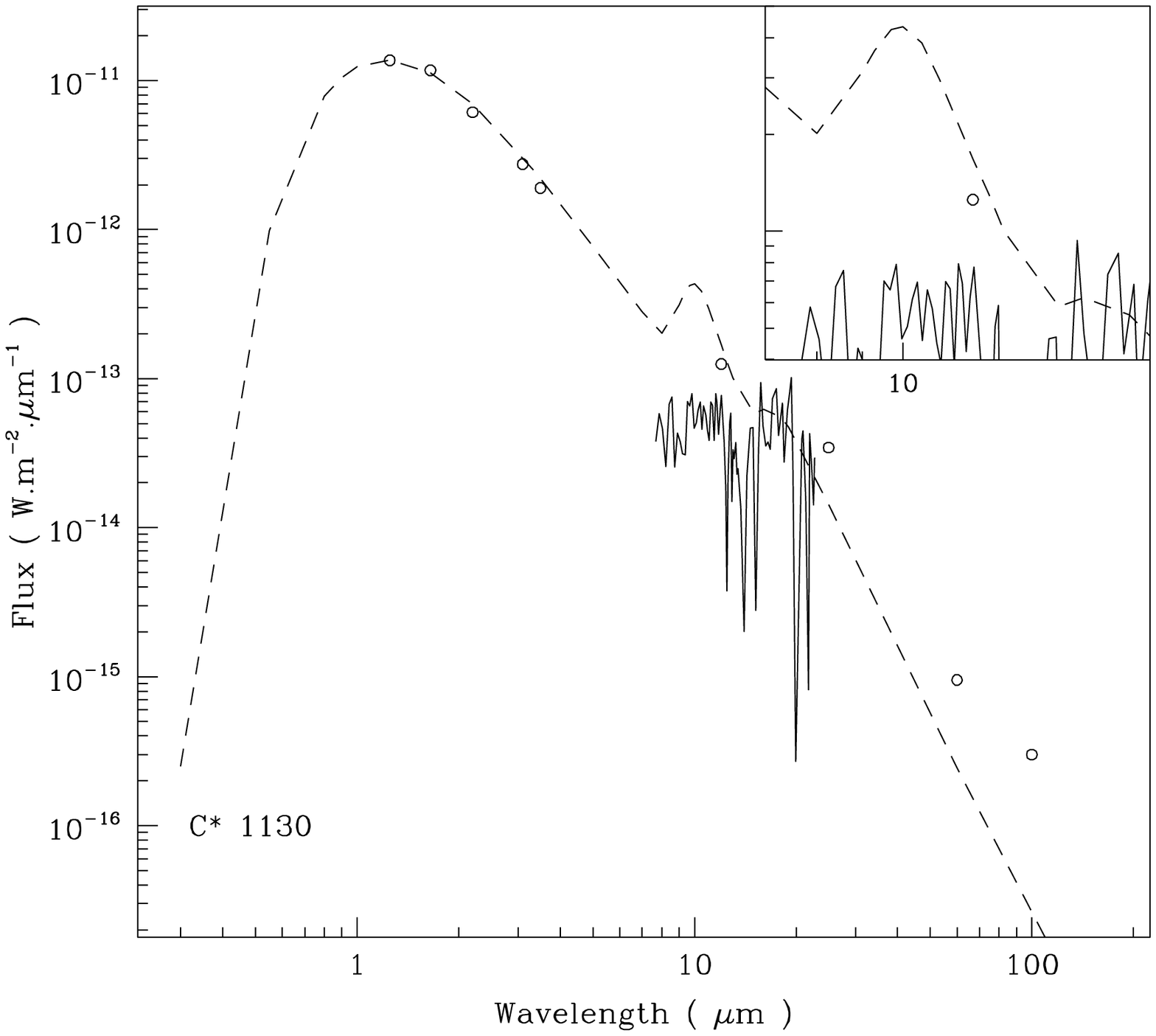} 
\plottwo{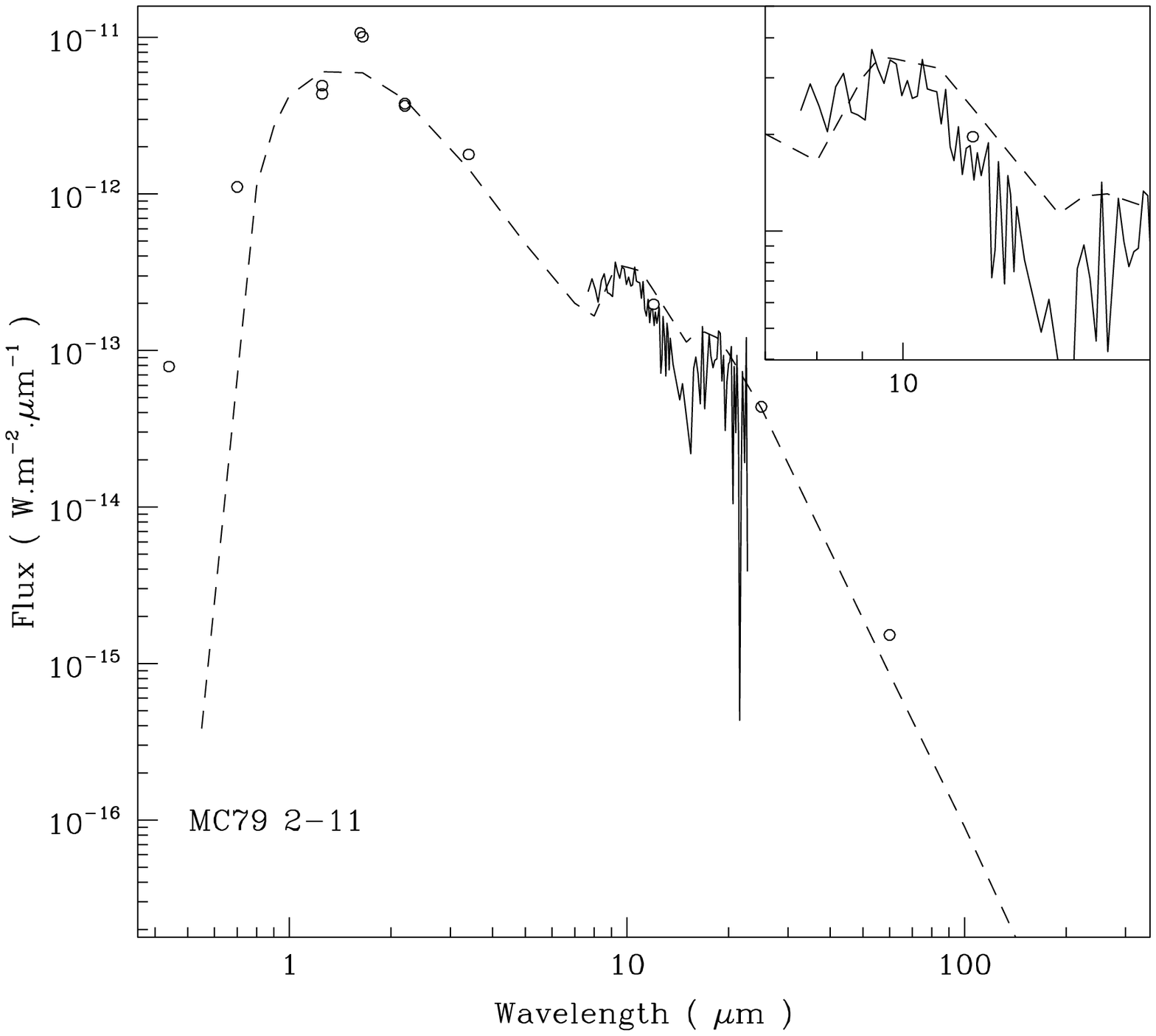}{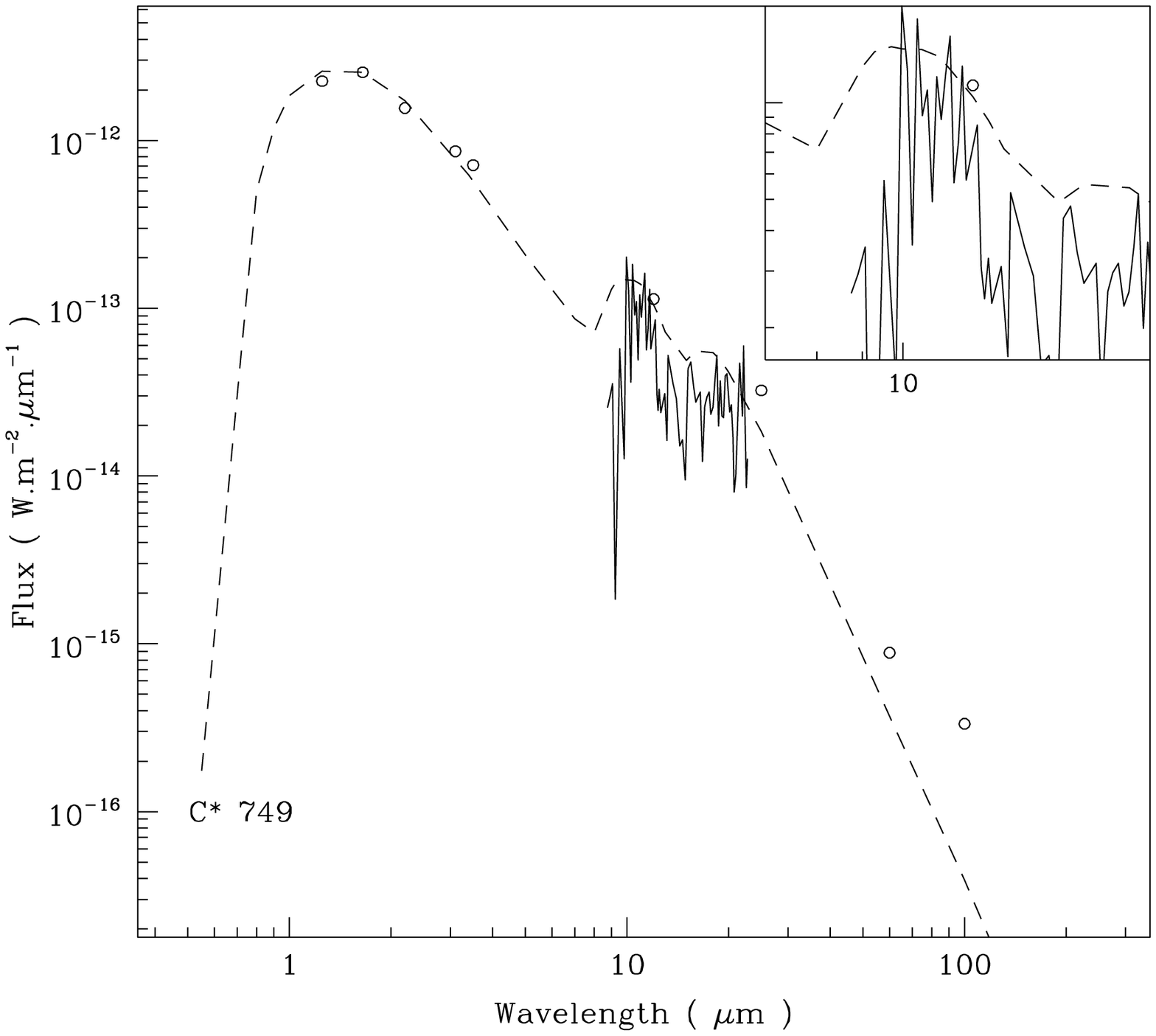}
\epsscale{0.37}
\plotone{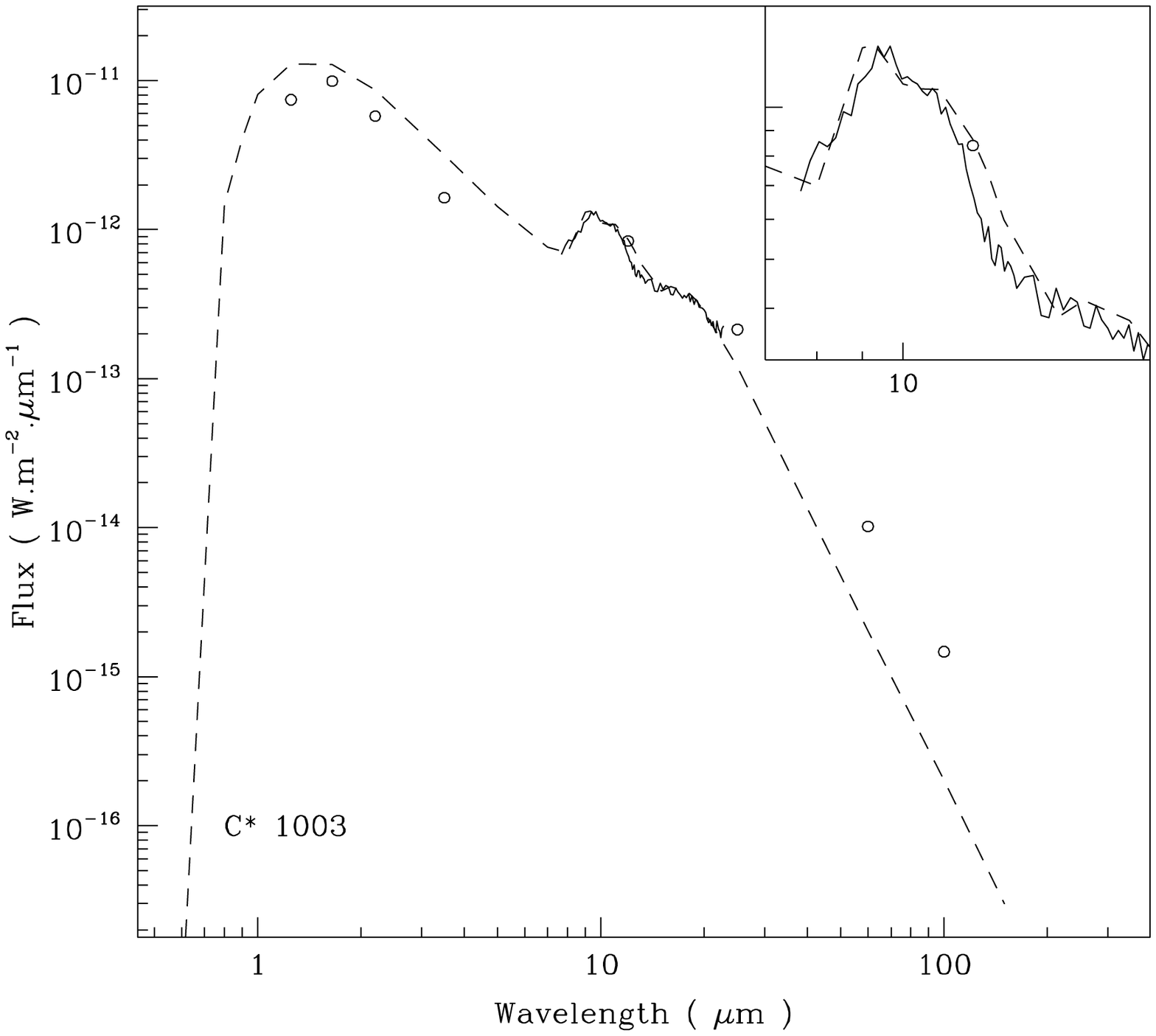}
\caption{Best fit model to the 9.8 $\mu$m feature for the sample} \label{modsil}
\end{center}
\end{figure}

\clearpage
\begin{figure}
\epsscale{0.9}
\begin{center}
\plotone{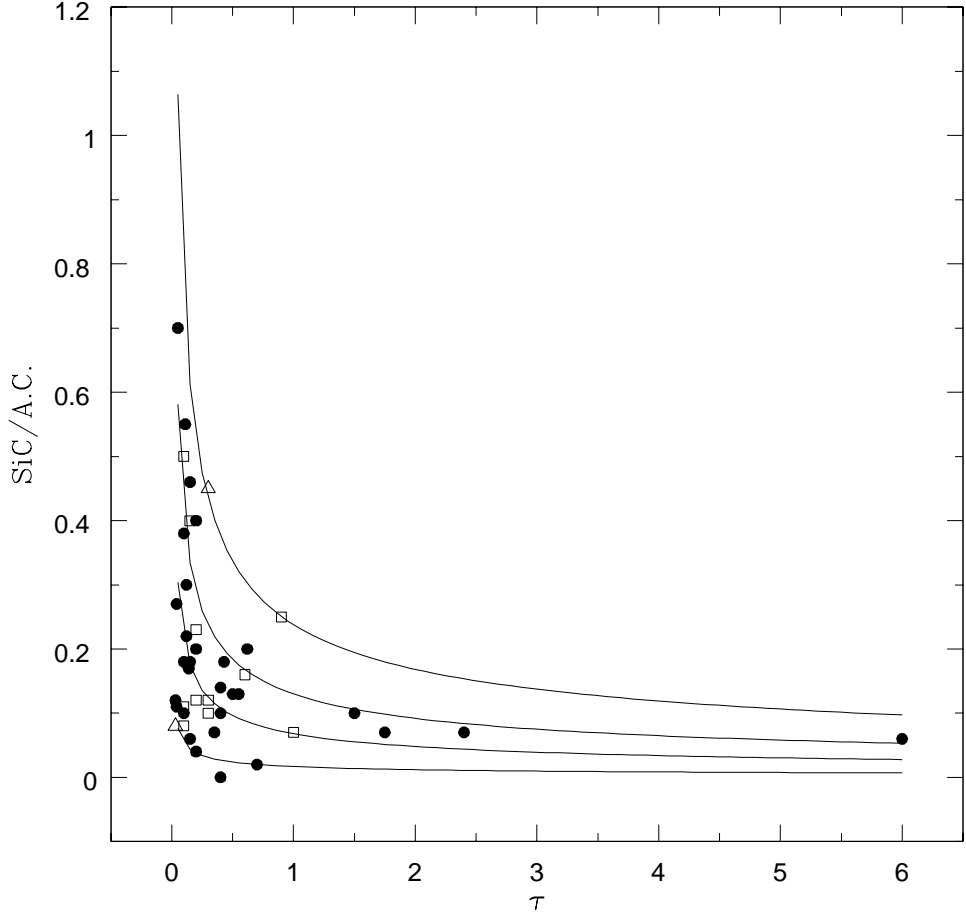}
\caption{The $SiC/A.C. \times \tau$ diagram shows the both parameters calculated through the radiative transfer model for the C-N (close circles), C-R (open triangles) and C-J stars (open squares) of the sample. Four evolutionary sequences are also represented by exponential functions with $\beta = 0.02, 0.07, 0.13$ and $0.24$ associated with $SiC/A.C._{initial} = 0.04, 0.14, 0.26$ and $0.48$, respectively. \label{fig9}}
\end{center}
\end{figure}

\clearpage
\begin{deluxetable}{rrrrr}
\tablecolumns{5}
\tablecaption{Ranges of the average indicator for each level of the \CC-index.\label{tbl-C2}}
\tablewidth{0pt}
\tablehead{\colhead{\CC X} & \colhead{$\bar{p}_{min} < \bar{p} < \bar{p}_{max}$} & \colhead{} & \colhead{\CC X} & \colhead{$\bar{p}_{min} < \bar{p} < \bar{p}_{max}$}}
\startdata
$X = 1$ & $0 < \bar{p} < 0.15$ 		& \nodata & $X = 5$ & $0.75 < \bar{p} < 0.95$ \\ 
$X = 2$ & $0.15 < \bar{p} < 0.35$ & \nodata & $X = 6$ & $0.95 < \bar{p} < 1.15$ \\
$X = 3$ & $0.35 < \bar{p} < 0.55$ & \nodata & $X = 7$ & $1.15 < \bar{p} < 1.35$ \\
$X = 4$ & $0.55 < \bar{p} < 0.75$ & \nodata & $X = 8$ & $1.35 < \bar{p} < 1.55$ \\ 
\enddata
\end{deluxetable}

\begin{deluxetable}{rrrrr}
\tablecolumns{5}
\tablecaption{Ranges of the average indicator for each level of the j-index.\label{tbl-j}}
\tablewidth{0pt}
\tablehead{\colhead{jX} & \colhead{$\bar{q}_{min} < \bar{q} < \bar{q}_{max}$} & \colhead{} & \colhead{jX} & \colhead{$\bar{q}_{min} < \bar{q} < \bar{q}_{max}$}}
\startdata
$X = 0$ & $0 < \bar{q} < 0.15$ 		& \nodata & $X = 4$ & $0.45 < \bar{q} < 0.55$ \\ 
$X = 1$ & $0.15 < \bar{q} < 0.25$ & \nodata & $X = 5$ & $0.55 < \bar{q} < 0.65$ \\
$X = 2$ & $0.25 < \bar{q} < 0.35$ & \nodata & $X = 6$ & $0.65 < \bar{q} < 0.75$ \\
$X = 3$ & $0.35 < \bar{q} < 0.45$ & \nodata & $X = 7$ & $0.75 < \bar{q} < 0.85$ \\ 
\enddata
\end{deluxetable}


\begin{deluxetable}{l|ccc}
\tablecolumns{4}
\tablecaption{Combined inspection for the spectral subclasses.\label{tbl-pop}}
\tablewidth{0pt}
\tablehead{ 																					& Strong 			& Weak/Absent & Unclear}
\startdata
BaII $\lambda$ 4554,4934\AA; SrI $\lambda$ 4607\AA\ 	& C-N or C-H 	& C-R 				& C-R or C-N \\ 
P-branch $\lambda$ 4352\AA\ 													& C-R or C-H 	& C-N 				& C-R or C-N \\
Blue-Violet Extention 																& C-R or C-H 	& C-N 				&   \\
\enddata
\end{deluxetable}


\begin{deluxetable}{ccc}
\tablecolumns{3}
\tablecaption{Ranges of each level of the $\tau$-index. The last column presents the groups from \citet{lml94}. \label{tbl-tau}}
\tablewidth{0pt}
\tablehead{\colhead{$\tau X$} & \colhead{$\tau_{min} < \tau < \tau_{max}$} & Opacity Level}
\startdata
$\tau = 0$ & $\tau \leq 0.10$ 				& Optically Thin Envelope (I)\\ 
$\tau = 1$ & $0.10 < \tau \leq 0.20$ 	& \\ 
\hline
$\tau = 2$ & $0.20 < \tau \leq 3.00$ 	& Intermediate Opacity (II)\\ 
\hline
$\tau = 3$ & $\tau > 3.00$ 						& Optically Thick Envelope (III)\\
\enddata
\end{deluxetable}


\begin{deluxetable}{rrrrr}
\tablecolumns{5}
\tablecaption{Ranges of each level of the SiC-index.\label{tbl-sic}}
\tablewidth{0pt}
\tablehead{\colhead{SiC X} & \colhead{$SiC_{min} < SiC < SiC_{max}$} & \colhead{} & \colhead{SiC X} & \colhead{$SiC_{min} < SiC < SiC_{max}$}}
\startdata
$X = 1$ 	& $SiC \leq 0.10$				 & \nodata & $X = 3$ & $0.20 < SiC \leq 0.30$ \\ 
$X = 2$ 	& $0.10 < SiC \leq 0.20$ & \nodata & $X = 4$ & $SiC > 0.30$ \\
\enddata
\end{deluxetable}


\begin{deluxetable}{rrrrr}
\tablecolumns{5}
\tablecaption{Ranges of each level of the Temperature-index based on the effective temperature of the central star.\label{tbl-temp}}
\tablewidth{0pt}
\tablehead{\colhead{T X} & \colhead{$T_{min} < T < T_{max}$} & \colhead{} & \colhead{T X} & \colhead{$T_{min} < T < T_{max}$}}
\startdata
$X = 3$ 	& $2600 K < T \leq 2800 K$ & \nodata & $X = 5$ & $2200 K < T \leq 2400 K$ \\ 
$X = 4$ 	& $2400 K < T \leq 2600 K$ & \nodata & $X = 6$ & $T \leq 2200 K$ \\
\enddata
\end{deluxetable}

\clearpage
\begin{deluxetable}{rrrrr}
\rotate
\tablecolumns{5}
\tablecaption{Comparison with Other Schemes\label{consist}}
\tablewidth{0pt}
\tablehead{\colhead{} & \colhead{\citep{Yama72,Yama75}}	& \colhead{\citep{Keenan93}} & \colhead{\citep{barn96}} & \colhead{\citep{sloan98}}}
\startdata
Spectral subclasses		&	\nodata		&	2 (4)		&		11 (11)		&	\nodata \\
\CC-index							&	10 (17)		&	2 (4)		&		11 (11)		& \nodata \\
j-index								& 12 (18)		&	1 (1)		&		2 (3)			& \nodata \\
MS-index							& 4 (6)			&	1 (1)		&		2 (3)			& \nodata \\
$\tau$-index					& \nodata		& \nodata	&	\nodata			& \nodata \\
SiC-index							& \nodata		& \nodata	&	\nodata			& 8 (12) \\
Temperature index			&	12 (18)		&	1 (2)		&		6 (10)		& \nodata \\
\enddata
\begin{flushleft}
number of stars with matching classification, followed by the total amount of stars in common with our sample, in parenthesis, for each published scheme
\end{flushleft}
\end{deluxetable}

\begin{deluxetable}{lrrclrlclrrclrl}
\rotate
\tablecolumns{15}
\tablecaption{Results: New Scheme of Classification of C-rich Stars .\label{tbl-final}}
\tablewidth{0pt}
\tablehead{\colhead{Target} & \colhead{} & \colhead{}	& \colhead{} & \colhead{}	& \colhead{} & \colhead{} & \colhead{} & \colhead{Target} & \colhead{} & \colhead{} & \colhead{} & \colhead{} & \colhead{} & \colhead{}}
\startdata
W Pic			&	C-J5	&	\CC6	&	j7&	MS3	&	$\tau$0	&	SiC1	& \nodata	&C* 1467		&	C-J:5	&	\CC4	&	j4&	MS1	&	$\tau$1	&	SiC3\\
Y Tau			&	C-N3	&	\CC4	&		&	MS1	&	$\tau$1	&	SiC2 	& \nodata	&XZ Vel			&	C-N:5	&	\CC5	&		&	MS3	&	$\tau$1	&	SiC3\\
AA Lep		&	C-N:4	&	\CC4	&		&			&	$\tau$2	&	SiC2 	& \nodata	&C* 1705		&	C-H:4	&	\CC4	&		&	MS3	&	$\tau$0	&	SiC4\\
BN Mon		&	C- 4	&				&		&			&	$\tau$1	&	SiC3 	& \nodata	&U Ant			&	C-N3	&	\CC5	&		&	MS1	&	$\tau$0	&	SiC2\\
BL Ori		&	C-N3	&	\CC3	&		&			&	$\tau$0	&	SiC2 	& \nodata	&U Hya			&	C-N3	&	\CC4	&		&	MS3	&	$\tau$2	&	SiC1\\
V636 Mon	&	C-N5	&	\CC5	&		&	MS3	&	$\tau$2	&	SiC1 	& \nodata	&TV Car			&	C-N		&				&		&	MS3	&	$\tau$2	&	SiC2\\
DH Gem		&	C-R2	&	\CC4	&		&			&	$\tau$2:& SiC4:	& \nodata	&TZ Car			&	C-J2	&	\CC6	&	j6&	MS3	&	$\tau$2:&SiC2:\\
CR Gem		&	C-N5	&	\CC3	&		&	MS1	&	$\tau$1	&	SiC2 	& \nodata	&V Hya $\dagger$&C-N:6&\CC5	&		&	MS3	&	$\tau$2	&	SiC1\\
GP CMa		&	C-N6	&	\CC2	&		&	MS3	&	$\tau$2	&	SiC1 	& \nodata	&S Cen			&	C-J2	&	\CC6	&	j7&	MS3	&	$\tau$1:&SiC4:\\
NP Pup		&	C-R3	&	\CC3	&		&			&	$\tau$0	&	SiC1 	& \nodata	&SS Vir			&	C-N:3	&	\CC5	&		&	MS3	&	$\tau$2	&	SiC1\\
CL Mon		&	C-J:6	&	\CC4	&	j4&			&	$\tau$2	&	SiC1 	& \nodata	&C* 2031		&	C-N:6	&	\CC5	&		&	MS3	&	$\tau$2	&	SiC2\\
HV CMa		&	C- 6	&				&		&			&	$\tau$3	&	SiC1 	& \nodata	&RX Cru			&	C-N:6	&	\CC5	&		&	MS3	&	$\tau$1	&	SiC4\\
RY Mon $\dagger$&C-N5&\CC5&		&	MS3	&	$\tau$0	&	SiC2 	& \nodata	&HD 113801	&	C-R:	&	\CC2	&		&	MS1	&					&			\\
W CMa $\dagger$&C-N:3&\CC4&		&			&	$\tau$0	&	SiC3 	& \nodata	&T Mus			&	C-J3	&	\CC7	&	j7&	MS3	&	$\tau$2	&	SiC2\\
BO CMa		&	C-N5	&	\CC5	&		&	MS2	&	$\tau$1	&	SiC4 	& \nodata	&V971 Cen		&	C-J:	&	\CC5	&	j4&	MS3	&					&			\\
BM Gem		&	C-J4	&	\CC5	&	j5&			&	$\tau$2	&	Jpec 	& \nodata	&RV Cen			&	C-N3	&	\CC4	&		&	MS1	&	$\tau$1	&	SiC1\\
BE CMa $\dagger$&C-J2&\CC5&	j5&	MS1	& $\tau$0:& SiC4:	& \nodata	&C* 2208		&	C-J:	&	\CC6	&	j7&	MS3	&					&			\\
C* 749		&	C-J5	&	\CC5	&	j6&	MS2	&	$\tau$2:&	Jpec 	& \nodata	&V Oph			&	C-N:4	&	\CC4	&		&	MS1	&	$\tau$0	&	SiC4\\
C* 849		&	C-H:6	&	\CC4	&		&			&	$\tau$2	&	SiC2 	& \nodata	&SU Sco			&	C-J5	&	\CC5	&	j4&	MS3	&	$\tau$0	&	SiC2\\
C* 846		&	C-N5	&	\CC5	&		&	MS3	&	$\tau$1	&	SiC4 	& \nodata	&V2309 Oph	&	C-N:5	&	\CC5	&		&	MS3	&	$\tau$1	&	SiC2\\
C*  1003	&	C-J4	&	\CC5	&	j4&			&	$\tau$3	&	Jpec 	& \nodata	&TW Oph $\dagger$&C-N:5&\CC5&		&	MS3	&	$\tau$0	&	SiC1\\
FK Pup		&	C-N5	&	\CC4	&		&	MS3	&	$\tau$1	&	SiC2 	& \nodata	&FO Ser			&	C-J3	&	\CC6	&	j7&	MS3	&	$\tau$2:&SiC1:\\
RY Hya $\dagger$&C-N4&\CC4&		&	MS3	&	$\tau$2	&	SiC2 	& \nodata	&C* 2535		&	C-N:6	&	\CC4	&		&	MS3	&	$\tau$2	&	SiC1\\
C* 1130		&	C-J:4	&	\CC5	&	j4&			&	$\tau$2:&	Jpec 	& \nodata	&UW Sgr			&	C-R:*	&	\CC3	&		&			&					& 	  \\
T Cnc			&	C-N5	&	\CC5	&		&	MS3	&	$\tau$1	&	SiC1 	& \nodata	&BI Cap			&	C-R:*	&	\CC4	&		&			&					& 	 \\
MC79 2-11	&	C-J6	&	\CC6	&	j6&			&	$\tau$2:&	JPec 	& \nodata	&						&				&				&		&			&					&		 \\
\enddata
\\
$\dagger$ representative standards of the NSCC
\end{deluxetable}


\end{document}